\documentclass[aps,prl,superscriptaddress,longbibliography]{revtex4-2}

\usepackage{dcolumn}
\usepackage{bm}
\usepackage{xcolor}
\usepackage{amsmath,amsfonts,amssymb}
\usepackage{graphicx}
\usepackage{siunitx}
\usepackage{natmove}
\usepackage{hyperref}
\usepackage[version=4]{mhchem}
\usepackage{multirow}

\DeclareMathOperator{\real}{Re}
\DeclareMathOperator{\imag}{Im}

\newcommand{\pan}[1]{\textbf{#1}.}

\begin{document}

\title{Lattice-tunable substituted iron garnets for low-temperature magnonics}

\author{William Legrand}
\email{william.legrand@neel.cnrs.fr}
\author{Yana Kemna}
\author{Stefan Sch{\"{a}}ren}
\author{Hanchen Wang}
\author{Davit Petrosyan}
\affiliation{Department of Materials, ETH Zurich, H{\"{o}}nggerbergring 64, 8093 Zurich, Switzerland}
\author{Luise Holder}
\author{Richard Schlitz}
\affiliation{Department of Physics, University of Konstanz, 78457 Konstanz, Germany}
\author{Myriam H.~Aguirre}
\affiliation{Department of Condensed Matter Physics, University of Zaragoza, E-50009 Zaragoza, Spain}
\affiliation{Institute of Nanoscience and Materials of Arag{\'{o}}n, UNIZAR-CSIC, E-50018 Zaragoza, Spain}
\affiliation{Laboratory of Advanced Microscopy, University of Zaragoza, E-50018 Zaragoza, Spain}
\author{Michaela Lammel}
\affiliation{Department of Physics, University of Konstanz, 78457 Konstanz, Germany}
\author{Pietro Gambardella}
\email{pietro.gambardella@mat.ethz.ch}
\affiliation{Department of Materials, ETH Zurich, H{\"{o}}nggerbergring 64, 8093 Zurich, Switzerland}

\begin{abstract}
The synthesis of nm-thick epitaxial films of iron garnets by physical vapor deposition has opened up exciting opportunities for the on-chip generation and processing of microwave signals encoded in magnons. However, iron garnet thin films suffer from demanding lattice-matching and stoichiometry requirements. Here a new approach to their synthesis is developed, enabling a precise and continuous tuning of iron garnet compositions based on the co-sputtering of binary oxides. By substituting a controlled proportion of iron with additional yttrium, \ce{Y_{3}(Y_{x}Fe_{5-x})O_{12}} films of high crystalline quality are obtained, combining a widely tunable lattice parameter and excellent magnetization dynamics. This enables iron garnet thin films suited for cryogenic applications, which have long remained impractical due to microwave losses caused by paramagnetic garnet substrates. Low-temperature ferromagnetic resonance confirms the elimination of substrate paramagnetic losses for \ce{Y_{3}(Y_{x}Fe_{5-x})O_{12}} films lattice-matched to \ce{Y_{3}Sc_{2}Ga_{3}O_{12}} (YSGG), a diamagnetic substrate. The \ce{Y_{3}(Y_{x}Fe_{5-x})O_{12}} system can be matched to other substrates such as \ce{(Gd,Y)_{3}Sc_{2}Ga_{3}O_{12}}. Bi-substituted films of \ce{(Bi_{0.8}Y_{2.2})Fe_{5}O_{12}} also have ideal lattice matching to YSGG, demonstrating the versatility of this approach. This opens unprecedented options for cation substitutions in iron garnet films, offering a promising avenue to new properties and quantum magnonic devices operating in low-temperature environments.
\end{abstract}

\maketitle

\section{Introduction}\label{sec:introduction}

Iron garnets combine unrivaled magnetic and optical properties with remarkably versatile compositions, forming the backbone of a vast area of technology. This class of magnetic insulators notably underpins high-end devices such as microwave circulators, tunable microwave filters and optical isolators. While the properties of their bulk single-crystal and ceramic forms have been comprehensively investigated over the past decades, the synthesis of iron garnets as high-quality epitaxial \si{\nano\meter}-thick films has only recently been demonstrated \cite{Heinrich2011,Sun2012,AllivyKelly2013,Wang2013b,Onbasli2014,Jermain2016,Hauser2016,Beaulieu2018,Dubs2020,Ding2020a,Schmidt2020}. This scaling brings numerous novel opportunities for exploiting the magnetization dynamics of iron garnets, as it enables the interaction of magnons with spin-polarized electronic currents \cite{Kajiwara2010}, connecting with the field of spintronics. It also appears particularly promising for device miniaturization, sparking a surge of interest in the field of magnonics. Of all known magnetic materials, the record lifetimes for the dynamical collective magnetic excitations in yttrium iron garnet (YIG, \ce{Y_{3}Fe_{5}O_{12}}) \cite{Spencer1959}, leading to outstanding propagation distances, qualify it as the system of choice to build magnonic devices. YIG is considered the ideal medium for both room-temperature coherent magnonics \cite{Pirro2021} and quantum magnonics \cite{Yuan2022}, as it offers room for perfect epitaxial growth, magnetic insulating behavior, great structural stability, and tailored magnetic interactions and nonlinearities through substitutions. In particular, highest-quality iron garnet films are required at cryogenic temperatures for realizing hybrid systems with superconducting resonators \cite{Li2019,Hou2019,Guo2023} or with nitrogen-vacancy spins in diamond \cite{Fukami2021}, and to investigate quantum states of magnons \cite{Yuan2022}.

The most limiting constraint affecting the epitaxial growth of iron garnet films is the need to find a suitable substrate with a lattice-matched garnet structure for each desired composition. Gadolinium gallium garnet (GGG, \ce{Gd_{3}Ga_{5}O_{12}}) is the only commercially available substrate whose lattice parameter is close enough to YIG, preventing gradual strain relaxation with a mismatch $<$~\SI{0.1}{\percent}. However, GGG has the major drawback of a strong paramagnetism. At cryogenic temperatures, the magnetic films grown on GGG couple dipolarly and interfacially to the paramagnetic \ce{Gd^{3+}} cations in the substrate, which alters their magnetic properties and considerably deteriorates the magnonic lifetimes. While it has been shown that through liquid phase epitaxy, YIG films grown on GGG with perfect interface separation can be almost free of additional magnetic damping \cite{Will-Cole2023}, the paramagnetism of GGG still causes strong microwave dissipation that prohibits the use of integrated circuitry. Therefore, low-temperature magnonics calls for new film/substrate combinations alternative to YIG/GGG, in order to grow \si{\nano\meter}-thick magnetic insulators with as small magnetic resonance linewidth as possible on non-paramagnetic substrates.

Recently, YIG has been purposely grown on lattice-mismatched yttrium scandium gallium garnet (YSGG, \ce{Y_{3}Sc_{2}Ga_{3}O_{12}}), a commercially available substrate that is nominally free of paramagnetic species, either directly as a substrate \cite{Guo2023} or as an intermediate layer between YIG and GGG \cite{Guo2022}. In particular, this first approach solves the issue of substrate paramagnetism, but it still faces intrinsic limitations arising from the unmatched lattice parameters, with detrimental consequences on the resonance linewidth. Alternative strategies to achieve lattice matching of iron garnets with diamagnetic substrates need to be explored.

The partial replacement of the rare earth \ce{Y^{3+}} and magnetic \ce{Fe^{3+}} with other cations offers a valuable control over the lattice parameter, magnetic and magneto-optical properties of iron garnets. Among the many substituting cations that can occupy the dodecahedral sites of the garnet lattice, \ce{Lu^{3+}} and \ce{Bi^{3+}} maintain long magnonic lifetimes, as they satisfy the shell-filling conditions to minimize the influence of the spin-orbit coupling. Substitution of the \ce{Fe^{3+}} cations in the tetrahedral and octahedral sites by larger or smaller non-magnetic cations of identical charge, typically \ce{Al^{3+}}, \ce{Ga^{3+}}, \ce{Sc^{3+}} and \ce{In^{3+}}, allows for further tuning of the magnetization and lattice parameter of the garnets \cite{Gilleo1958,Gilleo1958a}. In small amounts, these substitutions in the ferrimagnetically ordered tetrahedral and octahedral sub-lattices only induce moderate additional losses for magnetization dynamics.

\begin{figure*}
    \centering
    \includegraphics[clip,width=7in]{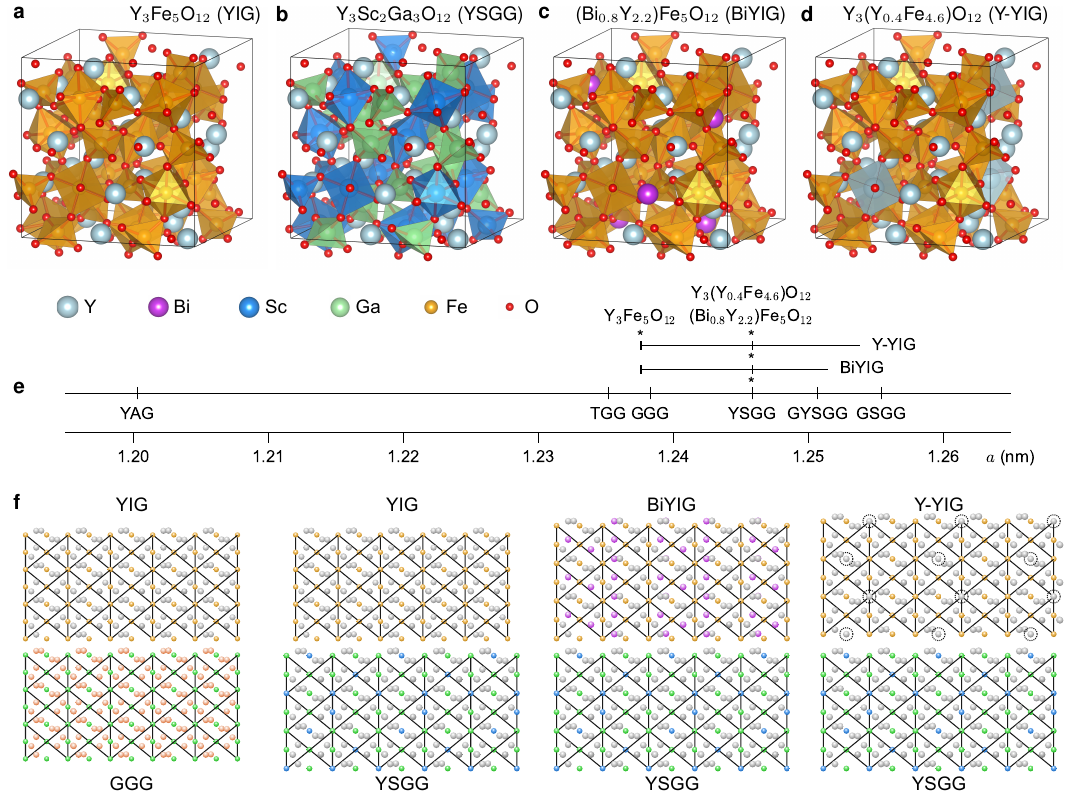}
    \caption{Principle of lattice-matching to non-paramagnetic YSGG. Representations of the unit cell of garnets \pan{a}\ \ce{Y_{3}Fe_{5}O_{12}} (YIG), \pan{b}\ \ce{Y_{3}Sc_{2}Ga_{3}O_{12}} (YSGG), partially substituted \pan{c}\ \ce{(Bi_{x}Y_{3-x})Fe_{5}O_{12}} (BiYIG) with $x\approx0.8$ and \pan{d}\ \ce{Y_{3}(Y_{x}Fe_{5-x})O_{12}} (Y-YIG) with $x\approx0.4$. \pan{e}\ Lattice parameters of different commercial garnet substrates and various iron garnets, with YAG: \ce{Y_{3}Al_{5}O_{12}}, TGG: \ce{Tb_{3}Ga_{5}O_{12}}, GGG: \ce{Gd_{3}Ga_{5}O_{12}}, YSGG, GYSGG: \ce{(Gd_{0.63}Y_{2.37})Sc_{2}Ga_{3}O_{12}}, and GSGG: \ce{Gd_{3}Sc_{2}Ga_{3}O_{12}}. The ranges accessible by \ce{Bi^{3+}} and \ce{Y^{3+}} substitutions are represented as segments. Stars identify the compounds represented above. \pan{f}\ Lattice-matching of (111)-oriented garnet substrates and films (lines represent the edges of the cubic unit cells, and only the cations located on the faces of each unit cell are represented). YIG is matched to paramagnetic GGG, but not to diamagnetic YSGG. BiYIG and Y-YIG are matched to YSGG at specific partial substitution parameters.}
    \label{fig:principle}
\end{figure*}

Previous investigations have shown that substitution with \ce{Bi^{3+}} cations in dodecahedral sites expands the lattice of iron garnets, by up to at least $\approx$~\SI{0.7}{\percent} \cite{Hansen1983} while maintaining a small resonance linewidth \cite{Vittoria1985,Tamada1988}. This can also be exploited to adjust the sign and strength of the magnetic anisotropy, through engineering of the epitaxial strain \cite{Soumah2018}. More recently, a substitution of \ce{Fe^{3+}} by excess \ce{Y^{3+}} cations has been obtained in intentionally off-stoichiometric iron garnet films grown with a chemical composition of the \ce{YFeO_{3}} orthoferrite \cite{Su2021}. The non-magnetic, filled electronic shells of \ce{Y^{3+}} suggest that excess yttrium would also not lead to a strong increase in resonance linewidth compared to unsubstituted YIG \cite{Santiso2023}. These two types of substitution thus appear adequate for adjusting the lattice parameter of iron garnets (as represented schematically in \textbf{Figure~\ref{fig:principle}}) while maintaining long magnonic lifetimes, potentially solving complications pertaining to the choice of the substrate. 

All of these approaches require a precise control over the stoichiometry of the iron garnet films, which is challenging for physical vapor deposition of thin film oxide materials in general. With magnetron sputtering in particular, the stoichiometry of the deposited oxide is the result of indirect factors related to plasma processes \cite{Yang2018d}, such as the position of the substrate relative to the center of the ignited target, gas pressure, plasma power, etc. In turn, this often requires a lengthy optimization of the deposition parameters, or to find a precise off-stoichiometric target composition, in order to compensate for the stoichiometry change from the target to the films.

Here we demonstrate an innovative approach to control the stoichiometry of garnet films, based on the co-sputtering of binary oxides. We apply this approach to the investigation of a wide range of compositions in \ce{Y_{3}(Y_{x}Fe_{5-x})O_{12}} deposited on GGG and YSGG. We show that these films have excellent magnetic resonance properties at room temperature, comparable to unsubstituted YIG grown in similar conditions. As this substitution increases the lattice parameter, an off-stoichiometric \ce{Y_{3}(Y_{x}Fe_{5-x})O_{12}} can be epitaxially matched to substrates with larger lattice parameters than GGG, in particular, diamagnetic YSGG. Comparing films grown on GGG and YSGG, we demonstrate that lattice-matched \ce{Y_{3}(Y_{x}Fe_{5-x})O_{12}} deposited on YSGG eliminates the issue of substrate paramagnetism. As an alternative, we also demonstrate the suitability of lattice-matched Bi-substituted YIG on YSGG to reduce low-temperature magnetic resonance losses. These different substrate/film combinations offer a very promising path toward ultrathin iron garnet films maintaining small magnetization dynamic losses at low temperatures, while the approach of co-sputtering binary oxides to precisely control film stoichiometries is expected to apply to other systems beyond garnets. 

\section{Results}

\subsection{Principle of the oxide co-sputtering approach}
The technique of off-axis radio-frequency (rf) magnetron sputtering, relying on a single target with a garnet formula composition, has been demonstrated in prior works to provide single-crystalline films of iron garnets with consistently high quality \cite{Wang2013b,Wang2014b,Gallagher2016,Yang2018d,Ding2020a}. Beyond this established approach, controlling the balance between deposition rates of different targets in a co-sputtering approach would provide a much more straightforward control of the deposited film stoichiometry. We exemplify this here with rf co-deposition from \ce{Y2O3} and \ce{Fe2O3} targets. In addition to greatly simplifying the optimization of the growth process, this approach offers a direct and continuous way of tuning the composition within a substitutional system.

We first briefly describe our technique to produce oxide films with rf co-sputtering. The iron garnet films investigated in this work are deposited at elevated temperatures in an rf-magnetron sputtering tool with base pressure better than \SI{4e-6}{\pascal}. The deposition atmosphere is a mixture of \ce{Ar} and \ce{O2} gases, where the partial pressure of \ce{O2} is adjusted to compensate for the oxygen depletion in the deposited films \cite{Dorsey1993,Kubota2013,Wang2013b,Yang2018d}. Two plasmas are ignited from two separate 2-inch magnetrons mounted respectively with a \ce{Y2O3} and a \ce{Fe2O3} target, and controlled individually by two rf sources. The rotating planetary stage holding the samples faces the midpoint between the targets, in an off-axis configuration with offset angle of \SI{55}{deg} \cite{Yang2018d} for each target. From different optimization series, a window of best growth conditions is obtained around \SI{1.0}{sccm} and \SI{100}{sccm} of \ce{O2} and {Ar} gas flows, a deposition pressure of \SI{0.5}{\pascal}, and \SI{3}{\watt\per\centi\meter\squared} and \SI{1.5}{\watt\per\centi\meter\squared} of respective powers applied to the \ce{Y2O3} and \ce{Fe2O3} targets. For these parameters, the combined deposition rate of the garnet is approximately \SI{0.25}{\nano\meter\per\minute}.

We are not aware of previous works having resorted to targets of binary oxides for the epitaxy of garnet thin films with magnetron sputtering. The approach followed here is motivated by prior efforts in obtaining epitaxial YIG films from nanolaminates of \ce{Fe2O3} and \ce{Y2O3} layers deposited in stoichiometric quantities by atomic layer deposition \cite{Lammel2022}. To guide the search for stoichiometric conditions, the thickness deposition rates in \si{\nano\meter\per\second} of each target can be converted into film compositions. To achieve 5 \ce{Fe2O3} + 3 \ce{Y2O3}, considering their respective mass densities and molar masses, a ratio of thickness deposition rates of $r_{\ce{Fe2O3}}$/$r_{\ce{Y2O3}}\approx$ 1.13 is desirable (see Experimental Section). 

\subsection{Structural characterization}
The crystalline quality of off-stoichiometric YIG with excess yttrium (Y-YIG in the following) is explored for varied amounts of substitution in a series of films with a thickness of about \SI{30}{\nano\meter}, deposited on GGG and YSGG substrates. The substitution parameter $x$ in \ce{Y_{3+x}Fe_{5-x}O_{12}} can be estimated from the ratio of deposition rates $y=r_{\ce{Fe2O3}}/r_{\ce{Y2O3}}$ following $x/5=\,(1.13/y-1)/(1.13/y+5/3)$, see Experimental Section. In off-stoichiometric conditions with $x>0$ and below some limit amount of substitution, the additional \ce{Y^{3+}} cations are expected to fit in the garnet lattice and to preferentially replace cations from the octahedral sites, as has been observed in bulk compounds, with the formation of single crystals of \ce{Y_{3}(Y_{x}Ga_{5-x})O_{12}} with $x$ up to 0.7 \cite{Geller1967}.

\begin{figure*}
    \centering
    \includegraphics[clip,width=6.99in]{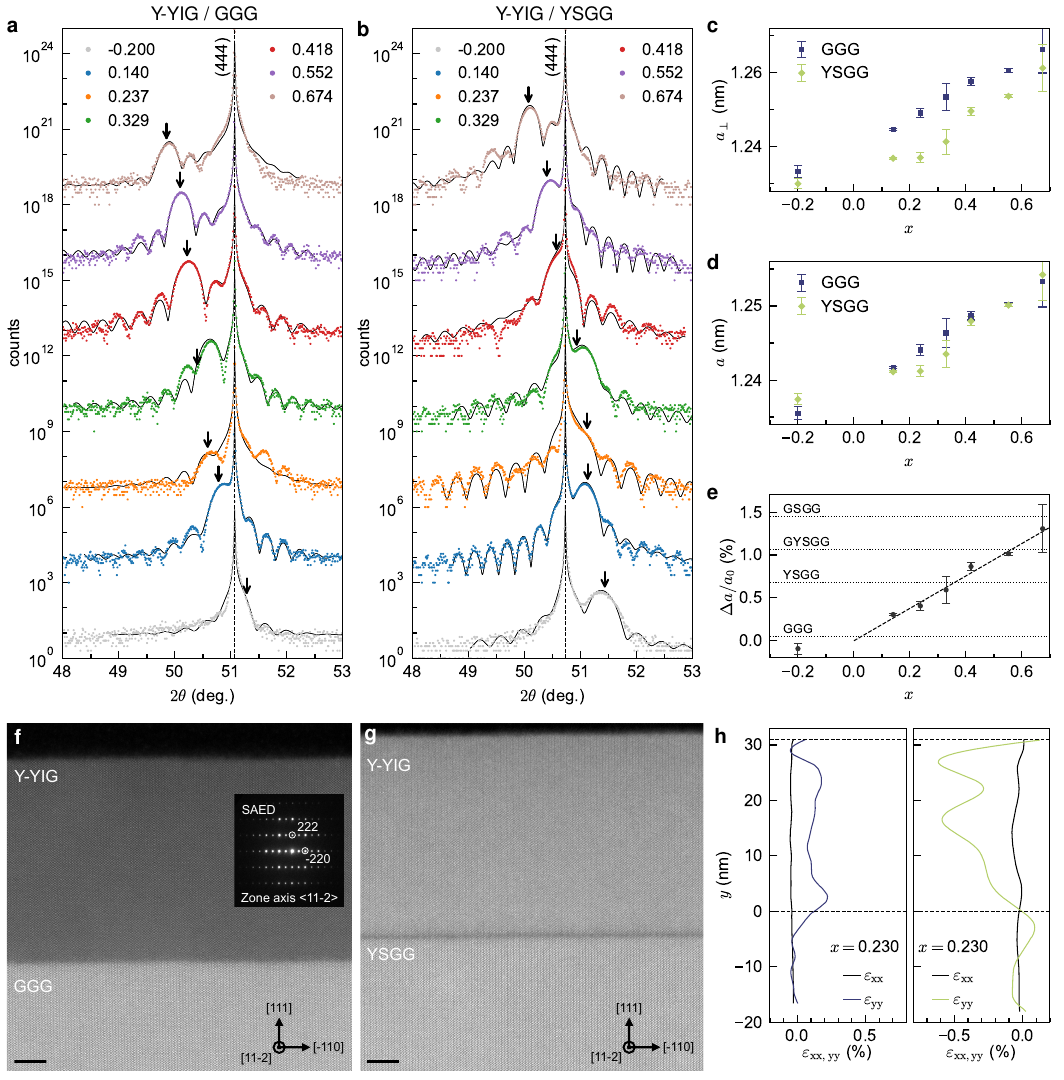}
    \caption{Structural characterization of Y-YIG epitaxial films. High-resolution XRD for \SI{30}{\nano\meter}-thick \ce{Y_{3+x}Fe_{5-x}O_{12}} films, labeled Y-YIG-1--7 in Table \ref{tab:samples}, with $2\theta-\omega$ diffractograms of the (444) peaks for films on \pan{a}\ GGG and \pan{b}\ YSGG substrates. The legend displays the estimated substitution $x$ for each composition. Black lines are fits to the data using an enhanced kinematical diffraction model (see Experimental Section). The dashed vertical lines locate the (444) diffraction peak of the substrate, the arrows indicate the film peak position as deduced from the fit, which may be offset from the data local maximum due to interference effects (see Experimental Section). \pan{c}\ Out-of-plane lattice parameter $a_{\perp}$ extracted from fits in panels a and b, \pan{d}\ corresponding lattice parameter of the unstrained film $a$, and \pan{e}\ relative change $\Delta{}a/a_0$ in lattice parameter of the unstrained film compared to YIG, obtained by combining GGG and YSGG values, as a function of estimated $x$. The dashed line corresponds to the expected evolution of $a$ in bulk single crystals, the dotted horizontal lines indicate the required change of $a$ for the films to be matched with some specific substrates. Scanning transmission electron microscopy in high angular annular dark-field mode images for a sample nominally identical to Y-YIG-3, on \pan{f}\ GGG and \pan{g}\ YSGG substrate. Scale bar is \SI{5}{\nano\meter}. Inset in f shows the diffraction pattern in selected area diffraction (SAED) on Y-YIG on GGG, confirming the zone axis orientation. \pan{h}\  Profiles along $y=[111]$ of the strain tensor components $\varepsilon_{\rm{xx,yy}}$ extracted from geometric phase analysis of the images in f and g, on GGG (left, blue) and on YSGG (right, green). Dashed lines locate both bottom and top interfaces of the Y-YIG layer along $x=[-110]$, and $y=0$ corresponds to the substrate/Y-YIG interface.}
    \label{fig:XRD}
\end{figure*}

\textbf{Figure~\ref{fig:XRD}} presents the structural characterization of the single-crystalline films with a range of estimated substitutions $x$ listed in \textbf{Table~\ref{tab:samples}}. In the X-ray diffraction (XRD) symmetrical $2\theta-\omega$ curves of Figure~\ref{fig:XRD}a,b, nearby the (444) diffraction peak of the substrate, the (444) peaks from the epitaxial Y-YIG films are indicated by arrows. These peaks are surrounded by marked Laue oscillations, corresponding to thickness fringes indicative of the high crystalline quality and lateral uniformity of the films. The absence of other crystalline phases than garnet in these samples is confirmed by XRD performed in the full angle range 2--\SI{150}{deg}, and reported in Supplementary Note II. For $x>0$, the film peaks in Figure~\ref{fig:XRD}a are found at smaller diffracting $2\theta$ angles than for unsubstituted YIG and GGG, shifting down to \SI{49.9}{deg} for the largest level of substitution investigated here, $x\approx0.7$. Being a direct measure of the vertical spacing between the (444) diffracting planes, this indicates that the lattice of Y-YIG expands with substitution of \ce{Fe^{3+}} by \ce{Y^{3+}} cations, consistent with their respective ionic radii. This is in line with a previous report on Y-YIG thin films obtained by pulsed laser deposition, where a nominal target composition of \ce{YFeO_{3}} ($x=1$) was used \cite{Su2021}. The XRD (444) film peaks on GGG were found at around $2\theta=$~\SI{49.6}{deg} \cite{Su2021}, consistent with a larger lattice parameter due to an even larger $x$ than in the present series of samples. For samples Y-YIG-1, the substitution is opposite with $x<0$ and the films are on the contrary Fe-rich. Since a combination of \ce{Fe^{3+}} cation substitution in dodecahedral sites and vacancies formation can occur, a different dependence of the lattice parameter on $x$ is expected for Fe-rich films, which is not always appreciably affected by the additional Fe \cite{Noun2010}.

To verify the type of growth in the present epitaxial films, reciprocal space maps have been recorded for Y-YIG-5 on GGG and YSGG, representative of the whole series, as reported in Supplementary Note III. These exhibit (624) diffraction peaks of substrate and film showing an equal in-plane component of the reciprocal vector. This demonstrates that the present Y-rich YIG films grow in a fully-strained state, free of structural relaxation despite the significant lattice mismatch and strain observed in some of them. Therefore, these epitaxial films undergo rhombohedral distortion, with their unit cell expanding along the body-diagonal aligned with the [111] axis, through a reduction of the rhombic base angle to accommodate for the lattice mismatch.

The out-of-plane lattice parameter $a_{\perp}$ of each film, incorporating this rhombohedral distortion, is deduced from a fit to the XRD curve (see Experimental Section), as shown in Figure~\ref{fig:XRD}c and listed in Table \ref{tab:samples}. Since the present Y-YIG films are fully strained, it is possible to deduce for each of them the lattice parameter $a$ of the cubic unit cell that would form in unstrained films of identical composition. Owing to the unknown Poisson coefficient of Y-YIG, the coefficient of unsubstituted YIG has been employed in this procedure (see Experimental Section). The resulting values of $a$ for unstrained films, consistent between films grown on GGG and YSGG, are shown in Figure~\ref{fig:XRD}d and listed in Table \ref{tab:samples}. The deduced values of $a$ are consistent with the off-stoichiometry in Y-YIG. The relative increase in the lattice parameter resulting from the substitution is shown as a function of $x$ in Figure~\ref{fig:XRD}e.

We further characterize the epitaxial quality and strain in the Y-YIG layers, by performing scanning transmission electron microscopy on cross-sectional electron-transparent lamellas, prepared from samples nominally identical to Y-YIG-3 ($x=0.230$, see Supplementary Note IV), grown on both GGG and YSGG. The images displayed in Figure~\ref{fig:XRD}f,g (additional images in Supplementary Note IV) confirm a high-quality, coherent epitaxial growth on either substrate, free of structural relaxation. Strain profiles obtained by geometric phase analysis (details in Experimental Section and Supplementary Note IV) are shown in Figure~\ref{fig:XRD}h. As expected, for this value of $x$ the epitaxial growth results in a positive vertical strain on GGG, as Y-YIG-3 has a larger $a$ than GGG, and conversely, in a negative vertical strain on YSGG, as Y-YIG-3 has a smaller $a$ than YSGG. We show in Supplementary Note IV that these strains are quantitatively consistent with the XRD analysis.

Density functional theory modeling of the unit cell of Y-rich YIG shows that the lattice parameter increases with \ce{Y^{3+}} substitution by $\Delta{}a/a_0=$~\SI{0.2394}{\percent} for $x=0.125$,  i.\,e., $\Delta{}a/a_0=$~\SI{1.92}{\percent\per{}}$x$ \cite{Su2021}. Although for a different cation \ce{Ga^{3+}} occupying the tetrahedral and octahedral sites, substitution to $x=0.5$ in \ce{Y_{3}(Y_{x}Ga_{5-x})O_{12}} leads to an increase in the lattice parameter of $\Delta{}a=$~\SI{0.0117}{\nano\meter} \cite{Geller1967}, converting into $\Delta{}a/a_0=$~\SI{1.89}{\percent\per{}}$x$ in the case of Y-rich YIG. In comparison, the fits of the XRD peaks from Y-YIG-7 estimate $a=$~\SI{1.2538}{\nano\meter} for $x=0.67$, i.\,e., $\Delta{}a/a_0=$~\SI{1.95}{\percent\per{}}$x$. The lattice expansion observed here for $x>0$ is thus in very good agreement, and consistent with Vegard's law (dashed line in Figure~\ref{fig:XRD}e). For $x<0$ instead, the lattice parameter found for Y-YIG-1 is not very different from the one of YIG, consistent with the discussion above. As a conclusion to this section, the lattice parameter of various common garnet substrates are reported as horizontal dashed lines in Figure~\ref{fig:XRD}e. The present results show that the approach of \ce{Y^{3+}} substitution enables high crystalline quality, non-relaxing Y-YIG films to be purposely matched with a diverse choice of substrates.

\begin{table*}[]
    \caption{Summary of the properties of single-crystalline Y-YIG films grown on GGG and YSGG substrates investigated in the present work. Sputtering plasma rectification voltages $U_{\ce{Y2O3}}$ and $U_{\ce{Fe2O3}}$ for each target, relative proportions of deposited oxides $y$ and corresponding stoichiometry $x$ in \ce{Y_{3+x}Fe_{5-x}O_{12}}; film thickness $t$ obtained from X-ray reflectivity measurements (included in Supplementary Note I, Y-YIG-1 curves could not be fit); and thickness $t$, out-of-plane lattice parameter $a_{\perp}$, unstrained film lattice parameter $a$ and relative lattice parameter change $\Delta{}a/a_0$, extracted from fits to the XRD measurements.}
    \label{tab:samples}
    \begin{tabular*}{\textwidth}{@{\extracolsep{\fill}}lS[table-format=3.1]S[table-format=3.1]S[table-format=1.3]S[table-format=1.2]S[table-format=2.1]S[table-format=2.1]S[table-format=1.4]S[table-format=1.4]S[table-format=1.2]}
        \hline\hline
        & \multicolumn{4}{c}{{Sputtering}} & {XRR} & \multicolumn{4}{c}{{XRD}} \\
        \cline{2-5}\cline{6-6}\cline{7-10}
        & {$U_{\ce{Y2O3}}$} & {$U_{\ce{Fe2O3}}$} & {$y$} & {$x$} & {$t$} & {$t$} & {$a_{\perp}$} & {$a$} & {$\Delta{}a/a_0$} \\
        Samples & {(\si{\volt})} & {(\si{\volt})} & {($\pm$0.020)} & {($\pm$0.040)} & {(\si{\nano\meter})} & {(\si{\nano\meter})} & {(\si{\nano\meter})} & {(\si{\nano\meter})} & {(\si{\percent})} \\
        \hline
        Y-YIG-1/GGG   & 190.8 & 106.4 & 1.260 & -0.200 & {\multirow{2}{*}{-}}      & {\multirow{2}{*}{26.6}} & 1.2333 & 1.2355 & -0.17 \\
        Y-YIG-1/YSGG & 190.8 & 106.4 & 1.260 & -0.200 &                                      &                                      & 1.2300 & 1.2374 & -0.02 \\
        Y-YIG-2/GGG   & 206.3 & 107.0 & 1.050 & 0.140 & {\multirow{2}{*}{33.9}} & {\multirow{2}{*}{33.8}} & 1.2447 & 1.2417 & 0.33 \\
        Y-YIG-2/YSGG & 206.3 & 107.0 & 1.050 & 0.140 &                                      &                                      & 1.2369 & 1.2411 & 0.28 \\
        Y-YIG-3/GGG   & 206.2 & 106.0 & 0.999 & 0.237 & {\multirow{2}{*}{32.4}} & {\multirow{2}{*}{30.7}} & 1.2491 & 1.2441 & 0.52 \\
        Y-YIG-3/YSGG & 206.2 & 106.0 & 0.999 & 0.237 &                                      &                                      & 1.2371 & 1.2412 & 0.29 \\
        Y-YIG-4/GGG   & 201.4 & 103.4 & 0.952 & 0.329 & {\multirow{2}{*}{28.4}} & {\multirow{2}{*}{29.9}} & 1.2534 & 1.2464 & 0.71 \\
        Y-YIG-4/YSGG & 201.4 & 103.4 & 0.952 & 0.329 &                                      &                                      & 1.2414 & 1.2435 & 0.48 \\
        Y-YIG-5/GGG   & 203.0 & 103.2 & 0.910 & 0.418 & {\multirow{2}{*}{29.8}} & {\multirow{2}{*}{29.2}} & 1.2577 & 1.2487 & 0.89 \\
        Y-YIG-5/YSGG & 203.0 & 103.2 & 0.910 & 0.418 &                                      &                                      & 1.2496 & 1.2480 & 0.84 \\
        Y-YIG-6/GGG   & 211.0 & 104.6 & 0.850 & 0.552 & {\multirow{2}{*}{34.8}} & {\multirow{2}{*}{34.9}} & 1.2606 & 1.2502 & 1.02 \\
        Y-YIG-6/YSGG & 211.0 & 104.6 & 0.850 & 0.552 &                                      &                                      & 1.2536 & 1.2501 & 1.01 \\
        Y-YIG-7/GGG   & 216.5 & 105.1 & 0.799 & 0.674 & {\multirow{2}{*}{36.3}} & {\multirow{2}{*}{36.6}} & 1.2664 & 1.2533 & 1.27 \\
        Y-YIG-7/YSGG & 216.5 & 105.1 & 0.799 & 0.674 &                                      &                                      & 1.2613 & 1.2542 & 1.34 \\
        \hline\hline
    \end{tabular*}

\end{table*}

\subsection{Temperature-dependent magnetic properties}
\begin{figure*}
    \centering
    \includegraphics[clip,width=6.99in]{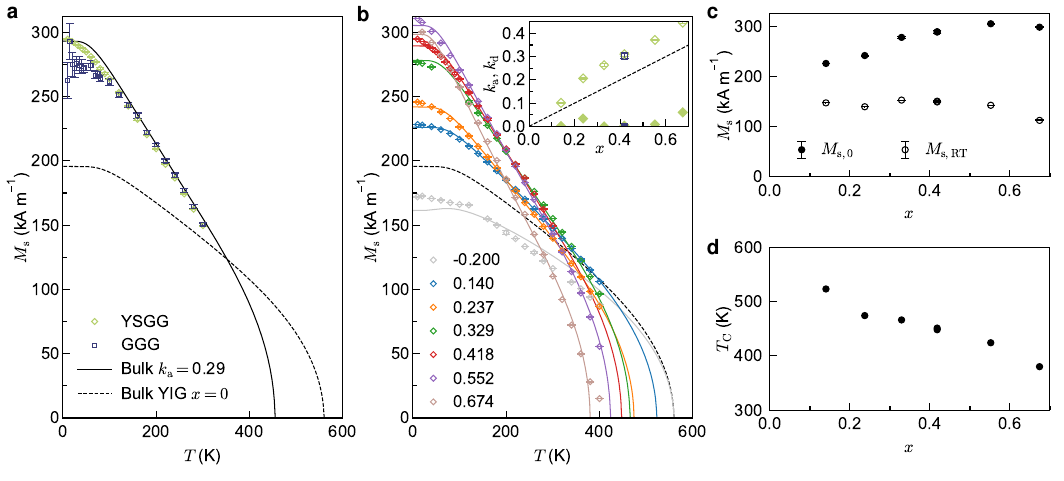}
    \caption{Saturation magnetization as a function of temperature for films Y-YIG-1--7. \pan{a}\ Comparison of $M_{\rm{s}}(T)$ for Y-YIG-5 on GGG and YSGG substrates, with fit (solid line) to molecular-field-coefficients model (see main text), providing octahedral sites substitution $k_{\rm{a}}=0.29$ and tetrahedral sites substitution $k_{\rm{d}}=0.00$. \pan{b}\ Evolution of $M_{\rm{s}}(T)$ on YSGG for different $x$, colored solid lines are fits to molecular-field-coefficients model with free parameters $k_{\rm{a}},k_{\rm{d}}$. The black dashed lines correspond to model values for unsubstituted bulk YIG, $k_{\rm{a}},k_{\rm{d}}$ = 0. The inset shows how $k_{\rm{a}}$ (open symbols) increases with $x>0$, while $k_{\rm{d}}$ (filled symbols) remains close to zero. The dashed line in the inset corresponds to $k_{\rm{a}}=x/2$ expected in case of exclusively octahedral site occupations with no defects or vacancies. \pan{c}\ Zero-temperature saturation magnetization ($M_{\rm{s,0}}$) and saturation magnetization at \SI{300}{\kelvin} ($M_{\rm{s,RT}}$), and \pan{d}\ Curie temperature extracted from the fits, as a function of $x>0$.}
    \label{fig:SQUID}
\end{figure*}

The saturation magnetization $M_{\rm{s}}$ of rare-earth-free iron garnets is due to the imbalance of magnetic cations located in tetrahedral and octahedral sites, and its dependence on temperature $T$ accurately reflects the site occupancy of substituants in the garnet structure \cite{Gilleo1958,Gilleo1958a,Geller1966a}. It is therefore relevant to compare the $M_{\rm{s}}(T)$ curves of the Y-YIG films grown on both GGG and YSGG to $M_{\rm{s}}(T)$ of unsubstituted bulk YIG \cite{Gilleo1958,Anderson1964}. The evolution of $M_{\rm{s}}$ in the range 10--300 or \SI{400}{\kelvin} for films Y-YIG-1--7 have been acquired by superconducting quantum interference device magnetometry (data analysis is detailed in Supplementary Note V, also comparing substrate backgrounds, related to the presence of paramagnetic species). A comparison is shown for the composition Y-YIG-5 on GGG and YSGG in \textbf{Figure~\ref{fig:SQUID}}a. A small discrepancy between the $M_{\rm{s}}(T)$ curves of Y-YIG-5 on GGG and YSGG appears at the lowest temperatures, which is likely due to the coupling to paramagnetic \ce{Gd^{3+}} in GGG, as we discuss further later on. Still, films grown on GGG and on YSGG behave very similarly as a function of $T>$~\SI{100}{\kelvin}. This confirms the nearly identical composition and substitutional behavior of the films grown on the two substrates, despite their different levels of strain.

We focus next on the films grown on diamagnetic YSGG, as their magnetometry data is more precise. In order to confirm the substitutional behavior of the excess yttrium in the garnet structure, $M_{\rm{s}}(T)$ has been measured for different $x$, as shown in Figure~\ref{fig:SQUID}b. The measurement points can be fit to a model based on molecular-field coefficients, established for iron garnets with diamagnetic substitutional atoms \cite{Dionne1970}, thereby providing quantitative information on the relative occupancy of the different cation sites \cite{Dionne1970,Roeschmann1981,Su2021,Gross2024}, as detailed in Experimental Section. Here, $k_{\rm{a}}$ is the fraction of \ce{Fe^{3+}} replaced by non-magnetic cations in octahedral sites and $k_{\rm{d}}$ in tetrahedral sites, shown in the inset of Figure~\ref{fig:SQUID}b. From this model, we find that the increase in zero temperature magnetization (Figure~\ref{fig:SQUID}c) and decrease in extrapolated Curie temperature (Figure~\ref{fig:SQUID}d) are the consequence of the \ce{Y^{3+}} cations occupying almost exclusively the larger octahedral sites, except for the largest value of $x$. Finally, the $M_{\rm{s}}(T)$ curve for Y-YIG-1, significantly below the curve for the bulk unsubstituted compound, confirms it has $x<0$. Within the model approximations, the values of $k_{\rm{a}}$ in inset of Figure~\ref{fig:SQUID}b confirm the results from the XRD analysis regarding the trend of the substitution, and suggest that the values of $x$ estimated from deposition rates underestimate the amount of octahedral \ce{Fe^{3+}} cations replaced by non-magnetic species or vacancies. From the analysis above, the nominal formula of Y-YIG when $x>0$ can be written \ce{Y_{3}(Y_{x}Fe_{5-x})O_{12}}. The actual Y/Fe stoichiometries are confirmed by X-ray photoelectron spectroscopy \cite{Rosenberg2021,Rosenberg2023} on samples Y-YIG-1--7, and energy-dispersive X-ray spectroscopy performed on the transmission electron microscopy samples, as presented in Supplementary Note VI. It also discusses, in more detail, some remaining indeterminacy of the defect chemistry of YIG and Y-YIG thin films \cite{Manuilov2009,Manuilov2010,Santiso2023,Rosenberg2023}.

\subsection{FMR linewidth and resonance field}
The quality factor of magnetization dynamics and the coherence of magnonic modes can be assessed by the ferromagnetic resonance (FMR) linewidth, defined as the width of the resonance of the dynamical magnetic susceptibility when sweeping an externally applied static magnetic field for a fixed excitation frequency. We consider the full-width at half-maximum linewidth, characterized in common cases by $\mu_0\Delta{}H=\mu_0\Delta{}H_0+2\alpha{}\omega/\gamma$ with an inhomogeneous broadening $\mu_0\Delta{}H_0$ that is frequency independent, and a term proportional to the frequency $\omega/(2\pi)$, where $\alpha$ is the phenomenological Gilbert damping parameter and $\gamma$ the gyromagnetic ratio. Achieving a satisfactorily low FMR linewidth $\mu_0\Delta{}H$ requires minimizing both inhomogeneous broadening and Gilbert damping terms. The impact of the excess yttrium on the resonance linewidth in Y-YIG is investigated by a series of FMR measurements performed with magnetic field applied in the film plane, first at room temperature (see Experimental Section), summarized in \textbf{Figure~\ref{fig:FMRroomT}}. Additional measurements for films grown under varied \ce{O2} gas flow are reported in Supplementary Note VII, showing that \SI{1.0}{sccm} of \ce{O2} for \SI{100}{sccm} of \ce{Ar} flow provides optimal magnetic resonance properties. The FMR linewidth $\mu_0\Delta{}H$ of these films regularly reaches below \SI{1}{\milli\tesla} at \SI{10}{\giga\hertz}, with $\mu_0\Delta{}H_0$ as low as \SI{0.2}{\milli\tesla} (Figure~\ref{fig:FMRroomT}e,f). This corresponds to very homogeneous, well-crystallized layers of iron garnet that place this series of films within the best results achieved by sputtering for their range of thicknesses \cite{Schmidt2020}, even with their off-stoichiometric Y/Fe ratio. The achievement of $\mu_0\Delta{}H<$~\SI{1}{\milli\tesla} at \SI{10}{\giga\hertz} notably indicates a low concentration $<$~\SI{1}{\percent} of non-\ce{Fe^{3+}} cations, see also Supplementary Note VI \cite{Judy1966,Epstein1967}.

The present sputtering parameters have been optimized to provide best results for off-stoichiometric depositions with $x\approx0.42$. The FMR linewidth of the films appears to degrade for $x>0.6$, see Figure~\ref{fig:FMRroomT}g. In addition to the impact of the substitution itself, this could also be due to an imbalance of the oxygen stoichiometry when moving away from locally optimal parameters. In Y-YIG-1 with $x<0$, the strong increase in $\mu_0\Delta{}H$ is expected, and arises either from excess iron in the dodecahedral sites or from the formation of vacancy and cationic charge defects. As can be seen in Figure~\ref{fig:FMRroomT}i, $\alpha$ appears scattered in the investigated range of $x>0$, and is likely determined by extrinsic defects, in addition to the expected impact of the impurity of the source materials.

\begin{figure*}
    \centering
    \includegraphics[clip,width=6.99in]{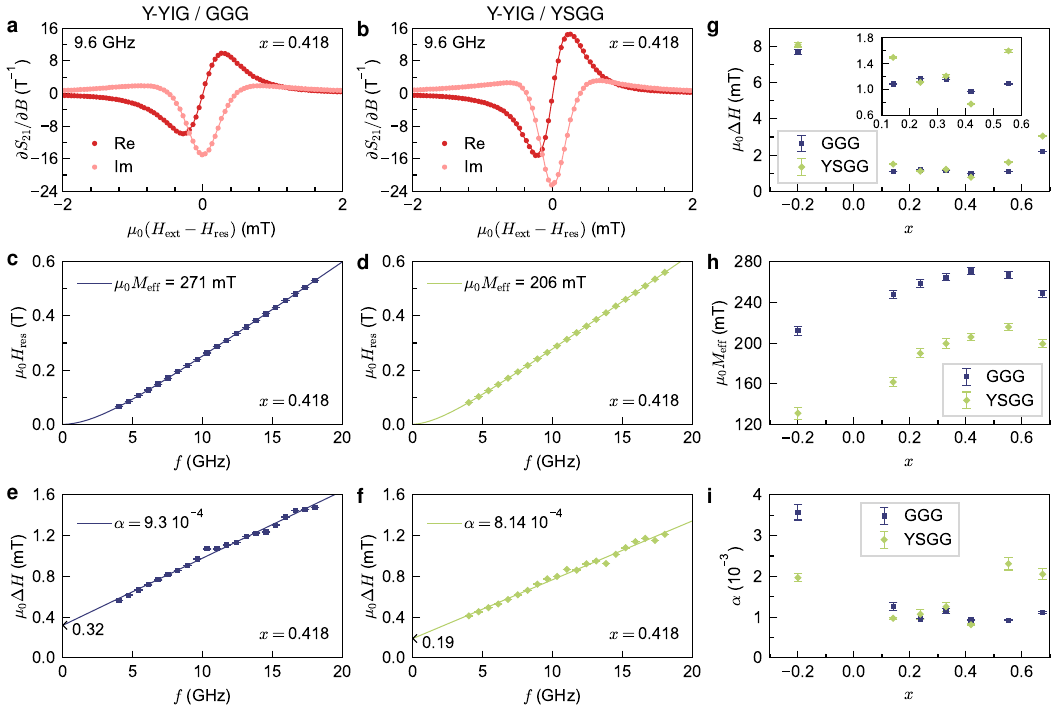}
    \caption{Ferromagnetic resonance (FMR) at room temperature for films Y-YIG-1--7. Derivative of the complex microwave transmission $\partial{}S_{21}/\partial{}B$ at \SI{9.6}{\giga\hertz} as a function of external magnetic field around resonance, for Y-YIG-5 on \pan{a}\ GGG and \pan{b}\ YSGG. Solid lines are fits to in-plane magnetic susceptibility $\chi$, with $\real{\partial{}S_{21}/\partial{}B}$ and $\imag{\partial{}S_{21}/\partial{}B}$ proportional to $\partial{}\chi''/\partial{}B$ and $\partial{}\chi'/\partial{}B$, respectively. Resonance field $\mu_0H_{\rm{res}}$ as a function of frequency, solid lines are fits to Kittel law to extract $\mu_0M_{\rm{eff}}$, for \pan{c}\ GGG and \pan{d}\ YSGG substrates. FMR linewidth $\mu_0\Delta{}H$ as a function of frequency and fit to $\mu_0\Delta{}H=\mu_0\Delta{}H_0+2\alpha{}\omega/\gamma$ (solid lines), for \pan{e}\ GGG and \pan{f}\ YSGG substrates. Comparison of \pan{g}\ FMR linewidth $\mu_0\Delta{}H$ at \SI{9.6}{\giga\hertz}, \pan{h}\ effective magnetization $\mu_0M_{\rm{eff}}$, and \pan{i}\ Gilbert damping $\alpha$, for all films on GGG and YSGG substrates.}
    \label{fig:FMRroomT}
\end{figure*}

The magnetic anisotropy in single-crystalline garnet thin films depends on various terms, summing two magnetocrystalline contributions with easy axis along [111], a magnetoelastic contribution induced by strain, and a magnetostatic uniaxial perpendicular anisotropy due to dipolar interactions in the extended film, added to potential contributions from growth-induced anisotropy and surface anisotropy. For in-plane FMR, the out-of-plane effective anisotropy value can be obtained by fitting the resonance fields to the Kittel formula $\omega=\gamma\mu_0\sqrt{H_{\rm{ext}}(H_{\rm{ext}}+M_{\rm{eff}})}$, as shown for films deposited on GGG (Figure~\ref{fig:FMRroomT}c) and on YSGG (Figure~\ref{fig:FMRroomT}d). Here, $M_{\rm{eff}}=M_{\rm{s}}-H_{\rm{ani}}$ is the effective magnetization, with $H_{\rm{ani}}$ a perpendicular anisotropy field along the normal to the film, excluding the shape anisotropy. We note that to achieve best fit results, an additional field shift, due to an in-plane anisotropy component, or a two-magnon scattering term \cite{Arias1999a,Jermain2016}, needs to be added to the Kittel formula. Since at room temperature this term is only a few percent of $M_{\rm{eff}}$ for the samples of interest, this shift can be ignored for the present purposes, and is discussed further in Supplementary Note VII. The magnetoelastic contribution is expected to dominate in the perpendicular anisotropy field, and the overall reduction of $M_{\rm{eff}}$ between films grown on GGG and YSGG is consistent with the magnetoelastic anisotropy induced by the strained growth, which has already been evidenced in the case of YIG on YSGG \cite{Gueckelhorn2021}. With increasing $x$, the unit cell of the off-stoichiometric YIG expands, which leads to less tensile and later more compressive in-plane strain for a given substrate. Considering the sign of the magnetoelastic coupling terms in YIG, the anisotropy induced by strain evolves toward a negative uniaxial anisotropy along the [111] growth direction, and thus enhances $M_{\rm{eff}}$ with increasing $x$, as seen in Figure~\ref{fig:FMRroomT}h. Films grown on GGG experience compressive strain for $x>0$, thus reinforcing $M_{\rm{eff}}$ above the shape anisotropy value given by $M_{\rm{s}}$. Films grown on YSGG experience tensile strain for $x<0.36$ reducing $M_{\rm{eff}}$ below the shape anisotropy, and compressive strain for $x>0.36$ (see Figure~\Ref{fig:XRD}e).

This series of Y-YIG films thus offers a unique compromise between tunability and quality, as it allows the composition and lattice parameter to be varied over a wide and continuous range, while maintaining a low FMR linewidth over this entire range. This is a distinctive advantage of rf co-sputtering synthesis, with respect to other approaches such as liquid phase epitaxy or single-source physical vapor deposition, which do not enable direct changes in composition and require a precise knowledge of the optimal melt or target stoichiometry to reach a desired film stoichiometry.

\subsection{Temperature-dependent FMR linewidth}
\begin{figure*}
    \centering
    \includegraphics[clip,width=6.99in]{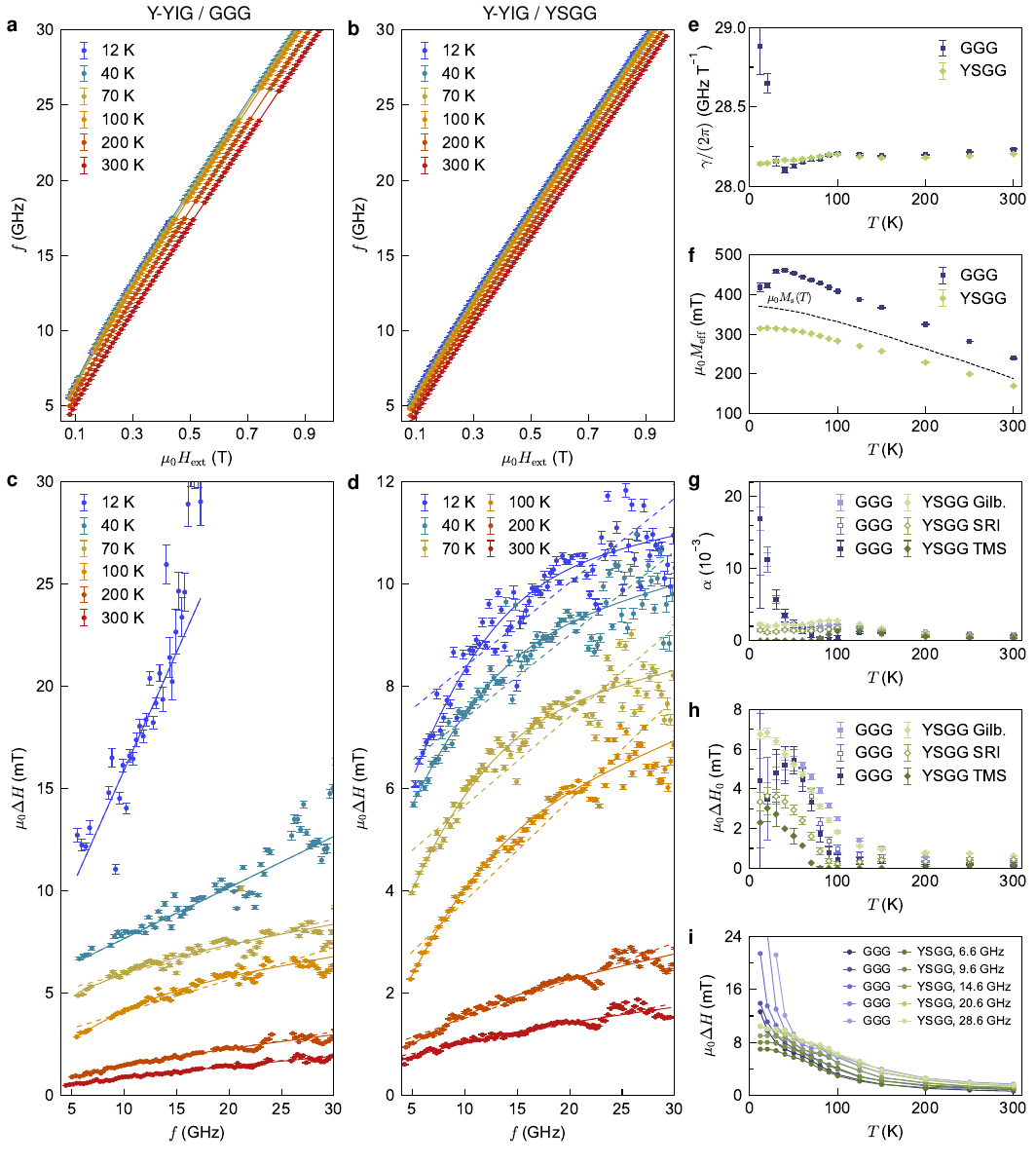}
    \caption{Ferromagnetic resonance (FMR) in the temperature range 10--\SI{300}{\kelvin} for films Y-YIG-5. Resonance frequency as a function of applied magnetic field for film on \pan{a}\ GGG and \pan{b}\ YSGG. Solid lines are fits to Kittel law with in-plane anisotropy term (see Experimental Section). FMR linewidth as a function of frequency for film on \pan{c}\ GGG and \pan{d}\ YSGG. Solid lines are fits to a slow-relaxer impurities model, dashed lines are fits to $\mu_0\Delta{}H=\mu_0\Delta{}H_0+2\alpha{}\omega/\gamma$. \pan{e}\ Gyromagnetic ratio $\gamma/(2\pi)$, \pan{f}\ effective magnetization $\mu_0M_{\rm{eff}}$, \pan{g}\ Gilbert damping $\alpha$, \pan{h}\ inhomogeneous broadening $\mu_0\Delta{}H_0$ and \pan{i}\ FMR linewidth $\mu_0\Delta{}H$ at several frequencies, as a function of temperature, for films on GGG and YSGG. The $M_{\rm{s}}(T)$ dependence from Figure~\ref{fig:SQUID}a is reported as a dashed line in panel f. The different values presented for the damping $\alpha$ in panel g and for the inhomogeneous broadening $\mu_0\Delta{}H_0$ in panel h correspond to results of the Gilbert (Gilb.), slow-relaxer impurities (SRI) and two-magnon scattering (TMS) models.}
    \label{fig:FMRlowT}
\end{figure*}

The films with the most promising composition Y-YIG-5, showing $x\approx0.42$ and a nearly ideal lattice matching with YSGG, have been measured by FMR as a function of temperature in the range 10--\SI{300}{\kelvin} (see Experimental Section). The results for the samples grown on GGG and YSGG are shown and compared in \textbf{Figure~\ref{fig:FMRlowT}}. We first discuss the evolution of the resonance frequency with magnetic field (Figure~\ref{fig:FMRlowT}a,b) providing the gyromagnetic ratio $\gamma$ (Figure~\ref{fig:FMRlowT}e) and effective magnetization $\mu_0M_{\rm{eff}}$ (Figure~\ref{fig:FMRlowT}f), directly related to magnetic anisotropy in each system. We then discuss the evolution of the FMR linewidth with frequency (Figure~\ref{fig:FMRlowT}c,d) and its deviations from the common model of Gilbert damping with inhomogeneous broadening (Figure~\ref{fig:FMRlowT}g--i). The complete dataset is presented in Supplementary Note VIII.

The dependence of the resonance frequency on the external field is fit with a phenomenological Kittel formula, including an additional in-plane component of the magnetocrystalline anisotropy $\mu_0H_{\rm{K}}$, see Experimental Section and Supplementary Note VIII. The sharp decrease in the effective anisotropy $\mu_0M_{\rm{eff}}$ detected at low temperatures for films grown on GGG (Figure~\ref{fig:FMRlowT}f) is consistent with the coupling to the paramagnetic moments of \ce{Gd^{3+}} \cite{Danilov1989,Mihalceanu2018,Roos2022,Knauer2023,Serha2024}, which is indeed absent on YSGG substrates. This effect becomes critical at low temperatures as the substrate paramagnetism follows a susceptibility law $\chi_{\rm{p}} = \mu_0NJ(J+1)(g\mu_{\rm{B}})^2/(3k_{\rm{B}}T)$ with $N$ the density of \ce{Gd^{3+}} cations per unit volume, $J=7/2$ and $g=2$ for \ce{Gd^{3+}}, $\mu_{\rm{B}}$ the Bohr magneton, and $k_{\rm{B}}$ the Boltzmann constant, which is overall inversely proportional to temperature $T$. Substrate paramagnetism is also expected to affect the apparent gyromagnetic ratio $\gamma$ in the YIG films, by counteracting the externally applied magnetic field with a proportional opposite dipolar field contribution, and by interfacial coupling. Consistently, we find that Y-YIG exhibits a much weaker temperature dependence of $\gamma$ when grown on YSGG, compared to GGG (Figure~\ref{fig:FMRlowT}e). Substrate paramagnetism can be assigned to the difference between the temperature dependence of $\gamma$ on GGG and YSGG, but not for the slight variations of $\gamma$ on YSGG, more pronounced than in bulk YIG \cite{Maier-Flaig2017}.

The coupling to the paramagnetic moments in GGG is at the origin of a dramatic broadening of the FMR linewidth at the lowest temperatures, evidenced by comparing the curves for GGG and YSGG substrates in Figure~\ref{fig:FMRlowT}c,d,i, which has also been observed in previous works \cite{Jermain2017,Guo2023}. In contrast, the main result from the present low-temperature measurements is that the cryogenic FMR linewidth degradation induced by substrate paramagnetism for $T<$~\SI{40}{\kelvin} can be eliminated successfully, following the approach of matching the lattice parameter of Y-YIG with YSGG by substitution of excess yttrium. 

The remaining increase in the FMR linewidth at cryogenic temperatures, also present on YSGG substrates, deserves further consideration. A similar evolution has been previously observed for YIG films grown by liquid phase epitaxy \cite{Beaulieu2018,Kosen2019,Wang2020d}, but not in all investigations \cite{Will-Cole2023} and not in bulk YIG single crystals \cite{Spencer1959,Maier-Flaig2017,Kosen2019}. On YSGG, where there is no \ce{Gd^{3+}} in the substrate, other mechanisms than substrate-induced paramagnetism are involved. Whereas near room temperature the FMR linewidth increases almost linearly with frequency ($T=$~\SI{300}{\kelvin} in Figure~\ref{fig:FMRlowT}b,d), corresponding to a Gilbert damping with inhomogeneous broadening behavior, at lower temperatures it deviates from linearity, with a steeper initial slope. For this reason, the evolution of the linewidth as a function of temperature is best analyzed at individual frequencies (Figure~\ref{fig:FMRlowT}i), and the values of damping and inhomogeneous broadening corresponding to the Gilbert model in Figure~\ref{fig:FMRlowT}g,h are only indicative and provided for comparison with other works.

Two different main mechanisms are expected to contribute to non-linear terms in the in-plane FMR linewidth of magnetic insulator thin films: a phenomenon in which the uniformly precessing magnetization scatters on defects to other spin-wave excitations within the magnetic lattice, known as two-magnon scattering \cite{Lenz2006}, and a process involving a coupling to the energy level structure of rare-earth impurities, known as slow-relaxer impurities \cite{Dillon1959,Dillon1962,Seiden1964}. Both terms exhibit a steep slope at low frequencies, but whereas two-magnon scattering saturates above a frequency related to $\gamma\mu_0M_{\rm{eff}}$, the slow-relaxer impurities term peaks at the inverse lifetime of the impurity excitations, before it decreases to zero (refer to Experimental Section for more discussion). Despite this difference, both models can satisfactorily fit the data at each temperature, without noticeable differences in the modeled linewidth dependence on frequency. We refer to Supplementary Note IX for an extended discussion of the data. Except if the density of defects increases sharply at low temperatures, a two magnon-scattering process is not expected to strongly depend on temperature as is observed here \cite{Jermain2017}. However, the linewidth evolution with frequency at each temperature is consistent with $\gamma\mu_0M_{\rm{eff}}(T)$. Possibly related to the finite frequency range of the present measurements, the relaxation time of the slow-relaxer impurity model extracted from fits to the data is found constant across all temperatures, while a strong temperature dependence is expected instead \cite{Seiden1964}. We conclude that both effects must be combined to produce the present linewidth behavior, and cannot be easily distinguished.

The monotonic evolution with temperature of the linewidth of Y-YIG-5 on YSGG (Figure~\ref{fig:FMRlowT}i) is reminiscent of the temperature dependence of the linewidth in bulk single-crystal spheres of \ce{Y_{3}Fe_{4.9}In_{0.1}O_{12}} \cite{Spencer1964}, where the role of \ce{In^{3+}} substituted for octahedral \ce{Fe^{3+}} cations may be similar to that of the additional \ce{Y^{3+}} cations in the present samples. The contribution of these \ce{In^{3+}} cations was found roughly 1--2 orders of magnitude smaller than here, as well as anisotropic. The larger degree of substitution ($x\approx0.42$) for a presumably larger ion \cite{Gilleo1958}, combined with the \si{\nano\meter}-range thickness of the films reinforcing the role of defects, could explain the larger increase in linewidth at low temperatures in the Y-YIG epitaxial films. In addition, when the temperature is lowered below $T\approx$~\SI{100}{\kelvin}, the gyromagnetic ratio starts to reduce (Figure~\ref{fig:FMRlowT}e) where the magnetic linewidth starts to noticeably increase. This suggests an effect related to spin-orbit coupling, a cause of both magnetic anisotropy and magnetic damping, which is analyzed further in Supplementary Note IX. Future investigations should therefore focus on minimizing these contributions to the cryogenic FMR linewidth.

\begin{figure*}
    \centering
    \includegraphics[clip,width=6.99in]{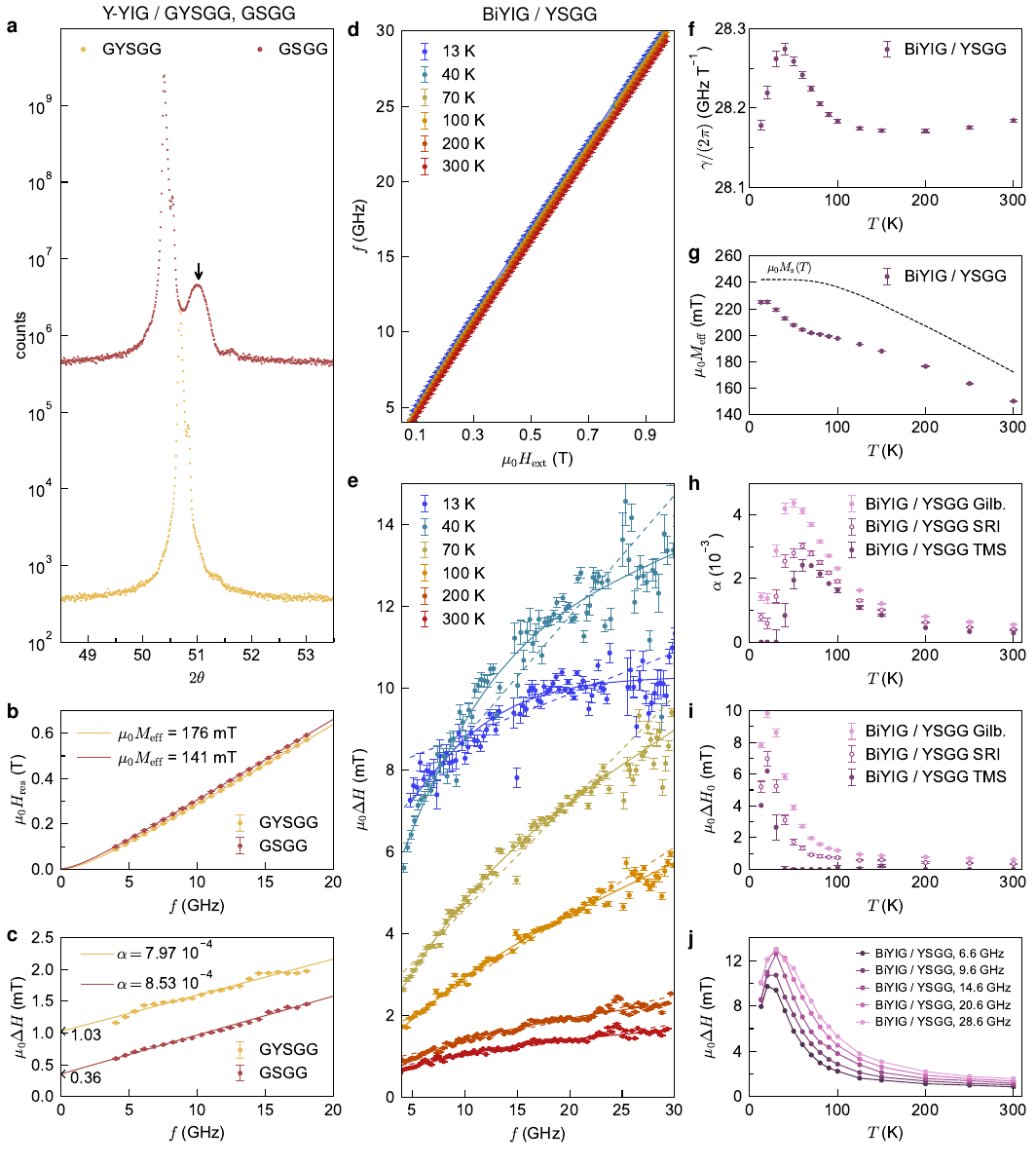}
    \caption{Alternative film/substrate combinations: \ce{Y_{3}(Y_{x}Fe_{5-x})O_{12}} on GYSGG and GSGG, and (Bi,Y)IG on YSGG. \pan{a}\ XRD $2\theta-\omega$ diffractograms of films of \ce{Y_{3}(Y_{x}Fe_{5-x})O_{12}} on GYSGG and GSGG substrates ($x=0.616$). The arrow locates the film peak on GSGG. At room temperature, \pan{b}\ resonance field $\mu_0H_{\rm{res}}$ as a function of frequency, solid lines are fits to Kittel law to extract $\mu_0M_{\rm{eff}}$ for GYSGG and GSGG substrates. \pan{c}\ FMR linewidth $\mu_0\Delta{}H$ as a function of frequency and fit to $\mu_0\Delta{}H=\mu_0\Delta{}H_0+2\alpha{}\omega/\gamma$ (solid lines), for GYSGG and GSGG substrates. Ferromagnetic resonance (FMR) in the temperature range 10--\SI{300}{\kelvin} for \SI{10}{\nano\meter}-thick \ce{(Bi_{0.8}Y_{2.2})Fe_{5}O_{12}} on YSGG substrate, with \pan{d}\ resonance frequency as a function of applied magnetic field (solid lines are fits to Kittel law with in-plane anisotropy term, see Experimental Section) and \pan{e}\ FMR linewidth as a function of frequency (solid lines are fits to a slow-relaxer impurities model, dashed lines are fits to $\mu_0\Delta{}H=\mu_0\Delta{}H_0+2\alpha{}\omega/\gamma$). \pan{f}\ Gyromagnetic ratio $\gamma/(2\pi)$, \pan{g}\ effective magnetization $\mu_0M_{\rm{eff}}$, \pan{h}\ Gilbert damping $\alpha$, \pan{i}\ inhomogeneous broadening $\mu_0\Delta{}H_0$ and \pan{j}\ FMR linewidth $\mu_0\Delta{}H$ at several frequencies, as a function of temperature. The $M_{\rm{s}}(T)$ dependence predicted from a molecular-field-coefficients model is reported as a dashed line in panel g. The different values presented for the damping $\alpha$ in panel h and for the inhomogeneous broadening $\Delta{}H_0$ in panel i correspond to results of the Gilbert (Gilb.), slow-relaxer impurities (SRI) and two-magnon scattering (TMS) models.}
    \label{fig:alternatives}
\end{figure*}

\subsection{Lattice matching with garnets having large mismatch to YIG}
Obtaining flexibility in the lattice parameter of iron garnets with narrow FMR is key in order to achieve the epitaxial growth of multilayers combining garnet layers of different compositions for enhanced properties, e.g., magneto-optical response or optical index. In the following, we demonstrate that for large values of $x$, the substitution of iron with excess yttrium enables matching the substrate lattice parameter of gadolinium yttrium scandium gallium garnet (GYSGG, \ce{(Gd_{0.63}Y_{2.37})Sc_{2}Ga_{3}O_{12}}), even reaching close to gadolinium scandium gallium garnet (GSGG, \ce{Gd_{3}Sc_{2}Ga_{3}O_{12}}), as suggested by Figure~\ref{fig:XRD}e. This matching is obtained while maintaining narrow FMR linewidth at room temperature. The XRD characterization for these single-crystalline films is reported in \textbf{Figure~\ref{fig:alternatives}}a, which evidences that for $x\approx0.6$, our Y-YIG has a lattice parameter perfectly matched with that of GYSGG at \SI{1.251}{\nano\meter}. The in-plane tensile strain for films grown on GSGG produces a positive out-of-plane magnetic anisotropy, which reduces $\mu_0M_{\rm{eff}}$ to \SI{141}{\milli\tesla}, as shown in Figure~\ref{fig:alternatives}b, slightly smaller than $\mu_0M_{\rm{s}}\approx$~\SI{160}{\milli\tesla}. The FMR linewidth at room temperature is shown as a function of frequency in Figure~\ref{fig:alternatives}c. For these Y-YIG films grown on substrates with a large lattice parameter mismatch with respect to YIG and GGG, the linewidth remains excellent for strained films obtained by magnetron sputtering. This approach circumvents the usual degradation of linewidth observed on substrates having a large mismatch to pure YIG \cite{Ding2020a}. It highlights the broad applicability of the present co-deposition of Y-YIG to enable a growth of iron garnets on a wide choice of substrates. Y-YIG with various substitutions $x$ can thus be combined with other garnet systems with varied lattice parameters, offering the possibility of relying on strain engineering to modulate the effective magnetic anisotropy in these films while maintaining a good FMR.

\subsection{Lattice-matched Bi-substituted YIG}
We now consider dodecahedral substitution as an alternative to octahedral substitution to match the lattice parameters of iron garnet films with YSGG. Instead of tuning the \ce{Y}/\ce{Fe} ratio, lattice matching can be achieved by replacing some \ce{Y^{3+}} cations by a larger, non-magnetic cation. With its filled $4f$- and $5d$-shells, Bi is not expected to be a significant source of magnetic damping in iron garnets. We have therefore investigated the low-temperature properties of a \ce{(Bi_{0.8}Y_{2.2})Fe_{5}O_{12}} film grown on lattice-matched YSGG. The partial substitution with Bi is also very interesting for the magneto-optical detection of the magnetization \cite{Hansen1983}. The film investigated here has an even lower thickness than previous films Y-YIG-1--7, with \SI{10}{\nano\meter}. It has been deposited in similar sputtering conditions as above, although relying on a single target, because we could not implement rf co-sputtering with three targets due to equipment limitations. FMR in the temperature range 10--\SI{300}{\kelvin} is reported in Figure~\ref{fig:alternatives}d--j. Several qualitative differences are observed between \ce{Bi_{0.8}Y_{2.2}Fe_{5}O_{12}} and \ce{Y_{3}(Y_{x}Fe_{5-x})O_{12}} films.

The gyromagnetic ratio $\gamma$ in the Bi-substituted film features a peak at around \SI{40}{\kelvin}, below which it tends toward the free electron value of $\gamma_{\rm{e}}=$~\SI{28.02}{\giga\hertz\per\tesla}. Conversely, the effective magnetization $\mu_0M_{\rm{eff}}(T)$ presents a different behavior compared to Y-YIG, with a negative peak at \SI{40}{\kelvin} added to a steady decrease with increasing temperature. This results overall in less relative variations with temperature. For comparison, $\mu_0M_{\rm{s}}(T)$ predicted for \ce{Bi_{0.8}Y_{2.2}Fe_{5}O_{12}} with the molecular-field-coefficients model is overlaid in Figure~\ref{fig:alternatives}g. The difference between $M_{\rm{eff}}(T)$ and $M_{\rm{s}}(T)$ corresponds to out-of-plane anisotropy contributions. By analyzing together Figure~\ref{fig:alternatives}f,g and j, the gyromagnetic ratio, the out-of-plane anisotropy and the linewidth again appear to include a common contribution. A more detailed comparison of the evolution of the different FMR parameters is provided in Supplementary Note IX, discussing this apparent correlation. 

Different from Y-YIG-5 on YSGG, the FMR linewidth in the investigated BiYIG film peaks near 30--\SI{40}{\kelvin}, but instead of saturating at a relatively high level compared to room temperature, it decreases again with further decreasing temperature, with the potential to recover near its room-temperature value when reaching a couple of \si{\kelvin}. This promising example demonstrates that controlling the composition to match the lattice parameter of epitaxial iron garnets to YSGG is not limited to octahedral \ce{Fe^{3+}} substitution by \ce{Y^{3+}}, but is also a valuable approach with dodecahedral cation substitutions such as with \ce{Bi^{3+}}. Further investigations shall aim at evaluating more comprehensively the advantages and the limits of these different options for substituting iron garnets in the context of cryogenic magnonics.

\section{Conclusion}

Producing microwave-compatible garnet thin films while continuously varying their composition with rf co-sputtering opens an exciting avenue for investigating new types of garnets that cannot be formed with growth techniques at equilibrium such as liquid phase epitaxy. The great potential of magnetron sputtering for industrial scale, reproducible growth of large homogeneous wafers is also an important aspect motivating further investigation of the proposed rf co-sputtering approach. It greatly simplifies the search for optimal deposition conditions in oxide growth by physical vapor deposition and avoids the difficulty of finding the ideal composition required of a single target. The improved precision on the cation stoichiometry is crucial for targeting ideal lattice matching or specific magnetic properties in multilayers. We expect that this proposed method, exemplified here for the growth of epitaxial iron garnets, could also be relevant to other classes of single-crystalline films such as perovskites and spinels.

The magnetic resonance properties of off-stoichiometric YIG with excess Y have long remained inaccessible due to difficulties in their bulk single-crystal synthesis. However, we show here that single-crystalline thin films can easily be grown relying on a physical vapor deposition technique \cite{Su2021}. We observe that a stoichiometry of \ce{Y}/\ce{Fe} $>0.6$ has a very moderate impact on the FMR linewidth at room temperature. These Y-YIG films display excellent magnetization dynamics, comparable to best results achieved at room-temperature for sputtered films, while additionally bringing unprecedented tunability in the lattice parameter. This feature enables lattice-matched growth on various garnet substrates and a strain-engineering approach to tune the magnetic anisotropy.

Our present investigation has focused on two types of cation substitution in YIG, \ce{Y^{3+}} in octahedral sites and \ce{Bi^{3+}} in dodecahedral sites, which increase the lattice parameter for matching with diamagnetic YSGG. This finds a direct use in producing \si{\nano\meter}-thick films with better magnetic resonance properties at low temperatures than commonly observed on GGG. Comparing the data measured at \SI{10}{\kelvin}, the present films are already within a factor 3 of the best linewidth results obtained with the long-established technique of single-target off-axis sputtering \cite{Guo2023}. This demonstrates that lattice-matched \ce{Y_{3}(Y_{x}Fe_{5-x})O_{12}} or \ce{(Y_{3-x}Bi_{x})Fe_{5}O_{12}} films on YSGG can provide magnetic properties that enable low-temperature magnonics. Further investigation along this promising route shall confirm whether aiming for purest films by minimizing interfacial intermixing and impurity concentrations, adjusting the composition, and exploring other cation substitutions \cite{Boettcher2022,Scheffler2023} could result in iron garnet thin films with low-temperature FMR linewidth even better than their linewidth at room temperature.

\section{Acknowledgements}
We acknowledge T.~Weber for assistance in acquiring and N.~Reyren for assistance in interpreting the X-ray diffractograms, as well as G.~Berthom\'{e} for assistance in acquiring X-ray photoelectron spectroscopy data. This research was partially supported by the Swiss National Science Foundation (Grant No. 200020-200465). W.L.~acknowledges the support of the ETH Zurich Postdoctoral Fellowship Program (21-1 FEL-48). H.W.~acknowledges the support of the China Scholarship Council (CSC, Grant No.\ 202206020091). We acknowledge the financial support of European Commission through Marie Skłodowska-Curie Actions H2020 RISE with the projects MELON (Grant No.\ 872631) and ULTIMATE-I (Grant No.\ 101007825), and access to the equipment of “Servicio General de Apoyo a la Investigación (SAI), Universidad de Zaragoza”.

\section{Author contributions}
W.L.~conceived and coordinated the project under P.G.'s supervision. W.L.,~Y.K.~and S.S.~developed the co-sputtering approach, and deposited the samples with H.W.~providing assistance. H.W.~collected the high-resolution XRD and reciprocal space maps, and W.L.~analyzed the XRD and reciprocal space maps data. M.H.A.~acquired and analyzed the transmission electron microscopy data. W.L.~performed the sample magnetometry and analyzed the data with the help of R.S.~for the molecular-field-coefficients model. D.P.~measured the room-temperature FMR with the help of S.S.~and H.W., and D.P.~and W.L.~analyzed the data.  L.S., R.S.~and M.L.~conducted the low-temperature FMR experiments with the help of H.W., and W.L.~analyzed the low-temperature FMR data. W.L.~and P.G.~wrote the manuscript. All the authors discussed the data and commented on the manuscript.

\section{Methods}

\subsection{Substrate preparation and details of the implementation of rf co-sputtering}
We employ substrates of GGG, GSGG purchased from SurfaceNet GmbH and substrates of YSGG, GYSGG purchased from MTI corporation, all polished to a $<$~\SI{0.5}{\nano\meter} quoted roughness on a (111) orientation. After dicing, the substrates are cleaned successively in acetone and isopropanol at \SI{50}{\celsius}, with ultra-sonication, for \SI{20}{mins} each. In the growth chamber, the single-crystalline substrates are first heated up to \SI{750}{\degreeCelsius} for 1 hour to release adsorbed impurities before deposition, in an \ce{O2} atmosphere at \SI{2.5}{\pascal}. A substrate holder temperature of \SI{750}{\degreeCelsius} is maintained over the whole deposition time and for 1 hour after the end of the deposition for further annealing of the epitaxial films. The substrates are then let to cool down in an \ce{O2} atmosphere at \SI{2.5}{\pascal}, reaching \SI{120}{\degreeCelsius} after 1 hour.

The \ce{Y2O3} and \ce{Fe2O3} targets used in this work feature moderate purities, respectively \SI{99.99}{\percent} and \SI{99.9}{\percent}. In the configuration presently investigated for co-sputtering, the two targets are coplanar and distant by roughly \SI{230}{\milli\meter}. The substrates are mounted on a rotating planetary holder facing the midpoint between the targets, at a sample height of \SI{80}{\milli\meter}, which corresponds to an off-axis configuration with offset angle of \SI{55}{deg} \cite{Yang2018d}.

\subsection{Estimation of stoichiometry from deposition rates}
We use tabulated mass densities $\rho_{\ce{Fe2O3}}=$~\SI{5.24e3}{\kilogram\per\meter\cubed} and $\rho_{\ce{Y2O3}}=$~\SI{5.01e3}{\kilogram\per\meter\cubed} for the powder forms, and molar masses $M_{\ce{Fe2O3}}=$~\SI{159.7}{\gram\per\mole} and $M_{\ce{Y2O3}}=$~\SI{225.8}{\gram\per\mole} for the two target materials. We write $n_{\ce{Fe}}=5$ and $n_{\ce{Y}}=3$ for their quantities per unit formula in YIG. The thickness deposition rates provide Y and Fe atoms in stoichiometric amounts for a ratio $r_{\ce{Fe2O3}}$/$r_{\ce{Y2O3}}=(n_{\ce{Fe}}/n_{\ce{Y}})(\rho_{\ce{Y2O3}}/\rho_{\ce{Fe2O3}})(M_{\ce{Fe2O3}}/M_{\ce{Y2O3}})=1.13$.

The thickness deposition rates for each target are calibrated independently by combining atomic force microscopy on a series of lift-off samples with the rates given by an in-situ piezo-crystal growth controller. For predicting the composition of the films, we have observed that the rectification voltages obtained in each plasma were a much more relevant measure of the respective deposition rates than the nominal or effective power applied by the rf sources during the growth. The sputtering yield is expected to be zero below a given threshold voltage, above which species start to be ejected from the sputtering target. The rates then grow linearly with rectified voltage in a good local approximation around reference values for which rates are calibrated, here about \SI{200}{\volt} for \ce{Y2O3} and \SI{105}{\volt} for \ce{Fe2O3}. Only for Y-YIG-1 for which $U_{\ce{Y2O3}}$ = \SI{190.8}{\volt}, significantly below the range of validity of a linear evolution of the deposition rates, this prediction underestimates $x$, which is why we use $x=-0.2$, as deduced from magnetometry.

\subsection{Measurement and analysis of X-ray diffraction data}
X-ray diffractograms have been acquired on a Panalytical X'pert MRD thin film diffractometer equipped with an Eulerian cradle, a Cu-tube, a parallel beam X-ray mirror, and a Ge(220) four-bounce Cu(K$\alpha$1) monochromator to suppress the Cu(K$\alpha$2,K$\beta$) radiation lines. Scans are performed in symmetric $2\theta-\omega$ geometry. The X-ray diffractograms on GYSGG and GSGG from Figure~\ref{fig:alternatives}a have been obtained on a Panalytical X'pert PRO MPD powder diffractometer in the Bragg--Brentano geometry. The instrument is equipped with a Johansson monochromator, which in addition to the dominant Cu(K$\alpha$1) line, also delivers some Cu(K$\alpha$2) radiation that results in weak peak doublets visible on the right of the peaks. All the analysis of the X-ray diffractograms has been performed in Python, partly relying on the package xrayutilities \cite{Kriegner2013}.

To extract the distance $d_{111}$ between diffracting planes of each sample, the film diffraction peak is fit relative to the substrate peak position, taking into account the interference between each plane of the film and substrate in a kinematic approximation. The latter is required because the present films have lattice parameters close to that of the substrate and a thickness of only a few tens of \si{\nano\meter}, providing a relatively weak signal compared to the substrate \cite{Pesquera2011}. The substrate response is modeled by a sum of diffracting planes in the kinematical approximation, accounting for a finite penetration depth of the X-rays of about \SI{3}{\micro\meter}, and corrected by a substrate-dependent phase factor obtained using a dynamical model. The film response is modeled by a sum of diffracting planes with constant form factor, allowing for a partial coverage of the last plane and for an interfacial layer with substrate of arbitrary composition, adjusting the precise film position and the relative phase of the film-diffracted beam. 

The response electric fields of substrate and film are summed \cite{Pesquera2011}, and added to an intensity background coming from the dark count rate of the detector and a diffuse scattering on atoms, much stronger in the case of substrates containing Gd. Note that due to variations with diffracting angle of the respective phases of the electric fields, the extracted film peak position can be different from the local maximum of signal in Figure~\ref{fig:XRD}, seen by a shift of the arrow. We take into account the finite-resolution broadening of the peaks by a convolution with the instrument function, approximated to be a Gaussian curve of full-width at half-maximum \SI{0.010}{deg}. An additional broadening of the film peaks due to inhomogeneities is introduced as a fitting parameter. Only data within $\pm$\SI{2.4}{deg} of $2\theta$ angle around the film peak is used for the fits. The averages of the measurements on our GGG and YSGG substrates provide lattice parameters of \SI{1.2382}{\nano\meter} and \SI{1.2460}{\nano\meter} $\pm$~\SI{0.0001}{\nano\meter}, respectively.

The out-of-plane lattice parameter of the rhombohedrally-distorted film, defined as the cubic lattice parameter having the same $d_{111}$ spacing between (111)-oriented planes, is given by $a_{\perp}=\sqrt{3}d_{111}$. In the absence of structural relaxation, the equivalent lattice parameter of the unstrained film is then given by $a=a_{\rm{subs}}-[(1-\mu_{111})/(1+\mu_{111})]\Delta{}a_{\perp}$, with $a_{\rm{subs}}$ the substrate lattice parameter, $\mu_{111}$ the Poisson coefficient for (111)-oriented films (0.29 for YIG \cite{Hansen1983}) and $\Delta{}a_{\perp}=a_{\rm{subs}}-a_{\perp}$ the misfit.

\subsection{Transmission Electron Microscopy}
Lamellas are prepared by Focused Ion Beam, on a Dual Beam Helios 650 equipment (FIB-SEM) from Thermofisher. The structural characterization is performed relying on High-Resolution Scanning Transmission Electron Microscopy with High Angular Annular dark field detector (HRSTEM-HAADF). The measurements are performed in a probe-corrected FEI Titan ranging 80--\SI{300}{\kilo\electronvolt}, complemented by the energy-dispersive X-ray spectrometer EDS Oxford Instruments Ultim Max TLE 100. The energy-dispersive X-ray spectroscopy analysis is performed in the Oxford Aztec software and the strain analysis is performed by Geometric Phase Analysis following the procedures reported in Refs.\ \cite{Hytch1998,Huee2005,Hytch2001}.

\subsection{Measurement of temperature-dependent magnetization and molecular-field-coefficients model}
Magnetometry measurements have been performed using an MPMS3 instrument from Quantum Design, operated in VSM-SQUID mode. Magnetic half-hysteresis loops are measured as a function of temperature in order to extract the saturation magnetization $M_{\rm{s}}(T)$ in the range 10--\SI{400}{\kelvin}. The field is swept in a range of $\pm$~\SI{31}{\milli\tesla} for measurements on GGG and $\pm$~\SI{735}{\milli\tesla} for measurements on YSGG substrates. The background subtraction from paramagnetic substrates such as GGG is a well known issue for garnet thin film magnetometry at low temperatures and has been performed carefully by taking into account the non-linearity of the field-current response of the superconducting coils from the magnetometer. Background data is presented in Supplementary Note V. Larger error bars for GGG substrate result from this background affecting the measurements. For films grown on YSGG, this subtraction becomes straightforward owing to the weakly diamagnetic behavior of the substrate. Noting that the magnetization in ultrathin films is often lower than in the bulk, due to various extrinsic reductions of the magnetic moments or to the formation of an interfacial non-magnetic layer, the possibility of a lower apparent magnetization needs to be kept in mind when comparing with the bulk magnetization values expected from a molecular-field-coefficients model.

The measurements are analyzed and fit with a molecular-field-coefficients model similar to previous implementations \cite{Gross2024}, after taking into account the evolution of the lattice parameter and the individual thickness-area product of each sample. For Y-YIG-2--7, the substitution levels by non-magnetic cations in octahedral sites $k_{\rm{a}}$ and in tetrahedral sites $k_{\rm{d}}$ are the free fitting parameters. For Y-YIG-1 with $x<0$, $k_{\rm{a}}$ and $k_{\rm{d}}$ are fixed at zero, while the substitution level by $\ce{Fe^{3+}}$ cations in dodecahedral sites $k_{\rm{cF}}$ is the free parameter. The temperature range used for the fits is 10--\SI{360}{\kelvin}, to avoid deviations at higher temperatures close to $T_{\rm{C}}$. Substitution in octahedral sites leads to an increase in the zero-temperature magnetization, as it strengthens the ferrimagnetic imbalance of opposite magnetic moments from the \ce{Fe^{3+}} cations of the two sub-lattices. They contribute a constant $15\,\mu_{\rm{B}}$ (Bohr magneton) per formula unit for cations in tetrahedral sites, and originally $-10\,\mu_{\rm{B}}$ per formula unit in octahedral sites, diluted by the diamagnetic atoms. A significant canting of the magnetic moments away from anti-collinear orientations is not expected for low levels of substitution up to $x\approx0.5$ \cite{Geller1966a,Dionne1970}.

\subsection{X-ray photoelectron spectroscopy}
X-ray photoelectron spectroscopy analyses have been carried out in a K-Alpha apparatus from Thermo Fisher Scientific. After cleaning of the surfaces with a soft Ar plasma, the samples are immediately introduced in a ultrahigh vacuum chamber ($10^{-9}\,$\si{\milli\bar} range). The X-ray source is Al K$\alpha$ radiation (\SI{1486.6}{\electronvolt}) and a flood gun is employed to minimize the effects of surface charging, providing spectra without charging artifacts. The irradiated area is an elongated ellipse focused on about \SI{400}{\micro\meter}. The ejected electrons are collected by a hemispherical analyzer at \SI{100}{\electronvolt} constant pass energy for survey scans and \SI{30}{\electronvolt} for detailed peak scans aimed for quantification. Detailed scans are repeated 10 times with a step size of \SI{0.1}{\electronvolt}. The energy scale is calibrated with the C 1s line from the adventitious carbon contamination at \SI{285.0}{\electronvolt}. Measurements are done at a constant angle, with normal incidence \SI{90}{deg} between the sample surface and the analyzer. Quantitative analysis is conducted on Fe 2p$_{3/2}$, O 1s, C 1s and Y 3d$_{5/2}$ peaks. A Shirley background is subtracted to each peak, for direct integration in the case of Fe 2p$_{3/2}$, or multiple-peak fitting for O 1s, C 1s and Y 3d$_{5/2}$ peaks, using Voigt profiles with \SI{80}{\percent} Gaussian and \SI{20}{\percent} Lorentzian. Sensitivities affecting respective peak amplitudes are found consistent with previous X-ray photoelectron characterization of iron garnet thin films \cite{Rosenberg2021}. 

\subsection{Setup for ferromagnetic resonance}
For measurements at room temperature, a custom FMR instrumentation is used to record changes in the microwave transmission $S_{21}$ of a coplanar waveguide (CPW). The garnet films are put in contact with an impedance-matched constriction in the CPW (line \SI{75}{\micro\meter}, gaps \SI{140}{\micro\meter}). A pair of frequency-detuned microwave signal generators and a mixer are used to record a demodulated signal at a MHz frequency, carrying the amplitude and phase of the microwave signal transmitted though the CPW. The microwave power is kept at \SI{-20}{dBm} or \SI{-10}{dBm} depending on linewidth, low enough to remain in a linear FMR excitation. The external magnetic field is applied in the sample plane, recorded by a Hall probe placed close to the magnetic sample, and swept at several fixed microwave frequencies. A pair of modulation coils attached to the main electromagnet is connected to a sinusoidal current source to provide a small AC varying magnetic field in addition to the resonance field. It is used for lock-in detection of microwave transmission with respect to the external magnetic field, enabling the detection of the field-derivative signals of both amplitude and phase, or equivalently, $\real{\partial{}S_{21}/\partial{}B}$ and $\imag{\partial{}S_{21}/\partial{}B}$. These signals have analytical lineshapes derived from the magnetic susceptibility, which can be fit to the data in order to extract resonance field and linewidth. To include the frequency shift required to best fit the data in room-temperature FMR measurements, the two-magnon scattering term is included in the Kittel formula as $\omega=\gamma\mu_0\sqrt{H_{\rm{ext}}(H_{\rm{ext}}+M_{\rm{eff}})}-\gamma\delta{}B_{\rm{TMS}}$ \cite{Arias1999a,Jermain2016}, where $\delta{}B_{\rm{TMS}}$ is the magnon frequency renormalization shift, see also Supplementary Note VII.

For measurements at low temperatures, the CPW is attached to a cryogenic measurement rod holding microwave lines and fitted in a variable temperature insert. The microwave transmission through the PCB is recorded using a Vector Network Analyzer (VNA). The $S_{21}$ transmission from the VNA is recorded as a dense $(B,f)$ map with field steps of \SI{10}{\milli\tesla}. This enables a numerical normalization of the $S_{21}$ data with respect to $S_{21}$ at $\pm{}B_{\rm{diff}}$ ($B_{\rm{diff}}$ larger than the FMR linewidth $\mu_0\Delta{}H$), removing the dependence of $S_{21}$ with field that is not related to the magnetic sample but to the instrumentation components, despite the use of non-magnetic assemblies and conductors as much as possible. The external magnetic field is applied in the film plane. For some frequencies, standing waves are formed due to reflection in microwave components, which affects phase and amplitude of the microwave signal and prevents a precise linewidth extraction. The corresponding data points are discarded.

The FMR peaks in $S_{21}(f)$ are then fit independently for the different values of the magnetic field, for each of which a resonance frequency $f_{\rm{res}}$ and a resonance linewidth $\mu_0\Delta{}H$ are obtained. At a fixed temperature, the dependence of $\omega(H_{\rm{ext}})$ on the external field is first fit with the Kittel law $\omega=\gamma\mu_0\sqrt{(H_{\rm{ext}}+H_{\rm{K}})(H_{\rm{ext}}+H_{\rm{K}}+M_{\rm{eff}})}$, containing a phenomenological term acting as an in-plane anisotropy, which becomes relevant at lower temperatures (see Supplementary Note VIII). This term can also include the potential small frequency shift due to two-magnon scattering processes. Fits resulting in either a too large uncertainty on the linewidth or a resonance frequency deviating from the fit line are excluded from the analysis and denoted by crosses in Figure~\ref{fig:FMRlowT}c,d and \ref{fig:alternatives}e. We present in Figure \ref{fig:FMRlowT}c,d and \ref{fig:alternatives}e different models for the linewidth: one including bare Gilbert damping, one adding a two-magnon scattering term and one adding a slow-relaxer impurities contribution. The two-magnon scattering contribution is added as a new term $\mu_0\Delta{}H_{\rm{TMS}}=\Gamma(T)\sin^{-1}[(\sqrt{\omega^2+(\omega_0/2)^2}-\omega_0/2)/(\sqrt{\omega^2+(\omega_0/2)^2}+\omega_0/2)]^{1/2}$, where $\Gamma(T)$ determines the limit contribution at large fields and $\omega_0=\gamma\mu_0M_{\rm{eff}}$ scales its frequency dependence \cite{Lenz2006}. The slow-relaxing impurities contribution is added as a new term $\mu_0\Delta{}H_{\rm{SRI}}=A(T)\omega\tau/[1+(\omega\tau)^2]$, where $A(T)$ determines the peak contribution to the linewidth and $\tau$ is the relaxation time for the impurities \cite{Seiden1964}. These two terms differ mainly in the high-frequency limit: the two-magnon scattering term monotonically increases and saturates at a finite value $\pi\Gamma(T)/2$, whereas the slow-relaxer impurities term decays to zero after reaching a peak for $\omega\tau=1$. At each $T$, we fit the $\mu_0\Delta{}H(\omega)$ data with two free parameters for a bare Gilbert model [$\mu_0\Delta{}H_0(T)$, $\alpha(T)$], three free parameters for the two-magnon scettering model [$\mu_0\Delta{}H_0(T)$, $\alpha(T)$, $\Gamma(T)$] and four free parameters for the slow-relaxer impurities model [$\mu_0\Delta{}H_0(T)$, $\alpha(T)$, $A(T)$ and $\tau(T)$]. All these parameters are constrained to remain positive, and the fit uncertainties are not considered when some parameters converge to zero. Since $\tau(T)$ varies only by a few percent and less than its estimated uncertainty over the whole temperature range, we then fix it in the model to its average value over all temperatures $\tau=$ \SI{8.16}{\pico\second}.

\clearpage
\bibliography{main}

\begin{thebibliography}{71}%
\makeatletter
\providecommand \@ifxundefined [1]{%
 \@ifx{#1\undefined}
}%
\providecommand \@ifnum [1]{%
 \ifnum #1\expandafter \@firstoftwo
 \else \expandafter \@secondoftwo
 \fi
}%
\providecommand \@ifx [1]{%
 \ifx #1\expandafter \@firstoftwo
 \else \expandafter \@secondoftwo
 \fi
}%
\providecommand \natexlab [1]{#1}%
\providecommand \enquote  [1]{``#1''}%
\providecommand \bibnamefont  [1]{#1}%
\providecommand \bibfnamefont [1]{#1}%
\providecommand \citenamefont [1]{#1}%
\providecommand \href@noop [0]{\@secondoftwo}%
\providecommand \href [0]{\begingroup \@sanitize@url \@href}%
\providecommand \@href[1]{\@@startlink{#1}\@@href}%
\providecommand \@@href[1]{\endgroup#1\@@endlink}%
\providecommand \@sanitize@url [0]{\catcode `\\12\catcode `\$12\catcode
  `\&12\catcode `\#12\catcode `\^12\catcode `\_12\catcode `\%12\relax}%
\providecommand \@@startlink[1]{}%
\providecommand \@@endlink[0]{}%
\providecommand \url  [0]{\begingroup\@sanitize@url \@url }%
\providecommand \@url [1]{\endgroup\@href {#1}{\urlprefix }}%
\providecommand \urlprefix  [0]{URL }%
\providecommand \Eprint [0]{\href }%
\providecommand \doibase [0]{https://doi.org/}%
\providecommand \selectlanguage [0]{\@gobble}%
\providecommand \bibinfo  [0]{\@secondoftwo}%
\providecommand \bibfield  [0]{\@secondoftwo}%
\providecommand \translation [1]{[#1]}%
\providecommand \BibitemOpen [0]{}%
\providecommand \bibitemStop [0]{}%
\providecommand \bibitemNoStop [0]{.\EOS\space}%
\providecommand \EOS [0]{\spacefactor3000\relax}%
\providecommand \BibitemShut  [1]{\csname bibitem#1\endcsname}%
\let\auto@bib@innerbib\@empty
\bibitem [{\citenamefont {Heinrich}\ \emph {et~al.}(2011)\citenamefont
  {Heinrich}, \citenamefont {Burrowes}, \citenamefont {Montoya}, \citenamefont
  {Kardasz}, \citenamefont {Girt}, \citenamefont {Song}, \citenamefont {Sun},\
  and\ \citenamefont {Wu}}]{Heinrich2011}%
  \BibitemOpen
  \bibfield  {author} {\bibinfo {author} {\bibfnamefont {B.}~\bibnamefont
  {Heinrich}}, \bibinfo {author} {\bibfnamefont {C.}~\bibnamefont {Burrowes}},
  \bibinfo {author} {\bibfnamefont {E.}~\bibnamefont {Montoya}}, \bibinfo
  {author} {\bibfnamefont {B.}~\bibnamefont {Kardasz}}, \bibinfo {author}
  {\bibfnamefont {E.}~\bibnamefont {Girt}}, \bibinfo {author} {\bibfnamefont
  {Y.-Y.}\ \bibnamefont {Song}}, \bibinfo {author} {\bibfnamefont
  {Y.}~\bibnamefont {Sun}},\ and\ \bibinfo {author} {\bibfnamefont
  {M.}~\bibnamefont {Wu}},\ }\bibfield  {title} {\bibinfo {title} {{Spin
  Pumping at the Magnetic Insulator (YIG)/Normal Metal (Au) Interfaces}},\
  }\href {https://doi.org/10.1103/physrevlett.107.066604} {\bibfield  {journal}
  {\bibinfo  {journal} {Phys. Rev. Lett.}\ }\textbf {\bibinfo {volume} {107}},\
  \bibinfo {pages} {066604} (\bibinfo {year} {2011})}\BibitemShut {NoStop}%
\bibitem [{\citenamefont {Sun}\ \emph {et~al.}(2012)\citenamefont {Sun},
  \citenamefont {Song}, \citenamefont {Chang}, \citenamefont {Kabatek},
  \citenamefont {Jantz}, \citenamefont {Schneider}, \citenamefont {Wu},
  \citenamefont {Schultheiss},\ and\ \citenamefont {Hoffmann}}]{Sun2012}%
  \BibitemOpen
  \bibfield  {author} {\bibinfo {author} {\bibfnamefont {Y.}~\bibnamefont
  {Sun}}, \bibinfo {author} {\bibfnamefont {Y.-Y.}\ \bibnamefont {Song}},
  \bibinfo {author} {\bibfnamefont {H.}~\bibnamefont {Chang}}, \bibinfo
  {author} {\bibfnamefont {M.}~\bibnamefont {Kabatek}}, \bibinfo {author}
  {\bibfnamefont {M.}~\bibnamefont {Jantz}}, \bibinfo {author} {\bibfnamefont
  {W.}~\bibnamefont {Schneider}}, \bibinfo {author} {\bibfnamefont
  {M.}~\bibnamefont {Wu}}, \bibinfo {author} {\bibfnamefont {H.}~\bibnamefont
  {Schultheiss}},\ and\ \bibinfo {author} {\bibfnamefont {A.}~\bibnamefont
  {Hoffmann}},\ }\bibfield  {title} {\bibinfo {title} {{Growth and
  ferromagnetic resonance properties of nanometer-thick yttrium iron garnet
  films}},\ }\href {https://doi.org/10.1063/1.4759039} {\bibfield  {journal}
  {\bibinfo  {journal} {Appl. Phys. Lett.}\ }\textbf {\bibinfo {volume}
  {101}},\ \bibinfo {pages} {152405} (\bibinfo {year} {2012})}\BibitemShut
  {NoStop}%
\bibitem [{\citenamefont {d'Allivy Kelly}\ \emph {et~al.}(2013)\citenamefont
  {d'Allivy Kelly}, \citenamefont {Anane}, \citenamefont {Bernard},
  \citenamefont {Ben~Youssef}, \citenamefont {Hahn}, \citenamefont
  {Molpeceres}, \citenamefont {Carr{\'{e}}t{\'{e}}ro}, \citenamefont {Jacquet},
  \citenamefont {Deranlot}, \citenamefont {Bortolotti}, \citenamefont
  {Lebourgeois}, \citenamefont {Mage}, \citenamefont {de~Loubens},
  \citenamefont {Klein}, \citenamefont {Cros},\ and\ \citenamefont
  {Fert}}]{AllivyKelly2013}%
  \BibitemOpen
  \bibfield  {author} {\bibinfo {author} {\bibfnamefont {O.}~\bibnamefont
  {d'Allivy Kelly}}, \bibinfo {author} {\bibfnamefont {A.}~\bibnamefont
  {Anane}}, \bibinfo {author} {\bibfnamefont {R.}~\bibnamefont {Bernard}},
  \bibinfo {author} {\bibfnamefont {J.}~\bibnamefont {Ben~Youssef}}, \bibinfo
  {author} {\bibfnamefont {C.}~\bibnamefont {Hahn}}, \bibinfo {author}
  {\bibfnamefont {A.~H.}\ \bibnamefont {Molpeceres}}, \bibinfo {author}
  {\bibfnamefont {C.}~\bibnamefont {Carr{\'{e}}t{\'{e}}ro}}, \bibinfo {author}
  {\bibfnamefont {E.}~\bibnamefont {Jacquet}}, \bibinfo {author} {\bibfnamefont
  {C.}~\bibnamefont {Deranlot}}, \bibinfo {author} {\bibfnamefont
  {P.}~\bibnamefont {Bortolotti}}, \bibinfo {author} {\bibfnamefont
  {R.}~\bibnamefont {Lebourgeois}}, \bibinfo {author} {\bibfnamefont {J.-C.}\
  \bibnamefont {Mage}}, \bibinfo {author} {\bibfnamefont {G.}~\bibnamefont
  {de~Loubens}}, \bibinfo {author} {\bibfnamefont {O.}~\bibnamefont {Klein}},
  \bibinfo {author} {\bibfnamefont {V.}~\bibnamefont {Cros}},\ and\ \bibinfo
  {author} {\bibfnamefont {A.}~\bibnamefont {Fert}},\ }\bibfield  {title}
  {\bibinfo {title} {{Inverse spin Hall effect in nanometer-thick yttrium iron
  garnet/Pt system}},\ }\href {https://doi.org/10.1063/1.4819157} {\bibfield
  {journal} {\bibinfo  {journal} {Appl. Phys. Lett.}\ }\textbf {\bibinfo
  {volume} {103}},\ \bibinfo {pages} {082408} (\bibinfo {year}
  {2013})}\BibitemShut {NoStop}%
\bibitem [{\citenamefont {Wang}\ \emph {et~al.}(2013)\citenamefont {Wang},
  \citenamefont {Du}, \citenamefont {Pu}, \citenamefont {Adur}, \citenamefont
  {Hammel},\ and\ \citenamefont {Yang}}]{Wang2013b}%
  \BibitemOpen
  \bibfield  {author} {\bibinfo {author} {\bibfnamefont {H.~L.}\ \bibnamefont
  {Wang}}, \bibinfo {author} {\bibfnamefont {C.~H.}\ \bibnamefont {Du}},
  \bibinfo {author} {\bibfnamefont {Y.}~\bibnamefont {Pu}}, \bibinfo {author}
  {\bibfnamefont {R.}~\bibnamefont {Adur}}, \bibinfo {author} {\bibfnamefont
  {P.~C.}\ \bibnamefont {Hammel}},\ and\ \bibinfo {author} {\bibfnamefont
  {F.~Y.}\ \bibnamefont {Yang}},\ }\bibfield  {title} {\bibinfo {title} {{Large
  spin pumping from epitaxial Y$_{3}$Fe$_{5}$O$_{12}$ thin films to Pt and W
  layers}},\ }\href {https://doi.org/10.1103/physrevb.88.100406} {\bibfield
  {journal} {\bibinfo  {journal} {Phys. Rev. B}\ }\textbf {\bibinfo {volume}
  {88}},\ \bibinfo {pages} {100406} (\bibinfo {year} {2013})}\BibitemShut
  {NoStop}%
\bibitem [{\citenamefont {Onbasli}\ \emph {et~al.}(2014)\citenamefont
  {Onbasli}, \citenamefont {Kehlberger}, \citenamefont {Kim}, \citenamefont
  {Jakob}, \citenamefont {Kl{\"{a}}ui}, \citenamefont {Chumak}, \citenamefont
  {Hillebrands},\ and\ \citenamefont {Ross}}]{Onbasli2014}%
  \BibitemOpen
  \bibfield  {author} {\bibinfo {author} {\bibfnamefont {M.~C.}\ \bibnamefont
  {Onbasli}}, \bibinfo {author} {\bibfnamefont {A.}~\bibnamefont {Kehlberger}},
  \bibinfo {author} {\bibfnamefont {D.~H.}\ \bibnamefont {Kim}}, \bibinfo
  {author} {\bibfnamefont {G.}~\bibnamefont {Jakob}}, \bibinfo {author}
  {\bibfnamefont {M.}~\bibnamefont {Kl{\"{a}}ui}}, \bibinfo {author}
  {\bibfnamefont {A.~V.}\ \bibnamefont {Chumak}}, \bibinfo {author}
  {\bibfnamefont {B.}~\bibnamefont {Hillebrands}},\ and\ \bibinfo {author}
  {\bibfnamefont {C.~A.}\ \bibnamefont {Ross}},\ }\bibfield  {title} {\bibinfo
  {title} {{Pulsed laser deposition of epitaxial yttrium iron garnet films with
  low Gilbert damping and bulk-like magnetization}},\ }\href
  {https://doi.org/10.1063/1.4896936} {\bibfield  {journal} {\bibinfo
  {journal} {APL Mater.}\ }\textbf {\bibinfo {volume} {2}},\ \bibinfo {pages}
  {106102} (\bibinfo {year} {2014})}\BibitemShut {NoStop}%
\bibitem [{\citenamefont {Jermain}\ \emph {et~al.}(2016)\citenamefont
  {Jermain}, \citenamefont {Paik}, \citenamefont {Aradhya}, \citenamefont
  {Buhrman}, \citenamefont {Schlom},\ and\ \citenamefont
  {Ralph}}]{Jermain2016}%
  \BibitemOpen
  \bibfield  {author} {\bibinfo {author} {\bibfnamefont {C.~L.}\ \bibnamefont
  {Jermain}}, \bibinfo {author} {\bibfnamefont {H.}~\bibnamefont {Paik}},
  \bibinfo {author} {\bibfnamefont {S.~V.}\ \bibnamefont {Aradhya}}, \bibinfo
  {author} {\bibfnamefont {R.~A.}\ \bibnamefont {Buhrman}}, \bibinfo {author}
  {\bibfnamefont {D.~G.}\ \bibnamefont {Schlom}},\ and\ \bibinfo {author}
  {\bibfnamefont {D.~C.}\ \bibnamefont {Ralph}},\ }\bibfield  {title} {\bibinfo
  {title} {{Low-damping sub-10-nm thin films of lutetium iron garnet grown by
  molecular-beam epitaxy}},\ }\href {https://doi.org/10.1063/1.4967695}
  {\bibfield  {journal} {\bibinfo  {journal} {Appl. Phys. Lett.}\ }\textbf
  {\bibinfo {volume} {109}},\ \bibinfo {pages} {192408} (\bibinfo {year}
  {2016})}\BibitemShut {NoStop}%
\bibitem [{\citenamefont {Hauser}\ \emph {et~al.}(2016)\citenamefont {Hauser},
  \citenamefont {Richter}, \citenamefont {Homonnay}, \citenamefont
  {Eisenschmidt}, \citenamefont {Qaid}, \citenamefont {Deniz}, \citenamefont
  {Hesse}, \citenamefont {Sawicki}, \citenamefont {Ebbinghaus},\ and\
  \citenamefont {Schmidt}}]{Hauser2016}%
  \BibitemOpen
  \bibfield  {author} {\bibinfo {author} {\bibfnamefont {C.}~\bibnamefont
  {Hauser}}, \bibinfo {author} {\bibfnamefont {T.}~\bibnamefont {Richter}},
  \bibinfo {author} {\bibfnamefont {N.}~\bibnamefont {Homonnay}}, \bibinfo
  {author} {\bibfnamefont {C.}~\bibnamefont {Eisenschmidt}}, \bibinfo {author}
  {\bibfnamefont {M.}~\bibnamefont {Qaid}}, \bibinfo {author} {\bibfnamefont
  {H.}~\bibnamefont {Deniz}}, \bibinfo {author} {\bibfnamefont
  {D.}~\bibnamefont {Hesse}}, \bibinfo {author} {\bibfnamefont
  {M.}~\bibnamefont {Sawicki}}, \bibinfo {author} {\bibfnamefont {S.~G.}\
  \bibnamefont {Ebbinghaus}},\ and\ \bibinfo {author} {\bibfnamefont
  {G.}~\bibnamefont {Schmidt}},\ }\bibfield  {title} {\bibinfo {title}
  {{Yttrium Iron Garnet Thin Films with Very Low Damping Obtained by
  Recrystallization of Amorphous Material}},\ }\href
  {https://doi.org/10.1038/srep20827} {\bibfield  {journal} {\bibinfo
  {journal} {Sci. Rep.}\ }\textbf {\bibinfo {volume} {6}},\ \bibinfo {pages}
  {20827} (\bibinfo {year} {2016})}\BibitemShut {NoStop}%
\bibitem [{\citenamefont {Beaulieu}\ \emph {et~al.}(2018)\citenamefont
  {Beaulieu}, \citenamefont {Kervarec}, \citenamefont {Thiery}, \citenamefont
  {Klein}, \citenamefont {Naletov}, \citenamefont {Hurdequint}, \citenamefont
  {de~Loubens}, \citenamefont {Youssef},\ and\ \citenamefont
  {Vukadinovic}}]{Beaulieu2018}%
  \BibitemOpen
  \bibfield  {author} {\bibinfo {author} {\bibfnamefont {N.}~\bibnamefont
  {Beaulieu}}, \bibinfo {author} {\bibfnamefont {N.}~\bibnamefont {Kervarec}},
  \bibinfo {author} {\bibfnamefont {N.}~\bibnamefont {Thiery}}, \bibinfo
  {author} {\bibfnamefont {O.}~\bibnamefont {Klein}}, \bibinfo {author}
  {\bibfnamefont {V.}~\bibnamefont {Naletov}}, \bibinfo {author} {\bibfnamefont
  {H.}~\bibnamefont {Hurdequint}}, \bibinfo {author} {\bibfnamefont
  {G.}~\bibnamefont {de~Loubens}}, \bibinfo {author} {\bibfnamefont {J.~B.}\
  \bibnamefont {Youssef}},\ and\ \bibinfo {author} {\bibfnamefont
  {N.}~\bibnamefont {Vukadinovic}},\ }\bibfield  {title} {\bibinfo {title}
  {{Temperature Dependence of Magnetic Properties of a Ultrathin Yttrium-Iron
  Garnet Film Grown by Liquid Phase Epitaxy: Effect of a Pt Overlayer}},\
  }\href {https://doi.org/10.1109/lmag.2018.2868700} {\bibfield  {journal}
  {\bibinfo  {journal} {IEEE Magn. Lett.}\ }\textbf {\bibinfo {volume} {9}},\
  \bibinfo {pages} {3706005} (\bibinfo {year} {2018})}\BibitemShut {NoStop}%
\bibitem [{\citenamefont {Dubs}\ \emph {et~al.}(2020)\citenamefont {Dubs},
  \citenamefont {Surzhenko}, \citenamefont {Thomas}, \citenamefont {Osten},
  \citenamefont {Schneider}, \citenamefont {Lenz}, \citenamefont {Grenzer},
  \citenamefont {H{\"{u}}bner},\ and\ \citenamefont {Wendler}}]{Dubs2020}%
  \BibitemOpen
  \bibfield  {author} {\bibinfo {author} {\bibfnamefont {C.}~\bibnamefont
  {Dubs}}, \bibinfo {author} {\bibfnamefont {O.}~\bibnamefont {Surzhenko}},
  \bibinfo {author} {\bibfnamefont {R.}~\bibnamefont {Thomas}}, \bibinfo
  {author} {\bibfnamefont {J.}~\bibnamefont {Osten}}, \bibinfo {author}
  {\bibfnamefont {T.}~\bibnamefont {Schneider}}, \bibinfo {author}
  {\bibfnamefont {K.}~\bibnamefont {Lenz}}, \bibinfo {author} {\bibfnamefont
  {J.}~\bibnamefont {Grenzer}}, \bibinfo {author} {\bibfnamefont
  {R.}~\bibnamefont {H{\"{u}}bner}},\ and\ \bibinfo {author} {\bibfnamefont
  {E.}~\bibnamefont {Wendler}},\ }\bibfield  {title} {\bibinfo {title} {{Low
  damping and microstructural perfection of sub-40nm-thin yttrium iron garnet
  films grown by liquid phase epitaxy}},\ }\href
  {https://doi.org/10.1103/physrevmaterials.4.024416} {\bibfield  {journal}
  {\bibinfo  {journal} {Phys. Rev. Mater.}\ }\textbf {\bibinfo {volume} {4}},\
  \bibinfo {pages} {024416} (\bibinfo {year} {2020})}\BibitemShut {NoStop}%
\bibitem [{\citenamefont {Ding}\ \emph {et~al.}(2020)\citenamefont {Ding},
  \citenamefont {Liu}, \citenamefont {Zhang}, \citenamefont {Erugu},
  \citenamefont {Quan}, \citenamefont {Yu}, \citenamefont {McCollum},
  \citenamefont {Mo}, \citenamefont {Yang}, \citenamefont {Ding}, \citenamefont
  {Xu}, \citenamefont {Tang}, \citenamefont {Yang},\ and\ \citenamefont
  {Wu}}]{Ding2020a}%
  \BibitemOpen
  \bibfield  {author} {\bibinfo {author} {\bibfnamefont {J.}~\bibnamefont
  {Ding}}, \bibinfo {author} {\bibfnamefont {C.}~\bibnamefont {Liu}}, \bibinfo
  {author} {\bibfnamefont {Y.}~\bibnamefont {Zhang}}, \bibinfo {author}
  {\bibfnamefont {U.}~\bibnamefont {Erugu}}, \bibinfo {author} {\bibfnamefont
  {Z.}~\bibnamefont {Quan}}, \bibinfo {author} {\bibfnamefont {R.}~\bibnamefont
  {Yu}}, \bibinfo {author} {\bibfnamefont {E.}~\bibnamefont {McCollum}},
  \bibinfo {author} {\bibfnamefont {S.}~\bibnamefont {Mo}}, \bibinfo {author}
  {\bibfnamefont {S.}~\bibnamefont {Yang}}, \bibinfo {author} {\bibfnamefont
  {H.}~\bibnamefont {Ding}}, \bibinfo {author} {\bibfnamefont {X.}~\bibnamefont
  {Xu}}, \bibinfo {author} {\bibfnamefont {J.}~\bibnamefont {Tang}}, \bibinfo
  {author} {\bibfnamefont {X.}~\bibnamefont {Yang}},\ and\ \bibinfo {author}
  {\bibfnamefont {M.}~\bibnamefont {Wu}},\ }\bibfield  {title} {\bibinfo
  {title} {{Nanometer-Thick Yttrium Iron Garnet Films with Perpendicular
  Anisotropy and Low Damping}},\ }\href
  {https://doi.org/10.1103/physrevapplied.14.014017} {\bibfield  {journal}
  {\bibinfo  {journal} {Phys. Rev. Appl.}\ }\textbf {\bibinfo {volume} {14}},\
  \bibinfo {pages} {014017} (\bibinfo {year} {2020})}\BibitemShut {NoStop}%
\bibitem [{\citenamefont {Schmidt}\ \emph {et~al.}(2020)\citenamefont
  {Schmidt}, \citenamefont {Hauser}, \citenamefont {Trempler}, \citenamefont
  {Paleschke},\ and\ \citenamefont {Papaioannou}}]{Schmidt2020}%
  \BibitemOpen
  \bibfield  {author} {\bibinfo {author} {\bibfnamefont {G.}~\bibnamefont
  {Schmidt}}, \bibinfo {author} {\bibfnamefont {C.}~\bibnamefont {Hauser}},
  \bibinfo {author} {\bibfnamefont {P.}~\bibnamefont {Trempler}}, \bibinfo
  {author} {\bibfnamefont {M.}~\bibnamefont {Paleschke}},\ and\ \bibinfo
  {author} {\bibfnamefont {E.~T.}\ \bibnamefont {Papaioannou}},\ }\bibfield
  {title} {\bibinfo {title} {{Ultra Thin Films of Yttrium Iron Garnet with Very
  Low Damping: A Review}},\ }\href {https://doi.org/10.1002/pssb.201900644}
  {\bibfield  {journal} {\bibinfo  {journal} {Phys. Status Solidi B}\ }\textbf
  {\bibinfo {volume} {257}},\ \bibinfo {pages} {1900644} (\bibinfo {year}
  {2020})}\BibitemShut {NoStop}%
\bibitem [{\citenamefont {Kajiwara}\ \emph {et~al.}(2010)\citenamefont
  {Kajiwara}, \citenamefont {Harii}, \citenamefont {Takahashi}, \citenamefont
  {Ohe}, \citenamefont {Uchida}, \citenamefont {Mizuguchi}, \citenamefont
  {Umezawa}, \citenamefont {Kawai}, \citenamefont {Ando}, \citenamefont
  {Takanashi}, \citenamefont {Maekawa},\ and\ \citenamefont
  {Saitoh}}]{Kajiwara2010}%
  \BibitemOpen
  \bibfield  {author} {\bibinfo {author} {\bibfnamefont {Y.}~\bibnamefont
  {Kajiwara}}, \bibinfo {author} {\bibfnamefont {K.}~\bibnamefont {Harii}},
  \bibinfo {author} {\bibfnamefont {S.}~\bibnamefont {Takahashi}}, \bibinfo
  {author} {\bibfnamefont {J.}~\bibnamefont {Ohe}}, \bibinfo {author}
  {\bibfnamefont {K.}~\bibnamefont {Uchida}}, \bibinfo {author} {\bibfnamefont
  {M.}~\bibnamefont {Mizuguchi}}, \bibinfo {author} {\bibfnamefont
  {H.}~\bibnamefont {Umezawa}}, \bibinfo {author} {\bibfnamefont
  {H.}~\bibnamefont {Kawai}}, \bibinfo {author} {\bibfnamefont
  {K.}~\bibnamefont {Ando}}, \bibinfo {author} {\bibfnamefont {K.}~\bibnamefont
  {Takanashi}}, \bibinfo {author} {\bibfnamefont {S.}~\bibnamefont {Maekawa}},\
  and\ \bibinfo {author} {\bibfnamefont {E.}~\bibnamefont {Saitoh}},\
  }\bibfield  {title} {\bibinfo {title} {{Transmission of electrical signals by
  spin-wave interconversion in a magnetic insulator}},\ }\href
  {https://doi.org/10.1038/nature08876} {\bibfield  {journal} {\bibinfo
  {journal} {Nature}\ }\textbf {\bibinfo {volume} {464}},\ \bibinfo {pages}
  {262} (\bibinfo {year} {2010})}\BibitemShut {NoStop}%
\bibitem [{\citenamefont {Spencer}\ \emph {et~al.}(1959)\citenamefont
  {Spencer}, \citenamefont {LeCraw},\ and\ \citenamefont
  {Clogston}}]{Spencer1959}%
  \BibitemOpen
  \bibfield  {author} {\bibinfo {author} {\bibfnamefont {E.~G.}\ \bibnamefont
  {Spencer}}, \bibinfo {author} {\bibfnamefont {R.~C.}\ \bibnamefont
  {LeCraw}},\ and\ \bibinfo {author} {\bibfnamefont {A.~M.}\ \bibnamefont
  {Clogston}},\ }\bibfield  {title} {\bibinfo {title} {{Low-Temperature
  Line-Width Maximum in Yttrium Iron Garnet}},\ }\href
  {https://doi.org/10.1103/physrevlett.3.32} {\bibfield  {journal} {\bibinfo
  {journal} {Phys. Rev. Lett.}\ }\textbf {\bibinfo {volume} {3}},\ \bibinfo
  {pages} {32} (\bibinfo {year} {1959})}\BibitemShut {NoStop}%
\bibitem [{\citenamefont {Pirro}\ \emph {et~al.}(2021)\citenamefont {Pirro},
  \citenamefont {Vasyuchka}, \citenamefont {Serga},\ and\ \citenamefont
  {Hillebrands}}]{Pirro2021}%
  \BibitemOpen
  \bibfield  {author} {\bibinfo {author} {\bibfnamefont {P.}~\bibnamefont
  {Pirro}}, \bibinfo {author} {\bibfnamefont {V.~I.}\ \bibnamefont
  {Vasyuchka}}, \bibinfo {author} {\bibfnamefont {A.~A.}\ \bibnamefont
  {Serga}},\ and\ \bibinfo {author} {\bibfnamefont {B.}~\bibnamefont
  {Hillebrands}},\ }\bibfield  {title} {\bibinfo {title} {{Advances in coherent
  magnonics}},\ }\href {https://doi.org/10.1038/s41578-021-00332-w} {\bibfield
  {journal} {\bibinfo  {journal} {Nat. Rev. Mater.}\ }\textbf {\bibinfo
  {volume} {6}},\ \bibinfo {pages} {1114} (\bibinfo {year} {2021})}\BibitemShut
  {NoStop}%
\bibitem [{\citenamefont {Yuan}\ \emph {et~al.}(2022)\citenamefont {Yuan},
  \citenamefont {Cao}, \citenamefont {Kamra}, \citenamefont {Duine},\ and\
  \citenamefont {Yan}}]{Yuan2022}%
  \BibitemOpen
  \bibfield  {author} {\bibinfo {author} {\bibfnamefont {H.~Y.}\ \bibnamefont
  {Yuan}}, \bibinfo {author} {\bibfnamefont {Y.}~\bibnamefont {Cao}}, \bibinfo
  {author} {\bibfnamefont {A.}~\bibnamefont {Kamra}}, \bibinfo {author}
  {\bibfnamefont {R.~A.}\ \bibnamefont {Duine}},\ and\ \bibinfo {author}
  {\bibfnamefont {P.}~\bibnamefont {Yan}},\ }\bibfield  {title} {\bibinfo
  {title} {{Quantum magnonics: When magnon spintronics meets quantum
  information science}},\ }\href
  {https://doi.org/10.1016/j.physrep.2022.03.002} {\bibfield  {journal}
  {\bibinfo  {journal} {Phys. Rep.}\ }\textbf {\bibinfo {volume} {965}},\
  \bibinfo {pages} {1} (\bibinfo {year} {2022})}\BibitemShut {NoStop}%
\bibitem [{\citenamefont {Li}\ \emph {et~al.}(2019)\citenamefont {Li},
  \citenamefont {Polakovic}, \citenamefont {Wang}, \citenamefont {Xu},
  \citenamefont {Lendinez}, \citenamefont {Zhang}, \citenamefont {Ding},
  \citenamefont {Khaire}, \citenamefont {Saglam}, \citenamefont {Divan},
  \citenamefont {Pearson}, \citenamefont {Kwok}, \citenamefont {Xiao},
  \citenamefont {Novosad}, \citenamefont {Hoffmann},\ and\ \citenamefont
  {Zhang}}]{Li2019}%
  \BibitemOpen
  \bibfield  {author} {\bibinfo {author} {\bibfnamefont {Y.}~\bibnamefont
  {Li}}, \bibinfo {author} {\bibfnamefont {T.}~\bibnamefont {Polakovic}},
  \bibinfo {author} {\bibfnamefont {Y.-L.}\ \bibnamefont {Wang}}, \bibinfo
  {author} {\bibfnamefont {J.}~\bibnamefont {Xu}}, \bibinfo {author}
  {\bibfnamefont {S.}~\bibnamefont {Lendinez}}, \bibinfo {author}
  {\bibfnamefont {Z.}~\bibnamefont {Zhang}}, \bibinfo {author} {\bibfnamefont
  {J.}~\bibnamefont {Ding}}, \bibinfo {author} {\bibfnamefont {T.}~\bibnamefont
  {Khaire}}, \bibinfo {author} {\bibfnamefont {H.}~\bibnamefont {Saglam}},
  \bibinfo {author} {\bibfnamefont {R.}~\bibnamefont {Divan}}, \bibinfo
  {author} {\bibfnamefont {J.}~\bibnamefont {Pearson}}, \bibinfo {author}
  {\bibfnamefont {W.-K.}\ \bibnamefont {Kwok}}, \bibinfo {author}
  {\bibfnamefont {Z.}~\bibnamefont {Xiao}}, \bibinfo {author} {\bibfnamefont
  {V.}~\bibnamefont {Novosad}}, \bibinfo {author} {\bibfnamefont
  {A.}~\bibnamefont {Hoffmann}},\ and\ \bibinfo {author} {\bibfnamefont
  {W.}~\bibnamefont {Zhang}},\ }\bibfield  {title} {\bibinfo {title} {{Strong
  Coupling between Magnons and Microwave Photons in On-Chip
  Ferromagnet-Superconductor Thin-Film Devices}},\ }\href
  {https://doi.org/10.1103/physrevlett.123.107701} {\bibfield  {journal}
  {\bibinfo  {journal} {Phys. Rev. Lett.}\ }\textbf {\bibinfo {volume} {123}},\
  \bibinfo {pages} {107701} (\bibinfo {year} {2019})}\BibitemShut {NoStop}%
\bibitem [{\citenamefont {Hou}\ and\ \citenamefont {Liu}(2019)}]{Hou2019}%
  \BibitemOpen
  \bibfield  {author} {\bibinfo {author} {\bibfnamefont {J.~T.}\ \bibnamefont
  {Hou}}\ and\ \bibinfo {author} {\bibfnamefont {L.}~\bibnamefont {Liu}},\
  }\bibfield  {title} {\bibinfo {title} {{Strong Coupling between Microwave
  Photons and Nanomagnet Magnons}},\ }\href
  {https://doi.org/10.1103/physrevlett.123.107702} {\bibfield  {journal}
  {\bibinfo  {journal} {Phys. Rev. Lett.}\ }\textbf {\bibinfo {volume} {123}},\
  \bibinfo {pages} {107702} (\bibinfo {year} {2019})}\BibitemShut {NoStop}%
\bibitem [{\citenamefont {Guo}\ \emph {et~al.}(2023)\citenamefont {Guo},
  \citenamefont {Russell}, \citenamefont {Lanier}, \citenamefont {Da},
  \citenamefont {Hammel},\ and\ \citenamefont {Yang}}]{Guo2023}%
  \BibitemOpen
  \bibfield  {author} {\bibinfo {author} {\bibfnamefont {S.}~\bibnamefont
  {Guo}}, \bibinfo {author} {\bibfnamefont {D.}~\bibnamefont {Russell}},
  \bibinfo {author} {\bibfnamefont {J.}~\bibnamefont {Lanier}}, \bibinfo
  {author} {\bibfnamefont {H.}~\bibnamefont {Da}}, \bibinfo {author}
  {\bibfnamefont {P.~C.}\ \bibnamefont {Hammel}},\ and\ \bibinfo {author}
  {\bibfnamefont {F.}~\bibnamefont {Yang}},\ }\bibfield  {title} {\bibinfo
  {title} {{Strong on-Chip Microwave Photon{\textendash}Magnon Coupling Using
  Ultralow-Damping Epitaxial Y$_{3}$Fe$_{5}$O$_{12}$ Films at 2 K}},\ }\href
  {https://doi.org/10.1021/acs.nanolett.3c00959} {\bibfield  {journal}
  {\bibinfo  {journal} {Nano Lett.}\ }\textbf {\bibinfo {volume} {23}},\
  \bibinfo {pages} {5055} (\bibinfo {year} {2023})}\BibitemShut {NoStop}%
\bibitem [{\citenamefont {Fukami}\ \emph {et~al.}(2021)\citenamefont {Fukami},
  \citenamefont {Candido}, \citenamefont {Awschalom},\ and\ \citenamefont
  {Flatt{\'{e}}}}]{Fukami2021}%
  \BibitemOpen
  \bibfield  {author} {\bibinfo {author} {\bibfnamefont {M.}~\bibnamefont
  {Fukami}}, \bibinfo {author} {\bibfnamefont {D.~R.}\ \bibnamefont {Candido}},
  \bibinfo {author} {\bibfnamefont {D.~D.}\ \bibnamefont {Awschalom}},\ and\
  \bibinfo {author} {\bibfnamefont {M.~E.}\ \bibnamefont {Flatt{\'{e}}}},\
  }\bibfield  {title} {\bibinfo {title} {{Opportunities for Long-Range
  Magnon-Mediated Entanglement of Spin Qubits via On- and Off-Resonant
  Coupling}},\ }\href {https://doi.org/10.1103/prxquantum.2.040314} {\bibfield
  {journal} {\bibinfo  {journal} {PRX Quantum}\ }\textbf {\bibinfo {volume}
  {2}},\ \bibinfo {pages} {040314} (\bibinfo {year} {2021})}\BibitemShut
  {NoStop}%
\bibitem [{\citenamefont {Will-Cole}\ \emph {et~al.}(2023)\citenamefont
  {Will-Cole}, \citenamefont {Hart}, \citenamefont {Lauter}, \citenamefont
  {Grutter}, \citenamefont {Dubs}, \citenamefont {Lindner}, \citenamefont
  {Reimann}, \citenamefont {Valdez}, \citenamefont {Pearce}, \citenamefont
  {Monson}, \citenamefont {Cha}, \citenamefont {Heiman},\ and\ \citenamefont
  {Sun}}]{Will-Cole2023}%
  \BibitemOpen
  \bibfield  {author} {\bibinfo {author} {\bibfnamefont {A.~R.}\ \bibnamefont
  {Will-Cole}}, \bibinfo {author} {\bibfnamefont {J.~L.}\ \bibnamefont {Hart}},
  \bibinfo {author} {\bibfnamefont {V.}~\bibnamefont {Lauter}}, \bibinfo
  {author} {\bibfnamefont {A.}~\bibnamefont {Grutter}}, \bibinfo {author}
  {\bibfnamefont {C.}~\bibnamefont {Dubs}}, \bibinfo {author} {\bibfnamefont
  {M.}~\bibnamefont {Lindner}}, \bibinfo {author} {\bibfnamefont
  {T.}~\bibnamefont {Reimann}}, \bibinfo {author} {\bibfnamefont {N.~R.}\
  \bibnamefont {Valdez}}, \bibinfo {author} {\bibfnamefont {C.~J.}\
  \bibnamefont {Pearce}}, \bibinfo {author} {\bibfnamefont {T.~C.}\
  \bibnamefont {Monson}}, \bibinfo {author} {\bibfnamefont {J.~J.}\
  \bibnamefont {Cha}}, \bibinfo {author} {\bibfnamefont {D.}~\bibnamefont
  {Heiman}},\ and\ \bibinfo {author} {\bibfnamefont {N.~X.}\ \bibnamefont
  {Sun}},\ }\bibfield  {title} {\bibinfo {title} {{Negligible magnetic losses
  at low temperatures in liquid phase epitaxy grown Y$_{3}$Fe$_{5}$O$_{12}$
  films}},\ }\href {https://doi.org/10.1103/physrevmaterials.7.054411}
  {\bibfield  {journal} {\bibinfo  {journal} {Phys. Rev. Mater.}\ }\textbf
  {\bibinfo {volume} {7}},\ \bibinfo {pages} {054411} (\bibinfo {year}
  {2023})}\BibitemShut {NoStop}%
\bibitem [{\citenamefont {Guo}\ \emph {et~al.}(2022)\citenamefont {Guo},
  \citenamefont {McCullian}, \citenamefont {Hammel},\ and\ \citenamefont
  {Yang}}]{Guo2022}%
  \BibitemOpen
  \bibfield  {author} {\bibinfo {author} {\bibfnamefont {S.}~\bibnamefont
  {Guo}}, \bibinfo {author} {\bibfnamefont {B.}~\bibnamefont {McCullian}},
  \bibinfo {author} {\bibfnamefont {P.~C.}\ \bibnamefont {Hammel}},\ and\
  \bibinfo {author} {\bibfnamefont {F.}~\bibnamefont {Yang}},\ }\bibfield
  {title} {\bibinfo {title} {{Low damping at few-K temperatures in
  Y$_{3}$Fe$_{5}$O$_{12}$ epitaxial films isolated from
  Gd$_{3}$Ga$_{5}$O$_{12}$ substrate using a diamagnetic
  Y$_{3}$Sc$_{2.5}$Al$_{2.5}$O$_{12}$ spacer}},\ }\href
  {https://doi.org/10.1016/j.jmmm.2022.169795} {\bibfield  {journal} {\bibinfo
  {journal} {J. Magn. Magn. Mater.}\ }\textbf {\bibinfo {volume} {562}},\
  \bibinfo {pages} {169795} (\bibinfo {year} {2022})}\BibitemShut {NoStop}%
\bibitem [{\citenamefont {Gilleo}\ and\ \citenamefont
  {Geller}(1958{\natexlab{a}})}]{Gilleo1958}%
  \BibitemOpen
  \bibfield  {author} {\bibinfo {author} {\bibfnamefont {M.~A.}\ \bibnamefont
  {Gilleo}}\ and\ \bibinfo {author} {\bibfnamefont {S.}~\bibnamefont
  {Geller}},\ }\bibfield  {title} {\bibinfo {title} {{Magnetic and
  Crystallographic Properties of Substituted Yttrium-Iron Garnet,
  3Y$_{2}$O$_{3}\cdot$$x$M$_{2}$O$_{3}\cdot$$(5-x)$Fe$_{2}$O$_{3}$}},\ }\href
  {https://doi.org/10.1103/PhysRev.110.73} {\bibfield  {journal} {\bibinfo
  {journal} {Phys. Rev.}\ }\textbf {\bibinfo {volume} {110}},\ \bibinfo {pages}
  {73} (\bibinfo {year} {1958}{\natexlab{a}})}\BibitemShut {NoStop}%
\bibitem [{\citenamefont {Gilleo}\ and\ \citenamefont
  {Geller}(1958{\natexlab{b}})}]{Gilleo1958a}%
  \BibitemOpen
  \bibfield  {author} {\bibinfo {author} {\bibfnamefont {M.~A.}\ \bibnamefont
  {Gilleo}}\ and\ \bibinfo {author} {\bibfnamefont {S.}~\bibnamefont
  {Geller}},\ }\bibfield  {title} {\bibinfo {title} {{Substitution for Iron in
  Ferrimagnetic Yttrium-Iron Garnet}},\ }\href
  {https://doi.org/10.1063/1.1723143} {\bibfield  {journal} {\bibinfo
  {journal} {J. Appl. Phys.}\ }\textbf {\bibinfo {volume} {29}},\ \bibinfo
  {pages} {380} (\bibinfo {year} {1958}{\natexlab{b}})}\BibitemShut {NoStop}%
\bibitem [{\citenamefont {Hansen}\ \emph {et~al.}(1983)\citenamefont {Hansen},
  \citenamefont {Witter},\ and\ \citenamefont {Tolksdorf}}]{Hansen1983}%
  \BibitemOpen
  \bibfield  {author} {\bibinfo {author} {\bibfnamefont {P.}~\bibnamefont
  {Hansen}}, \bibinfo {author} {\bibfnamefont {K.}~\bibnamefont {Witter}},\
  and\ \bibinfo {author} {\bibfnamefont {W.}~\bibnamefont {Tolksdorf}},\
  }\bibfield  {title} {\bibinfo {title} {{Magnetic and magneto-optic properties
  of lead- and bismuth-substituted yttrium iron garnet films}},\ }\href
  {https://doi.org/10.1103/physrevb.27.6608} {\bibfield  {journal} {\bibinfo
  {journal} {Phys. Rev. B}\ }\textbf {\bibinfo {volume} {27}},\ \bibinfo
  {pages} {6608} (\bibinfo {year} {1983})}\BibitemShut {NoStop}%
\bibitem [{\citenamefont {Vittoria}\ \emph {et~al.}(1985)\citenamefont
  {Vittoria}, \citenamefont {Lubitz}, \citenamefont {Hansen},\ and\
  \citenamefont {Tolksdorf}}]{Vittoria1985}%
  \BibitemOpen
  \bibfield  {author} {\bibinfo {author} {\bibfnamefont {C.}~\bibnamefont
  {Vittoria}}, \bibinfo {author} {\bibfnamefont {P.}~\bibnamefont {Lubitz}},
  \bibinfo {author} {\bibfnamefont {P.}~\bibnamefont {Hansen}},\ and\ \bibinfo
  {author} {\bibfnamefont {W.}~\bibnamefont {Tolksdorf}},\ }\bibfield  {title}
  {\bibinfo {title} {{{FMR} linewidth measurements in bismuth-substituted
  {YIG}}},\ }\href {https://doi.org/10.1063/1.334994} {\bibfield  {journal}
  {\bibinfo  {journal} {J. Appl. Phys.}\ }\textbf {\bibinfo {volume} {57}},\
  \bibinfo {pages} {3699} (\bibinfo {year} {1985})}\BibitemShut {NoStop}%
\bibitem [{\citenamefont {Tamada}\ \emph {et~al.}(1988)\citenamefont {Tamada},
  \citenamefont {Kaneko},\ and\ \citenamefont {Okamoto}}]{Tamada1988}%
  \BibitemOpen
  \bibfield  {author} {\bibinfo {author} {\bibfnamefont {H.}~\bibnamefont
  {Tamada}}, \bibinfo {author} {\bibfnamefont {K.}~\bibnamefont {Kaneko}},\
  and\ \bibinfo {author} {\bibfnamefont {T.}~\bibnamefont {Okamoto}},\
  }\bibfield  {title} {\bibinfo {title} {{Bi-Substituted LPE Garnet Films with
  FMR Linewidth as Small as YIG}},\ }\href
  {https://doi.org/10.1109/tjmj.1988.4563654} {\bibfield  {journal} {\bibinfo
  {journal} {IEEE Transl. J. Magn. Jpn.}\ }\textbf {\bibinfo {volume} {3}},\
  \bibinfo {pages} {98} (\bibinfo {year} {1988})}\BibitemShut {NoStop}%
\bibitem [{\citenamefont {Soumah}\ \emph {et~al.}(2018)\citenamefont {Soumah},
  \citenamefont {Beaulieu}, \citenamefont {Qassym}, \citenamefont
  {Carr{\'{e}}t{\'{e}}ro}, \citenamefont {Jacquet}, \citenamefont
  {Lebourgeois}, \citenamefont {Ben~Youssef}, \citenamefont {Bortolotti},
  \citenamefont {Cros},\ and\ \citenamefont {Anane}}]{Soumah2018}%
  \BibitemOpen
  \bibfield  {author} {\bibinfo {author} {\bibfnamefont {L.}~\bibnamefont
  {Soumah}}, \bibinfo {author} {\bibfnamefont {N.}~\bibnamefont {Beaulieu}},
  \bibinfo {author} {\bibfnamefont {L.}~\bibnamefont {Qassym}}, \bibinfo
  {author} {\bibfnamefont {C.}~\bibnamefont {Carr{\'{e}}t{\'{e}}ro}}, \bibinfo
  {author} {\bibfnamefont {E.}~\bibnamefont {Jacquet}}, \bibinfo {author}
  {\bibfnamefont {R.}~\bibnamefont {Lebourgeois}}, \bibinfo {author}
  {\bibfnamefont {J.}~\bibnamefont {Ben~Youssef}}, \bibinfo {author}
  {\bibfnamefont {P.}~\bibnamefont {Bortolotti}}, \bibinfo {author}
  {\bibfnamefont {V.}~\bibnamefont {Cros}},\ and\ \bibinfo {author}
  {\bibfnamefont {A.}~\bibnamefont {Anane}},\ }\bibfield  {title} {\bibinfo
  {title} {{Ultra-low damping insulating magnetic thin films get
  perpendicular}},\ }\href {https://doi.org/10.1038/s41467-018-05732-1}
  {\bibfield  {journal} {\bibinfo  {journal} {Nat. Commun.}\ }\textbf {\bibinfo
  {volume} {9}},\ \bibinfo {pages} {3355} (\bibinfo {year} {2018})}\BibitemShut
  {NoStop}%
\bibitem [{\citenamefont {Su}\ \emph {et~al.}(2021)\citenamefont {Su},
  \citenamefont {Ning}, \citenamefont {Cho},\ and\ \citenamefont
  {Ross}}]{Su2021}%
  \BibitemOpen
  \bibfield  {author} {\bibinfo {author} {\bibfnamefont {T.}~\bibnamefont
  {Su}}, \bibinfo {author} {\bibfnamefont {S.}~\bibnamefont {Ning}}, \bibinfo
  {author} {\bibfnamefont {E.}~\bibnamefont {Cho}},\ and\ \bibinfo {author}
  {\bibfnamefont {C.~A.}\ \bibnamefont {Ross}},\ }\bibfield  {title} {\bibinfo
  {title} {{Magnetism and site occupancy in epitaxial Y-rich yttrium iron
  garnet films}},\ }\href {https://doi.org/10.1103/physrevmaterials.5.094403}
  {\bibfield  {journal} {\bibinfo  {journal} {Phys. Rev. Mater.}\ }\textbf
  {\bibinfo {volume} {5}},\ \bibinfo {pages} {094403} (\bibinfo {year}
  {2021})}\BibitemShut {NoStop}%
\bibitem [{\citenamefont {Santiso}\ \emph {et~al.}(2023)\citenamefont
  {Santiso}, \citenamefont {Garc{\'{i}}a}, \citenamefont {Romanque},
  \citenamefont {Henry}, \citenamefont {Bernier}, \citenamefont {Bagu{\'{e}}s},
  \citenamefont {Caicedo}, \citenamefont {Valvidares},\ and\ \citenamefont
  {Sandiumenge}}]{Santiso2023}%
  \BibitemOpen
  \bibfield  {author} {\bibinfo {author} {\bibfnamefont {J.}~\bibnamefont
  {Santiso}}, \bibinfo {author} {\bibfnamefont {C.}~\bibnamefont
  {Garc{\'{i}}a}}, \bibinfo {author} {\bibfnamefont {C.}~\bibnamefont
  {Romanque}}, \bibinfo {author} {\bibfnamefont {L.}~\bibnamefont {Henry}},
  \bibinfo {author} {\bibfnamefont {N.}~\bibnamefont {Bernier}}, \bibinfo
  {author} {\bibfnamefont {N.}~\bibnamefont {Bagu{\'{e}}s}}, \bibinfo {author}
  {\bibfnamefont {J.~M.}\ \bibnamefont {Caicedo}}, \bibinfo {author}
  {\bibfnamefont {M.}~\bibnamefont {Valvidares}},\ and\ \bibinfo {author}
  {\bibfnamefont {F.}~\bibnamefont {Sandiumenge}},\ }\bibfield  {title}
  {\bibinfo {title} {{Antisite Defects and Chemical Expansion in Low‐damping,
  High‐magnetization Yttrium Iron Garnet Films}},\ }\href
  {https://doi.org/10.1002/cnma.202200495} {\bibfield  {journal} {\bibinfo
  {journal} {ChemNanoMat}\ }\textbf {\bibinfo {volume} {9}},\ \bibinfo {pages}
  {202200495} (\bibinfo {year} {2023})}\BibitemShut {NoStop}%
\bibitem [{\citenamefont {Yang}\ and\ \citenamefont
  {Hammel}(2018)}]{Yang2018d}%
  \BibitemOpen
  \bibfield  {author} {\bibinfo {author} {\bibfnamefont {F.}~\bibnamefont
  {Yang}}\ and\ \bibinfo {author} {\bibfnamefont {P.~C.}\ \bibnamefont
  {Hammel}},\ }\bibfield  {title} {\bibinfo {title} {{{FMR}-driven spin pumping
  in Y$_{3}$Fe$_{5}$O$_{12}$-based structures}},\ }\href
  {https://doi.org/10.1088/1361-6463/aac249} {\bibfield  {journal} {\bibinfo
  {journal} {J. Phys. D: Appl. Phys.}\ }\textbf {\bibinfo {volume} {51}},\
  \bibinfo {pages} {253001} (\bibinfo {year} {2018})}\BibitemShut {NoStop}%
\bibitem [{\citenamefont {Wang}\ \emph {et~al.}(2014)\citenamefont {Wang},
  \citenamefont {Du}, \citenamefont {Hammel},\ and\ \citenamefont
  {Yang}}]{Wang2014b}%
  \BibitemOpen
  \bibfield  {author} {\bibinfo {author} {\bibfnamefont {H.}~\bibnamefont
  {Wang}}, \bibinfo {author} {\bibfnamefont {C.}~\bibnamefont {Du}}, \bibinfo
  {author} {\bibfnamefont {P.~C.}\ \bibnamefont {Hammel}},\ and\ \bibinfo
  {author} {\bibfnamefont {F.}~\bibnamefont {Yang}},\ }\bibfield  {title}
  {\bibinfo {title} {{Strain-tunable magnetocrystalline anisotropy in epitaxial
  Y$_{3}$Fe$_{5}$O$_{12}$ thin films}},\ }\href
  {https://doi.org/10.1103/physrevb.89.134404} {\bibfield  {journal} {\bibinfo
  {journal} {Phys. Rev. B}\ }\textbf {\bibinfo {volume} {89}},\ \bibinfo
  {pages} {134404} (\bibinfo {year} {2014})}\BibitemShut {NoStop}%
\bibitem [{\citenamefont {Gallagher}\ \emph {et~al.}(2016)\citenamefont
  {Gallagher}, \citenamefont {Yang}, \citenamefont {Brangham}, \citenamefont
  {Esser}, \citenamefont {White}, \citenamefont {Page}, \citenamefont {Meng},
  \citenamefont {Yu}, \citenamefont {Adur}, \citenamefont {Ruane},
  \citenamefont {Dunsiger}, \citenamefont {McComb}, \citenamefont {Yang},\ and\
  \citenamefont {Hammel}}]{Gallagher2016}%
  \BibitemOpen
  \bibfield  {author} {\bibinfo {author} {\bibfnamefont {J.~C.}\ \bibnamefont
  {Gallagher}}, \bibinfo {author} {\bibfnamefont {A.~S.}\ \bibnamefont {Yang}},
  \bibinfo {author} {\bibfnamefont {J.~T.}\ \bibnamefont {Brangham}}, \bibinfo
  {author} {\bibfnamefont {B.~D.}\ \bibnamefont {Esser}}, \bibinfo {author}
  {\bibfnamefont {S.~P.}\ \bibnamefont {White}}, \bibinfo {author}
  {\bibfnamefont {M.~R.}\ \bibnamefont {Page}}, \bibinfo {author}
  {\bibfnamefont {K.-Y.}\ \bibnamefont {Meng}}, \bibinfo {author}
  {\bibfnamefont {S.}~\bibnamefont {Yu}}, \bibinfo {author} {\bibfnamefont
  {R.}~\bibnamefont {Adur}}, \bibinfo {author} {\bibfnamefont {W.}~\bibnamefont
  {Ruane}}, \bibinfo {author} {\bibfnamefont {S.~R.}\ \bibnamefont {Dunsiger}},
  \bibinfo {author} {\bibfnamefont {D.~W.}\ \bibnamefont {McComb}}, \bibinfo
  {author} {\bibfnamefont {F.}~\bibnamefont {Yang}},\ and\ \bibinfo {author}
  {\bibfnamefont {P.~C.}\ \bibnamefont {Hammel}},\ }\bibfield  {title}
  {\bibinfo {title} {{Exceptionally high magnetization of stoichiometric
  Y$_{3}$Fe$_{5}$O$_{12}$ epitaxial films grown on Gd$_{3}$Ga$_{5}$O$_{12}$}},\
  }\href {https://doi.org/10.1063/1.4961371} {\bibfield  {journal} {\bibinfo
  {journal} {Appl. Phys. Lett.}\ }\textbf {\bibinfo {volume} {109}},\ \bibinfo
  {pages} {072401} (\bibinfo {year} {2016})}\BibitemShut {NoStop}%
\bibitem [{\citenamefont {Dorsey}\ \emph {et~al.}(1993)\citenamefont {Dorsey},
  \citenamefont {Bushnell}, \citenamefont {Seed},\ and\ \citenamefont
  {Vittoria}}]{Dorsey1993}%
  \BibitemOpen
  \bibfield  {author} {\bibinfo {author} {\bibfnamefont {P.~C.}\ \bibnamefont
  {Dorsey}}, \bibinfo {author} {\bibfnamefont {S.~E.}\ \bibnamefont
  {Bushnell}}, \bibinfo {author} {\bibfnamefont {R.~G.}\ \bibnamefont {Seed}},\
  and\ \bibinfo {author} {\bibfnamefont {C.}~\bibnamefont {Vittoria}},\
  }\bibfield  {title} {\bibinfo {title} {{Epitaxial yttrium iron garnet films
  grown by pulsed laser deposition}},\ }\href
  {https://doi.org/10.1063/1.354927} {\bibfield  {journal} {\bibinfo  {journal}
  {J. Appl. Phys.}\ }\textbf {\bibinfo {volume} {74}},\ \bibinfo {pages} {1242}
  (\bibinfo {year} {1993})}\BibitemShut {NoStop}%
\bibitem [{\citenamefont {Kubota}\ \emph {et~al.}(2013)\citenamefont {Kubota},
  \citenamefont {Shibuya}, \citenamefont {Tokunaga}, \citenamefont {Kagawa},
  \citenamefont {Tsukazaki}, \citenamefont {Tokura},\ and\ \citenamefont
  {Kawasaki}}]{Kubota2013}%
  \BibitemOpen
  \bibfield  {author} {\bibinfo {author} {\bibfnamefont {M.}~\bibnamefont
  {Kubota}}, \bibinfo {author} {\bibfnamefont {K.}~\bibnamefont {Shibuya}},
  \bibinfo {author} {\bibfnamefont {Y.}~\bibnamefont {Tokunaga}}, \bibinfo
  {author} {\bibfnamefont {F.}~\bibnamefont {Kagawa}}, \bibinfo {author}
  {\bibfnamefont {A.}~\bibnamefont {Tsukazaki}}, \bibinfo {author}
  {\bibfnamefont {Y.}~\bibnamefont {Tokura}},\ and\ \bibinfo {author}
  {\bibfnamefont {M.}~\bibnamefont {Kawasaki}},\ }\bibfield  {title} {\bibinfo
  {title} {{Systematic control of stress-induced anisotropy in pseudomorphic
  iron garnet thin films}},\ }\href
  {https://doi.org/10.1016/j.jmmm.2013.02.045} {\bibfield  {journal} {\bibinfo
  {journal} {J. Magn. Magn. Mater.}\ }\textbf {\bibinfo {volume} {339}},\
  \bibinfo {pages} {63} (\bibinfo {year} {2013})}\BibitemShut {NoStop}%
\bibitem [{\citenamefont {Lammel}\ \emph {et~al.}(2022)\citenamefont {Lammel},
  \citenamefont {Scheffler}, \citenamefont {Pohl}, \citenamefont {Swekis},
  \citenamefont {Reitzig}, \citenamefont {Piontek}, \citenamefont {Reichlova},
  \citenamefont {Schlitz}, \citenamefont {Geishendorf}, \citenamefont {Siegl},
  \citenamefont {Rellinghaus}, \citenamefont {Eng}, \citenamefont {Nielsch},
  \citenamefont {Goennenwein},\ and\ \citenamefont {Thomas}}]{Lammel2022}%
  \BibitemOpen
  \bibfield  {author} {\bibinfo {author} {\bibfnamefont {M.}~\bibnamefont
  {Lammel}}, \bibinfo {author} {\bibfnamefont {D.}~\bibnamefont {Scheffler}},
  \bibinfo {author} {\bibfnamefont {D.}~\bibnamefont {Pohl}}, \bibinfo {author}
  {\bibfnamefont {P.}~\bibnamefont {Swekis}}, \bibinfo {author} {\bibfnamefont
  {S.}~\bibnamefont {Reitzig}}, \bibinfo {author} {\bibfnamefont
  {S.}~\bibnamefont {Piontek}}, \bibinfo {author} {\bibfnamefont
  {H.}~\bibnamefont {Reichlova}}, \bibinfo {author} {\bibfnamefont
  {R.}~\bibnamefont {Schlitz}}, \bibinfo {author} {\bibfnamefont
  {K.}~\bibnamefont {Geishendorf}}, \bibinfo {author} {\bibfnamefont
  {L.}~\bibnamefont {Siegl}}, \bibinfo {author} {\bibfnamefont
  {B.}~\bibnamefont {Rellinghaus}}, \bibinfo {author} {\bibfnamefont {L.~M.}\
  \bibnamefont {Eng}}, \bibinfo {author} {\bibfnamefont {K.}~\bibnamefont
  {Nielsch}}, \bibinfo {author} {\bibfnamefont {S.~T.~B.}\ \bibnamefont
  {Goennenwein}},\ and\ \bibinfo {author} {\bibfnamefont {A.}~\bibnamefont
  {Thomas}},\ }\bibfield  {title} {\bibinfo {title} {{Atomic layer deposition
  of yttrium iron garnet thin films}},\ }\href
  {https://doi.org/10.1103/physrevmaterials.6.044411} {\bibfield  {journal}
  {\bibinfo  {journal} {Phys. Rev. Mater.}\ }\textbf {\bibinfo {volume} {6}},\
  \bibinfo {pages} {044411} (\bibinfo {year} {2022})}\BibitemShut {NoStop}%
\bibitem [{\citenamefont {Geller}(1967)}]{Geller1967}%
  \BibitemOpen
  \bibfield  {author} {\bibinfo {author} {\bibfnamefont {S.}~\bibnamefont
  {Geller}},\ }\bibfield  {title} {\bibinfo {title} {{Crystal chemistry of the
  garnets}},\ }\href {https://doi.org/10.1524/zkri.1967.125.16.1} {\bibfield
  {journal} {\bibinfo  {journal} {Z. Kristallogr.}\ }\textbf {\bibinfo {volume}
  {125}},\ \bibinfo {pages} {1} (\bibinfo {year} {1967})}\BibitemShut {NoStop}%
\bibitem [{\citenamefont {Noun}\ \emph {et~al.}(2010)\citenamefont {Noun},
  \citenamefont {Popova}, \citenamefont {Bardelli}, \citenamefont {Dumont},
  \citenamefont {Bertacco}, \citenamefont {Tagliaferri}, \citenamefont
  {Tessier}, \citenamefont {Guyot}, \citenamefont {Berini},\ and\ \citenamefont
  {Keller}}]{Noun2010}%
  \BibitemOpen
  \bibfield  {author} {\bibinfo {author} {\bibfnamefont {W.}~\bibnamefont
  {Noun}}, \bibinfo {author} {\bibfnamefont {E.}~\bibnamefont {Popova}},
  \bibinfo {author} {\bibfnamefont {F.}~\bibnamefont {Bardelli}}, \bibinfo
  {author} {\bibfnamefont {Y.}~\bibnamefont {Dumont}}, \bibinfo {author}
  {\bibfnamefont {R.}~\bibnamefont {Bertacco}}, \bibinfo {author}
  {\bibfnamefont {A.}~\bibnamefont {Tagliaferri}}, \bibinfo {author}
  {\bibfnamefont {M.}~\bibnamefont {Tessier}}, \bibinfo {author} {\bibfnamefont
  {M.}~\bibnamefont {Guyot}}, \bibinfo {author} {\bibfnamefont
  {B.}~\bibnamefont {Berini}},\ and\ \bibinfo {author} {\bibfnamefont
  {N.}~\bibnamefont {Keller}},\ }\bibfield  {title} {\bibinfo {title}
  {{Determination of yttrium iron garnet superexchange parameters as a function
  of oxygen and cation stoichiometry}},\ }\href
  {https://doi.org/10.1103/physrevb.81.054411} {\bibfield  {journal} {\bibinfo
  {journal} {Phys. Rev. B}\ }\textbf {\bibinfo {volume} {81}},\ \bibinfo
  {pages} {054411} (\bibinfo {year} {2010})}\BibitemShut {NoStop}%
\bibitem [{\citenamefont {Geller}(1966)}]{Geller1966a}%
  \BibitemOpen
  \bibfield  {author} {\bibinfo {author} {\bibfnamefont {S.}~\bibnamefont
  {Geller}},\ }\bibfield  {title} {\bibinfo {title} {{Magnetic Behavior of
  Substituted Ferrimagnetic Garnets}},\ }\href
  {https://doi.org/10.1063/1.1708495} {\bibfield  {journal} {\bibinfo
  {journal} {J. Appl. Phys.}\ }\textbf {\bibinfo {volume} {37}},\ \bibinfo
  {pages} {1408} (\bibinfo {year} {1966})}\BibitemShut {NoStop}%
\bibitem [{\citenamefont {Anderson}(1964)}]{Anderson1964}%
  \BibitemOpen
  \bibfield  {author} {\bibinfo {author} {\bibfnamefont {E.~E.}\ \bibnamefont
  {Anderson}},\ }\bibfield  {title} {\bibinfo {title} {{Molecular Field Model
  and the Magnetization of YIG}},\ }\href
  {https://doi.org/10.1103/PhysRev.134.A1581} {\bibfield  {journal} {\bibinfo
  {journal} {Phys. Rev.}\ }\textbf {\bibinfo {volume} {134}},\ \bibinfo {pages}
  {A1581} (\bibinfo {year} {1964})}\BibitemShut {NoStop}%
\bibitem [{\citenamefont {Dionne}(1970)}]{Dionne1970}%
  \BibitemOpen
  \bibfield  {author} {\bibinfo {author} {\bibfnamefont {G.~F.}\ \bibnamefont
  {Dionne}},\ }\bibfield  {title} {\bibinfo {title} {{Molecular Field
  Coefficients of Substituted Yttrium Iron Garnets}},\ }\href
  {https://doi.org/10.1063/1.1658555} {\bibfield  {journal} {\bibinfo
  {journal} {J. Appl. Phys.}\ }\textbf {\bibinfo {volume} {41}},\ \bibinfo
  {pages} {4874} (\bibinfo {year} {1970})}\BibitemShut {NoStop}%
\bibitem [{\citenamefont {R{\"{o}}schmann}\ and\ \citenamefont
  {Hansen}(1981)}]{Roeschmann1981}%
  \BibitemOpen
  \bibfield  {author} {\bibinfo {author} {\bibfnamefont {P.}~\bibnamefont
  {R{\"{o}}schmann}}\ and\ \bibinfo {author} {\bibfnamefont {P.}~\bibnamefont
  {Hansen}},\ }\bibfield  {title} {\bibinfo {title} {{Molecular field
  coefficients and cation distribution of substituted yttrium iron garnets}},\
  }\href {https://doi.org/10.1063/1.328569} {\bibfield  {journal} {\bibinfo
  {journal} {J. Appl. Phys.}\ }\textbf {\bibinfo {volume} {52}},\ \bibinfo
  {pages} {6257} (\bibinfo {year} {1981})}\BibitemShut {NoStop}%
\bibitem [{\citenamefont {Gross}\ \emph {et~al.}(2024)\citenamefont {Gross},
  \citenamefont {Su}, \citenamefont {Bauer},\ and\ \citenamefont
  {Ross}}]{Gross2024}%
  \BibitemOpen
  \bibfield  {author} {\bibinfo {author} {\bibfnamefont {M.~J.}\ \bibnamefont
  {Gross}}, \bibinfo {author} {\bibfnamefont {T.}~\bibnamefont {Su}}, \bibinfo
  {author} {\bibfnamefont {J.~J.}\ \bibnamefont {Bauer}},\ and\ \bibinfo
  {author} {\bibfnamefont {C.~A.}\ \bibnamefont {Ross}},\ }\bibfield  {title}
  {\bibinfo {title} {{Molecular-field-coefficient modeling of
  temperature-dependent ferrimagnetism in a complex oxide}},\ }\href
  {https://doi.org/10.1103/physrevapplied.21.014060} {\bibfield  {journal}
  {\bibinfo  {journal} {Phys. Rev. Appl.}\ }\textbf {\bibinfo {volume} {21}},\
  \bibinfo {pages} {014060} (\bibinfo {year} {2024})}\BibitemShut {NoStop}%
\bibitem [{\citenamefont {Rosenberg}\ \emph {et~al.}(2021)\citenamefont
  {Rosenberg}, \citenamefont {Litzius}, \citenamefont {Shaw}, \citenamefont
  {Riley}, \citenamefont {Beach}, \citenamefont {Nembach},\ and\ \citenamefont
  {Ross}}]{Rosenberg2021}%
  \BibitemOpen
  \bibfield  {author} {\bibinfo {author} {\bibfnamefont {E.~R.}\ \bibnamefont
  {Rosenberg}}, \bibinfo {author} {\bibfnamefont {K.}~\bibnamefont {Litzius}},
  \bibinfo {author} {\bibfnamefont {J.~M.}\ \bibnamefont {Shaw}}, \bibinfo
  {author} {\bibfnamefont {G.~A.}\ \bibnamefont {Riley}}, \bibinfo {author}
  {\bibfnamefont {G.~S.~D.}\ \bibnamefont {Beach}}, \bibinfo {author}
  {\bibfnamefont {H.~T.}\ \bibnamefont {Nembach}},\ and\ \bibinfo {author}
  {\bibfnamefont {C.~A.}\ \bibnamefont {Ross}},\ }\bibfield  {title} {\bibinfo
  {title} {{Magnetic Properties and Growth-Induced Anisotropy in Yttrium
  Thulium Iron Garnet Thin Films}},\ }\href
  {https://doi.org/10.1002/aelm.202100452} {\bibfield  {journal} {\bibinfo
  {journal} {Adv. Electron. Mater.}\ }\textbf {\bibinfo {volume} {7}},\
  \bibinfo {pages} {2100452} (\bibinfo {year} {2021})}\BibitemShut {NoStop}%
\bibitem [{\citenamefont {Rosenberg}\ \emph {et~al.}(2023)\citenamefont
  {Rosenberg}, \citenamefont {Bauer}, \citenamefont {Cho}, \citenamefont
  {Kumar}, \citenamefont {Pelliciari}, \citenamefont {Occhialini},
  \citenamefont {Ning}, \citenamefont {Kaczmarek}, \citenamefont {Rosenberg},
  \citenamefont {Freeland}, \citenamefont {Chen}, \citenamefont {Wang},
  \citenamefont {LeBeau}, \citenamefont {Comin}, \citenamefont {de~Groot},\
  and\ \citenamefont {Ross}}]{Rosenberg2023}%
  \BibitemOpen
  \bibfield  {author} {\bibinfo {author} {\bibfnamefont {E.}~\bibnamefont
  {Rosenberg}}, \bibinfo {author} {\bibfnamefont {J.}~\bibnamefont {Bauer}},
  \bibinfo {author} {\bibfnamefont {E.}~\bibnamefont {Cho}}, \bibinfo {author}
  {\bibfnamefont {A.}~\bibnamefont {Kumar}}, \bibinfo {author} {\bibfnamefont
  {J.}~\bibnamefont {Pelliciari}}, \bibinfo {author} {\bibfnamefont {C.~A.}\
  \bibnamefont {Occhialini}}, \bibinfo {author} {\bibfnamefont
  {S.}~\bibnamefont {Ning}}, \bibinfo {author} {\bibfnamefont {A.}~\bibnamefont
  {Kaczmarek}}, \bibinfo {author} {\bibfnamefont {R.}~\bibnamefont
  {Rosenberg}}, \bibinfo {author} {\bibfnamefont {J.~W.}\ \bibnamefont
  {Freeland}}, \bibinfo {author} {\bibfnamefont {Y.}~\bibnamefont {Chen}},
  \bibinfo {author} {\bibfnamefont {J.}~\bibnamefont {Wang}}, \bibinfo {author}
  {\bibfnamefont {J.}~\bibnamefont {LeBeau}}, \bibinfo {author} {\bibfnamefont
  {R.}~\bibnamefont {Comin}}, \bibinfo {author} {\bibfnamefont {F.~M.~F.}\
  \bibnamefont {de~Groot}},\ and\ \bibinfo {author} {\bibfnamefont {C.~A.}\
  \bibnamefont {Ross}},\ }\bibfield  {title} {\bibinfo {title} {{Revealing Site
  Occupancy in a Complex Oxide: Terbium Iron Garnet}},\ }\href
  {https://doi.org/10.1002/smll.202300824} {\bibfield  {journal} {\bibinfo
  {journal} {Small}\ }\textbf {\bibinfo {volume} {19}},\ \bibinfo {pages}
  {2300824} (\bibinfo {year} {2023})}\BibitemShut {NoStop}%
\bibitem [{\citenamefont {Manuilov}\ \emph {et~al.}(2009)\citenamefont
  {Manuilov}, \citenamefont {Khartsev},\ and\ \citenamefont
  {Grishin}}]{Manuilov2009}%
  \BibitemOpen
  \bibfield  {author} {\bibinfo {author} {\bibfnamefont {S.~A.}\ \bibnamefont
  {Manuilov}}, \bibinfo {author} {\bibfnamefont {S.~I.}\ \bibnamefont
  {Khartsev}},\ and\ \bibinfo {author} {\bibfnamefont {A.~M.}\ \bibnamefont
  {Grishin}},\ }\bibfield  {title} {\bibinfo {title} {{Pulsed laser deposited
  Y3Fe5O12 films: Nature of magnetic anisotropy I}},\ }\href
  {https://doi.org/10.1063/1.3272731} {\bibfield  {journal} {\bibinfo
  {journal} {J. Appl. Phys.}\ }\textbf {\bibinfo {volume} {106}},\ \bibinfo
  {pages} {123917} (\bibinfo {year} {2009})}\BibitemShut {NoStop}%
\bibitem [{\citenamefont {Manuilov}\ and\ \citenamefont
  {Grishin}(2010)}]{Manuilov2010}%
  \BibitemOpen
  \bibfield  {author} {\bibinfo {author} {\bibfnamefont {S.~A.}\ \bibnamefont
  {Manuilov}}\ and\ \bibinfo {author} {\bibfnamefont {A.~M.}\ \bibnamefont
  {Grishin}},\ }\bibfield  {title} {\bibinfo {title} {{Pulsed laser deposited
  Y3Fe5O12 films: Nature of magnetic anisotropy II}},\ }\href
  {https://doi.org/10.1063/1.3446840} {\bibfield  {journal} {\bibinfo
  {journal} {J. Appl. Phys.}\ }\textbf {\bibinfo {volume} {108}},\ \bibinfo
  {pages} {013902} (\bibinfo {year} {2010})}\BibitemShut {NoStop}%
\bibitem [{\citenamefont {Judy}(1966)}]{Judy1966}%
  \BibitemOpen
  \bibfield  {author} {\bibinfo {author} {\bibfnamefont {J.~H.}\ \bibnamefont
  {Judy}},\ }\bibfield  {title} {\bibinfo {title} {{Anisotropic Magnetic
  Resonance and Relaxation in Silicon-Substituted YIG}},\ }\href
  {https://doi.org/10.1063/1.1708455} {\bibfield  {journal} {\bibinfo
  {journal} {J. Appl. Phys.}\ }\textbf {\bibinfo {volume} {37}},\ \bibinfo
  {pages} {1328} (\bibinfo {year} {1966})}\BibitemShut {NoStop}%
\bibitem [{\citenamefont {Epstein}\ and\ \citenamefont
  {Tocci}(1967)}]{Epstein1967}%
  \BibitemOpen
  \bibfield  {author} {\bibinfo {author} {\bibfnamefont {D.~J.}\ \bibnamefont
  {Epstein}}\ and\ \bibinfo {author} {\bibfnamefont {L.}~\bibnamefont
  {Tocci}},\ }\bibfield  {title} {\bibinfo {title} {{High Temperature Resonance
  Losses in Silicon-Doped Yttrium--Iron Garnet (YIG)}},\ }\href
  {https://doi.org/10.1063/1.1755026} {\bibfield  {journal} {\bibinfo
  {journal} {Appl. Phys. Lett.}\ }\textbf {\bibinfo {volume} {11}},\ \bibinfo
  {pages} {55} (\bibinfo {year} {1967})}\BibitemShut {NoStop}%
\bibitem [{\citenamefont {Arias}\ and\ \citenamefont
  {Mills}(1999)}]{Arias1999a}%
  \BibitemOpen
  \bibfield  {author} {\bibinfo {author} {\bibfnamefont {R.}~\bibnamefont
  {Arias}}\ and\ \bibinfo {author} {\bibfnamefont {D.~L.}\ \bibnamefont
  {Mills}},\ }\bibfield  {title} {\bibinfo {title} {Extrinsic contributions to
  the ferromagnetic resonance response of ultrathin films},\ }\href
  {https://doi.org/10.1103/physrevb.60.7395} {\bibfield  {journal} {\bibinfo
  {journal} {Phys. Rev. B}\ }\textbf {\bibinfo {volume} {60}},\ \bibinfo
  {pages} {7395} (\bibinfo {year} {1999})}\BibitemShut {NoStop}%
\bibitem [{\citenamefont {G{\"{u}}ckelhorn}\ \emph {et~al.}(2021)\citenamefont
  {G{\"{u}}ckelhorn}, \citenamefont {Wimmer}, \citenamefont {M{\"{u}}ller},
  \citenamefont {Gepr{\"{a}}gs}, \citenamefont {Huebl}, \citenamefont {Gross},\
  and\ \citenamefont {Althammer}}]{Gueckelhorn2021}%
  \BibitemOpen
  \bibfield  {author} {\bibinfo {author} {\bibfnamefont {J.}~\bibnamefont
  {G{\"{u}}ckelhorn}}, \bibinfo {author} {\bibfnamefont {T.}~\bibnamefont
  {Wimmer}}, \bibinfo {author} {\bibfnamefont {M.}~\bibnamefont
  {M{\"{u}}ller}}, \bibinfo {author} {\bibfnamefont {S.}~\bibnamefont
  {Gepr{\"{a}}gs}}, \bibinfo {author} {\bibfnamefont {H.}~\bibnamefont
  {Huebl}}, \bibinfo {author} {\bibfnamefont {R.}~\bibnamefont {Gross}},\ and\
  \bibinfo {author} {\bibfnamefont {M.}~\bibnamefont {Althammer}},\ }\bibfield
  {title} {\bibinfo {title} {{Magnon transport in Y$_{3}$Fe$_{5}$O$_{12}$/Pt
  nanostructures with reduced effective magnetization}},\ }\href
  {https://doi.org/10.1103/physrevb.104.l180410} {\bibfield  {journal}
  {\bibinfo  {journal} {Phys. Rev. B}\ }\textbf {\bibinfo {volume} {104}},\
  \bibinfo {pages} {L180410} (\bibinfo {year} {2021})}\BibitemShut {NoStop}%
\bibitem [{\citenamefont {Danilov}\ \emph {et~al.}(1989)\citenamefont
  {Danilov}, \citenamefont {Lyfar'}, \citenamefont {Lyubon'ko}, \citenamefont
  {Nechiporuk},\ and\ \citenamefont {Ryabchenko}}]{Danilov1989}%
  \BibitemOpen
  \bibfield  {author} {\bibinfo {author} {\bibfnamefont {V.~V.}\ \bibnamefont
  {Danilov}}, \bibinfo {author} {\bibfnamefont {D.~L.}\ \bibnamefont {Lyfar'}},
  \bibinfo {author} {\bibfnamefont {Y.~V.}\ \bibnamefont {Lyubon'ko}}, \bibinfo
  {author} {\bibfnamefont {A.~Y.}\ \bibnamefont {Nechiporuk}},\ and\ \bibinfo
  {author} {\bibfnamefont {S.~M.}\ \bibnamefont {Ryabchenko}},\ }\bibfield
  {title} {\bibinfo {title} {{Low-temperature ferromagnetic resonance in
  epitaxial garnet films on paramagnetic substrates}},\ }\href
  {https://doi.org/10.1007/bf00897267} {\bibfield  {journal} {\bibinfo
  {journal} {Sov. Phys. J.}\ }\textbf {\bibinfo {volume} {32}},\ \bibinfo
  {pages} {276} (\bibinfo {year} {1989})}\BibitemShut {NoStop}%
\bibitem [{\citenamefont {Mihalceanu}\ \emph {et~al.}(2018)\citenamefont
  {Mihalceanu}, \citenamefont {Vasyuchka}, \citenamefont {Bozhko},
  \citenamefont {Langner}, \citenamefont {Nechiporuk}, \citenamefont
  {Romanyuk}, \citenamefont {Hillebrands},\ and\ \citenamefont
  {Serga}}]{Mihalceanu2018}%
  \BibitemOpen
  \bibfield  {author} {\bibinfo {author} {\bibfnamefont {L.}~\bibnamefont
  {Mihalceanu}}, \bibinfo {author} {\bibfnamefont {V.~I.}\ \bibnamefont
  {Vasyuchka}}, \bibinfo {author} {\bibfnamefont {D.~A.}\ \bibnamefont
  {Bozhko}}, \bibinfo {author} {\bibfnamefont {T.}~\bibnamefont {Langner}},
  \bibinfo {author} {\bibfnamefont {A.~Y.}\ \bibnamefont {Nechiporuk}},
  \bibinfo {author} {\bibfnamefont {V.~F.}\ \bibnamefont {Romanyuk}}, \bibinfo
  {author} {\bibfnamefont {B.}~\bibnamefont {Hillebrands}},\ and\ \bibinfo
  {author} {\bibfnamefont {A.~A.}\ \bibnamefont {Serga}},\ }\bibfield  {title}
  {\bibinfo {title} {{Temperature-dependent relaxation of dipole-exchange
  magnons in yttrium iron garnet films}},\ }\href
  {https://doi.org/10.1103/physrevb.97.214405} {\bibfield  {journal} {\bibinfo
  {journal} {Phys. Rev. B}\ }\textbf {\bibinfo {volume} {97}},\ \bibinfo
  {pages} {214405} (\bibinfo {year} {2018})}\BibitemShut {NoStop}%
\bibitem [{\citenamefont {Roos}\ \emph {et~al.}(2022)\citenamefont {Roos},
  \citenamefont {Quarterman}, \citenamefont {Ding}, \citenamefont {Wu},
  \citenamefont {Kirby},\ and\ \citenamefont {Zink}}]{Roos2022}%
  \BibitemOpen
  \bibfield  {author} {\bibinfo {author} {\bibfnamefont {M.~J.}\ \bibnamefont
  {Roos}}, \bibinfo {author} {\bibfnamefont {P.}~\bibnamefont {Quarterman}},
  \bibinfo {author} {\bibfnamefont {J.}~\bibnamefont {Ding}}, \bibinfo {author}
  {\bibfnamefont {M.}~\bibnamefont {Wu}}, \bibinfo {author} {\bibfnamefont
  {B.~J.}\ \bibnamefont {Kirby}},\ and\ \bibinfo {author} {\bibfnamefont
  {B.~L.}\ \bibnamefont {Zink}},\ }\bibfield  {title} {\bibinfo {title}
  {{Magnetization and antiferromagnetic coupling of the interface between a
  20 nm Y$_{3}$Fe$_{5}$O$_{12}$ film and Gd$_{3}$Ga$_{5}$O$_{12}$
  substrate}},\ }\href {https://doi.org/10.1103/physrevmaterials.6.034401}
  {\bibfield  {journal} {\bibinfo  {journal} {Phys. Rev. Mater.}\ }\textbf
  {\bibinfo {volume} {6}},\ \bibinfo {pages} {034401} (\bibinfo {year}
  {2022})}\BibitemShut {NoStop}%
\bibitem [{\citenamefont {Knauer}\ \emph {et~al.}(2023)\citenamefont {Knauer},
  \citenamefont {Dav{\'{i}}dkov{\'{a}}}, \citenamefont {Schmoll}, \citenamefont
  {Serha}, \citenamefont {Voronov}, \citenamefont {Wang}, \citenamefont
  {Verba}, \citenamefont {Dobrovolskiy}, \citenamefont {Lindner}, \citenamefont
  {Reimann}, \citenamefont {Dubs}, \citenamefont {Urb{\'{a}}nek},\ and\
  \citenamefont {Chumak}}]{Knauer2023}%
  \BibitemOpen
  \bibfield  {author} {\bibinfo {author} {\bibfnamefont {S.}~\bibnamefont
  {Knauer}}, \bibinfo {author} {\bibfnamefont {K.}~\bibnamefont
  {Dav{\'{i}}dkov{\'{a}}}}, \bibinfo {author} {\bibfnamefont {D.}~\bibnamefont
  {Schmoll}}, \bibinfo {author} {\bibfnamefont {R.~O.}\ \bibnamefont {Serha}},
  \bibinfo {author} {\bibfnamefont {A.}~\bibnamefont {Voronov}}, \bibinfo
  {author} {\bibfnamefont {Q.}~\bibnamefont {Wang}}, \bibinfo {author}
  {\bibfnamefont {R.}~\bibnamefont {Verba}}, \bibinfo {author} {\bibfnamefont
  {O.~V.}\ \bibnamefont {Dobrovolskiy}}, \bibinfo {author} {\bibfnamefont
  {M.}~\bibnamefont {Lindner}}, \bibinfo {author} {\bibfnamefont
  {T.}~\bibnamefont {Reimann}}, \bibinfo {author} {\bibfnamefont
  {C.}~\bibnamefont {Dubs}}, \bibinfo {author} {\bibfnamefont {M.}~\bibnamefont
  {Urb{\'{a}}nek}},\ and\ \bibinfo {author} {\bibfnamefont {A.~V.}\
  \bibnamefont {Chumak}},\ }\bibfield  {title} {\bibinfo {title} {{Propagating
  spin-wave spectroscopy in a liquid-phase epitaxial nanometer-thick YIG film
  at millikelvin temperatures}},\ }\href {https://doi.org/10.1063/5.0137437}
  {\bibfield  {journal} {\bibinfo  {journal} {J. Appl. Phys.}\ }\textbf
  {\bibinfo {volume} {133}},\ \bibinfo {pages} {143905} (\bibinfo {year}
  {2023})}\BibitemShut {NoStop}%
\bibitem [{\citenamefont {Serha}\ \emph {et~al.}(2024)\citenamefont {Serha},
  \citenamefont {Voronov}, \citenamefont {Schmoll}, \citenamefont {Verba},
  \citenamefont {Levchenko}, \citenamefont {Koraltan}, \citenamefont
  {Dav{\'{i}}dkov{\'{a}}}, \citenamefont {Budinsk{\'{a}}}, \citenamefont
  {Wang}, \citenamefont {Dobrovolskiy}, \citenamefont {Urb{\'{a}}nek},
  \citenamefont {Lindner}, \citenamefont {Reimann}, \citenamefont {Dubs},
  \citenamefont {Gonzalez-Ballestero}, \citenamefont {Abert}, \citenamefont
  {Suess}, \citenamefont {Bozhko}, \citenamefont {Knauer},\ and\ \citenamefont
  {Chumak}}]{Serha2024}%
  \BibitemOpen
  \bibfield  {author} {\bibinfo {author} {\bibfnamefont {R.~O.}\ \bibnamefont
  {Serha}}, \bibinfo {author} {\bibfnamefont {A.~A.}\ \bibnamefont {Voronov}},
  \bibinfo {author} {\bibfnamefont {D.}~\bibnamefont {Schmoll}}, \bibinfo
  {author} {\bibfnamefont {R.}~\bibnamefont {Verba}}, \bibinfo {author}
  {\bibfnamefont {K.~O.}\ \bibnamefont {Levchenko}}, \bibinfo {author}
  {\bibfnamefont {S.}~\bibnamefont {Koraltan}}, \bibinfo {author}
  {\bibfnamefont {K.}~\bibnamefont {Dav{\'{i}}dkov{\'{a}}}}, \bibinfo {author}
  {\bibfnamefont {B.}~\bibnamefont {Budinsk{\'{a}}}}, \bibinfo {author}
  {\bibfnamefont {Q.}~\bibnamefont {Wang}}, \bibinfo {author} {\bibfnamefont
  {O.~V.}\ \bibnamefont {Dobrovolskiy}}, \bibinfo {author} {\bibfnamefont
  {M.}~\bibnamefont {Urb{\'{a}}nek}}, \bibinfo {author} {\bibfnamefont
  {M.}~\bibnamefont {Lindner}}, \bibinfo {author} {\bibfnamefont
  {T.}~\bibnamefont {Reimann}}, \bibinfo {author} {\bibfnamefont
  {C.}~\bibnamefont {Dubs}}, \bibinfo {author} {\bibfnamefont {C.}~\bibnamefont
  {Gonzalez-Ballestero}}, \bibinfo {author} {\bibfnamefont {C.}~\bibnamefont
  {Abert}}, \bibinfo {author} {\bibfnamefont {D.}~\bibnamefont {Suess}},
  \bibinfo {author} {\bibfnamefont {D.~A.}\ \bibnamefont {Bozhko}}, \bibinfo
  {author} {\bibfnamefont {S.}~\bibnamefont {Knauer}},\ and\ \bibinfo {author}
  {\bibfnamefont {A.~V.}\ \bibnamefont {Chumak}},\ }\bibfield  {title}
  {\bibinfo {title} {{Magnetic anisotropy and GGG substrate stray field in YIG
  films down to millikelvin temperatures}},\ }\href
  {https://doi.org/10.1038/s44306-024-00030-7} {\bibfield  {journal} {\bibinfo
  {journal} {npj Spintronics}\ }\textbf {\bibinfo {volume} {2}},\ \bibinfo
  {pages} {29} (\bibinfo {year} {2024})}\BibitemShut {NoStop}%
\bibitem [{\citenamefont {Maier-Flaig}\ \emph {et~al.}(2017)\citenamefont
  {Maier-Flaig}, \citenamefont {Klingler}, \citenamefont {Dubs}, \citenamefont
  {Surzhenko}, \citenamefont {Gross}, \citenamefont {Weiler}, \citenamefont
  {Huebl},\ and\ \citenamefont {Goennenwein}}]{Maier-Flaig2017}%
  \BibitemOpen
  \bibfield  {author} {\bibinfo {author} {\bibfnamefont {H.}~\bibnamefont
  {Maier-Flaig}}, \bibinfo {author} {\bibfnamefont {S.}~\bibnamefont
  {Klingler}}, \bibinfo {author} {\bibfnamefont {C.}~\bibnamefont {Dubs}},
  \bibinfo {author} {\bibfnamefont {O.}~\bibnamefont {Surzhenko}}, \bibinfo
  {author} {\bibfnamefont {R.}~\bibnamefont {Gross}}, \bibinfo {author}
  {\bibfnamefont {M.}~\bibnamefont {Weiler}}, \bibinfo {author} {\bibfnamefont
  {H.}~\bibnamefont {Huebl}},\ and\ \bibinfo {author} {\bibfnamefont
  {S.~T.~B.}\ \bibnamefont {Goennenwein}},\ }\bibfield  {title} {\bibinfo
  {title} {{Temperature-dependent magnetic damping of yttrium iron garnet
  spheres}},\ }\href {https://doi.org/10.1103/physrevb.95.214423} {\bibfield
  {journal} {\bibinfo  {journal} {Phys. Rev. B}\ }\textbf {\bibinfo {volume}
  {95}},\ \bibinfo {pages} {214423} (\bibinfo {year} {2017})}\BibitemShut
  {NoStop}%
\bibitem [{\citenamefont {Jermain}\ \emph {et~al.}(2017)\citenamefont
  {Jermain}, \citenamefont {Aradhya}, \citenamefont {Reynolds}, \citenamefont
  {Buhrman}, \citenamefont {Brangham}, \citenamefont {Page}, \citenamefont
  {Hammel}, \citenamefont {Yang},\ and\ \citenamefont {Ralph}}]{Jermain2017}%
  \BibitemOpen
  \bibfield  {author} {\bibinfo {author} {\bibfnamefont {C.~L.}\ \bibnamefont
  {Jermain}}, \bibinfo {author} {\bibfnamefont {S.~V.}\ \bibnamefont
  {Aradhya}}, \bibinfo {author} {\bibfnamefont {N.~D.}\ \bibnamefont
  {Reynolds}}, \bibinfo {author} {\bibfnamefont {R.~A.}\ \bibnamefont
  {Buhrman}}, \bibinfo {author} {\bibfnamefont {J.~T.}\ \bibnamefont
  {Brangham}}, \bibinfo {author} {\bibfnamefont {M.~R.}\ \bibnamefont {Page}},
  \bibinfo {author} {\bibfnamefont {P.~C.}\ \bibnamefont {Hammel}}, \bibinfo
  {author} {\bibfnamefont {F.~Y.}\ \bibnamefont {Yang}},\ and\ \bibinfo
  {author} {\bibfnamefont {D.~C.}\ \bibnamefont {Ralph}},\ }\bibfield  {title}
  {\bibinfo {title} {{Increased low-temperature damping in yttrium iron garnet
  thin films}},\ }\href {https://doi.org/10.1103/physrevb.95.174411} {\bibfield
   {journal} {\bibinfo  {journal} {Phys. Rev. B}\ }\textbf {\bibinfo {volume}
  {95}},\ \bibinfo {pages} {174411} (\bibinfo {year} {2017})}\BibitemShut
  {NoStop}%
\bibitem [{\citenamefont {Kosen}\ \emph {et~al.}(2019)\citenamefont {Kosen},
  \citenamefont {van Loo}, \citenamefont {Bozhko}, \citenamefont {Mihalceanu},\
  and\ \citenamefont {Karenowska}}]{Kosen2019}%
  \BibitemOpen
  \bibfield  {author} {\bibinfo {author} {\bibfnamefont {S.}~\bibnamefont
  {Kosen}}, \bibinfo {author} {\bibfnamefont {A.~F.}\ \bibnamefont {van Loo}},
  \bibinfo {author} {\bibfnamefont {D.~A.}\ \bibnamefont {Bozhko}}, \bibinfo
  {author} {\bibfnamefont {L.}~\bibnamefont {Mihalceanu}},\ and\ \bibinfo
  {author} {\bibfnamefont {A.~D.}\ \bibnamefont {Karenowska}},\ }\bibfield
  {title} {\bibinfo {title} {{Microwave magnon damping in YIG films at
  millikelvin temperatures}},\ }\href {https://doi.org/10.1063/1.5115266}
  {\bibfield  {journal} {\bibinfo  {journal} {APL Mater.}\ }\textbf {\bibinfo
  {volume} {7}},\ \bibinfo {pages} {101120} (\bibinfo {year}
  {2019})}\BibitemShut {NoStop}%
\bibitem [{\citenamefont {Wang}\ \emph {et~al.}(2020)\citenamefont {Wang},
  \citenamefont {Lu}, \citenamefont {Zhao}, \citenamefont {Zhang},
  \citenamefont {Chen}, \citenamefont {Tian}, \citenamefont {Yan},
  \citenamefont {Bai},\ and\ \citenamefont {Harder}}]{Wang2020d}%
  \BibitemOpen
  \bibfield  {author} {\bibinfo {author} {\bibfnamefont {L.}~\bibnamefont
  {Wang}}, \bibinfo {author} {\bibfnamefont {Z.}~\bibnamefont {Lu}}, \bibinfo
  {author} {\bibfnamefont {X.}~\bibnamefont {Zhao}}, \bibinfo {author}
  {\bibfnamefont {W.}~\bibnamefont {Zhang}}, \bibinfo {author} {\bibfnamefont
  {Y.}~\bibnamefont {Chen}}, \bibinfo {author} {\bibfnamefont {Y.}~\bibnamefont
  {Tian}}, \bibinfo {author} {\bibfnamefont {S.}~\bibnamefont {Yan}}, \bibinfo
  {author} {\bibfnamefont {L.}~\bibnamefont {Bai}},\ and\ \bibinfo {author}
  {\bibfnamefont {M.}~\bibnamefont {Harder}},\ }\bibfield  {title} {\bibinfo
  {title} {{Magnetization coupling in a YIG/GGG structure}},\ }\href
  {https://doi.org/10.1103/physrevb.102.144428} {\bibfield  {journal} {\bibinfo
   {journal} {Phys. Rev. B}\ }\textbf {\bibinfo {volume} {102}},\ \bibinfo
  {pages} {144428} (\bibinfo {year} {2020})}\BibitemShut {NoStop}%
\bibitem [{\citenamefont {Lenz}\ \emph {et~al.}(2006)\citenamefont {Lenz},
  \citenamefont {Wende}, \citenamefont {Kuch}, \citenamefont {Baberschke},
  \citenamefont {Nagy},\ and\ \citenamefont {J{\'{a}}nossy}}]{Lenz2006}%
  \BibitemOpen
  \bibfield  {author} {\bibinfo {author} {\bibfnamefont {K.}~\bibnamefont
  {Lenz}}, \bibinfo {author} {\bibfnamefont {H.}~\bibnamefont {Wende}},
  \bibinfo {author} {\bibfnamefont {W.}~\bibnamefont {Kuch}}, \bibinfo {author}
  {\bibfnamefont {K.}~\bibnamefont {Baberschke}}, \bibinfo {author}
  {\bibfnamefont {K.}~\bibnamefont {Nagy}},\ and\ \bibinfo {author}
  {\bibfnamefont {A.}~\bibnamefont {J{\'{a}}nossy}},\ }\bibfield  {title}
  {\bibinfo {title} {{Two-magnon scattering and viscous Gilbert damping in
  ultrathin ferromagnets}},\ }\href
  {https://doi.org/10.1103/physrevb.73.144424} {\bibfield  {journal} {\bibinfo
  {journal} {Phys. Rev. B}\ }\textbf {\bibinfo {volume} {73}},\ \bibinfo
  {pages} {144424} (\bibinfo {year} {2006})}\BibitemShut {NoStop}%
\bibitem [{\citenamefont {Dillon}\ and\ \citenamefont
  {Nielsen}(1959)}]{Dillon1959}%
  \BibitemOpen
  \bibfield  {author} {\bibinfo {author} {\bibfnamefont {J.~F.}\ \bibnamefont
  {Dillon}}\ and\ \bibinfo {author} {\bibfnamefont {J.~W.}\ \bibnamefont
  {Nielsen}},\ }\bibfield  {title} {\bibinfo {title} {{Effects of Rare Earth
  Impurities on Ferrimagnetic Resonance in Yttrium Iron Garnet}},\ }\href
  {https://doi.org/10.1103/PhysRevLett.3.30} {\bibfield  {journal} {\bibinfo
  {journal} {Phys. Rev. Lett.}\ }\textbf {\bibinfo {volume} {3}},\ \bibinfo
  {pages} {30} (\bibinfo {year} {1959})}\BibitemShut {NoStop}%
\bibitem [{\citenamefont {Dillon}(1962)}]{Dillon1962}%
  \BibitemOpen
  \bibfield  {author} {\bibinfo {author} {\bibfnamefont {J.~F.}\ \bibnamefont
  {Dillon}},\ }\bibfield  {title} {\bibinfo {title} {{Ferrimagnetic Resonance
  in Rare-Earth-Doped Yttrium Iron Garnet. {III}. Linewidth}},\ }\href
  {https://doi.org/10.1103/PhysRev.127.1495} {\bibfield  {journal} {\bibinfo
  {journal} {Phys. Rev.}\ }\textbf {\bibinfo {volume} {127}},\ \bibinfo {pages}
  {1495} (\bibinfo {year} {1962})}\BibitemShut {NoStop}%
\bibitem [{\citenamefont {Seiden}(1964)}]{Seiden1964}%
  \BibitemOpen
  \bibfield  {author} {\bibinfo {author} {\bibfnamefont {P.~E.}\ \bibnamefont
  {Seiden}},\ }\bibfield  {title} {\bibinfo {title} {{Ferrimagnetic Resonance
  Relaxation in Rare-Earth Iron Garnets}},\ }\href
  {https://doi.org/10.1103/PhysRev.133.A728} {\bibfield  {journal} {\bibinfo
  {journal} {Phys. Rev.}\ }\textbf {\bibinfo {volume} {133}},\ \bibinfo {pages}
  {A728} (\bibinfo {year} {1964})}\BibitemShut {NoStop}%
\bibitem [{\citenamefont {Spencer}\ \emph {et~al.}(1964)\citenamefont
  {Spencer}, \citenamefont {Remeika},\ and\ \citenamefont
  {Lenzo}}]{Spencer1964}%
  \BibitemOpen
  \bibfield  {author} {\bibinfo {author} {\bibfnamefont {E.~G.}\ \bibnamefont
  {Spencer}}, \bibinfo {author} {\bibfnamefont {J.~P.}\ \bibnamefont
  {Remeika}},\ and\ \bibinfo {author} {\bibfnamefont {P.~V.}\ \bibnamefont
  {Lenzo}},\ }\bibfield  {title} {\bibinfo {title} {{Ferromagnetic Resonance
  Losses in Indium-substituted Yttrium Iron Garnet}},\ }\href
  {https://doi.org/10.1063/1.1753921} {\bibfield  {journal} {\bibinfo
  {journal} {Appl. Phys. Lett.}\ }\textbf {\bibinfo {volume} {4}},\ \bibinfo
  {pages} {171} (\bibinfo {year} {1964})}\BibitemShut {NoStop}%
\bibitem [{\citenamefont {B{\"{o}}ttcher}\ \emph {et~al.}(2022)\citenamefont
  {B{\"{o}}ttcher}, \citenamefont {Ruhwedel}, \citenamefont {Levchenko},
  \citenamefont {Wang}, \citenamefont {Chumak}, \citenamefont {Popov},
  \citenamefont {Zavislyak}, \citenamefont {Dubs}, \citenamefont {Surzhenko},
  \citenamefont {Hillebrands}, \citenamefont {Chumak},\ and\ \citenamefont
  {Pirro}}]{Boettcher2022}%
  \BibitemOpen
  \bibfield  {author} {\bibinfo {author} {\bibfnamefont {T.}~\bibnamefont
  {B{\"{o}}ttcher}}, \bibinfo {author} {\bibfnamefont {M.}~\bibnamefont
  {Ruhwedel}}, \bibinfo {author} {\bibfnamefont {K.~O.}\ \bibnamefont
  {Levchenko}}, \bibinfo {author} {\bibfnamefont {Q.}~\bibnamefont {Wang}},
  \bibinfo {author} {\bibfnamefont {H.~L.}\ \bibnamefont {Chumak}}, \bibinfo
  {author} {\bibfnamefont {M.~A.}\ \bibnamefont {Popov}}, \bibinfo {author}
  {\bibfnamefont {I.~V.}\ \bibnamefont {Zavislyak}}, \bibinfo {author}
  {\bibfnamefont {C.}~\bibnamefont {Dubs}}, \bibinfo {author} {\bibfnamefont
  {O.}~\bibnamefont {Surzhenko}}, \bibinfo {author} {\bibfnamefont
  {B.}~\bibnamefont {Hillebrands}}, \bibinfo {author} {\bibfnamefont {A.~V.}\
  \bibnamefont {Chumak}},\ and\ \bibinfo {author} {\bibfnamefont
  {P.}~\bibnamefont {Pirro}},\ }\bibfield  {title} {\bibinfo {title} {{Fast
  long-wavelength exchange spin waves in partially compensated Ga:YIG}},\
  }\href {https://doi.org/10.1063/5.0082724} {\bibfield  {journal} {\bibinfo
  {journal} {Appl. Phys. Lett.}\ }\textbf {\bibinfo {volume} {120}},\ \bibinfo
  {pages} {102401} (\bibinfo {year} {2022})}\BibitemShut {NoStop}%
\bibitem [{\citenamefont {Scheffler}\ \emph {et~al.}(2023)\citenamefont
  {Scheffler}, \citenamefont {Steuer}, \citenamefont {Zhou}, \citenamefont
  {Siegl}, \citenamefont {Goennenwein},\ and\ \citenamefont
  {Lammel}}]{Scheffler2023}%
  \BibitemOpen
  \bibfield  {author} {\bibinfo {author} {\bibfnamefont {D.}~\bibnamefont
  {Scheffler}}, \bibinfo {author} {\bibfnamefont {O.}~\bibnamefont {Steuer}},
  \bibinfo {author} {\bibfnamefont {S.}~\bibnamefont {Zhou}}, \bibinfo {author}
  {\bibfnamefont {L.}~\bibnamefont {Siegl}}, \bibinfo {author} {\bibfnamefont
  {S.~T.~B.}\ \bibnamefont {Goennenwein}},\ and\ \bibinfo {author}
  {\bibfnamefont {M.}~\bibnamefont {Lammel}},\ }\bibfield  {title} {\bibinfo
  {title} {{Aluminium substituted yttrium iron garnet thin films with reduced
  Curie temperature}},\ }\href
  {https://doi.org/10.1103/physrevmaterials.7.094405} {\bibfield  {journal}
  {\bibinfo  {journal} {Phys. Rev. Mater.}\ }\textbf {\bibinfo {volume} {7}},\
  \bibinfo {pages} {094405} (\bibinfo {year} {2023})}\BibitemShut {NoStop}%
\bibitem [{\citenamefont {Kriegner}\ \emph {et~al.}(2013)\citenamefont
  {Kriegner}, \citenamefont {Wintersberger},\ and\ \citenamefont
  {Stangl}}]{Kriegner2013}%
  \BibitemOpen
  \bibfield  {author} {\bibinfo {author} {\bibfnamefont {D.}~\bibnamefont
  {Kriegner}}, \bibinfo {author} {\bibfnamefont {E.}~\bibnamefont
  {Wintersberger}},\ and\ \bibinfo {author} {\bibfnamefont {J.}~\bibnamefont
  {Stangl}},\ }\bibfield  {title} {\bibinfo {title} {{xrayutilities: a
  versatile tool for reciprocal space conversion of scattering data recorded
  with linear and area detectors}},\ }\href
  {https://doi.org/10.1107/s0021889813017214} {\bibfield  {journal} {\bibinfo
  {journal} {J. Appl. Crystallogr.}\ }\textbf {\bibinfo {volume} {46}},\
  \bibinfo {pages} {1162} (\bibinfo {year} {2013})}\BibitemShut {NoStop}%
\bibitem [{\citenamefont {Pesquera}\ \emph {et~al.}(2011)\citenamefont
  {Pesquera}, \citenamefont {Marti}, \citenamefont {Holy}, \citenamefont
  {Bachelet}, \citenamefont {Herranz},\ and\ \citenamefont
  {Fontcuberta}}]{Pesquera2011}%
  \BibitemOpen
  \bibfield  {author} {\bibinfo {author} {\bibfnamefont {D.}~\bibnamefont
  {Pesquera}}, \bibinfo {author} {\bibfnamefont {X.}~\bibnamefont {Marti}},
  \bibinfo {author} {\bibfnamefont {V.}~\bibnamefont {Holy}}, \bibinfo {author}
  {\bibfnamefont {R.}~\bibnamefont {Bachelet}}, \bibinfo {author}
  {\bibfnamefont {G.}~\bibnamefont {Herranz}},\ and\ \bibinfo {author}
  {\bibfnamefont {J.}~\bibnamefont {Fontcuberta}},\ }\bibfield  {title}
  {\bibinfo {title} {{X-ray interference effects on the determination of
  structural data in ultrathin La$_{2/3}$Sr$_{1/3}$MnO$_{3}$ epitaxial thin
  films}},\ }\href {https://doi.org/10.1063/1.3663574} {\bibfield  {journal}
  {\bibinfo  {journal} {Appl. Phys. Lett.}\ }\textbf {\bibinfo {volume} {99}},\
  \bibinfo {pages} {221901} (\bibinfo {year} {2011})}\BibitemShut {NoStop}%
\bibitem [{\citenamefont {H{\"{y}}tch}\ \emph {et~al.}(1998)\citenamefont
  {H{\"{y}}tch}, \citenamefont {Snoeck},\ and\ \citenamefont
  {Kilaas}}]{Hytch1998}%
  \BibitemOpen
  \bibfield  {author} {\bibinfo {author} {\bibfnamefont {M.}~\bibnamefont
  {H{\"{y}}tch}}, \bibinfo {author} {\bibfnamefont {E.}~\bibnamefont
  {Snoeck}},\ and\ \bibinfo {author} {\bibfnamefont {R.}~\bibnamefont
  {Kilaas}},\ }\bibfield  {title} {\bibinfo {title} {{Quantitative measurement
  of displacement and strain fields from HREM micrographs}},\ }\href
  {https://doi.org/10.1016/s0304-3991(98)00035-7} {\bibfield  {journal}
  {\bibinfo  {journal} {Ultramicroscopy}\ }\textbf {\bibinfo {volume} {74}},\
  \bibinfo {pages} {131} (\bibinfo {year} {1998})}\BibitemShut {NoStop}%
\bibitem [{\citenamefont {H{\"{u}}e}\ \emph {et~al.}(2005)\citenamefont
  {H{\"{u}}e}, \citenamefont {Johnson}, \citenamefont {Lartigue-Korinek},
  \citenamefont {Wang}, \citenamefont {Buseck},\ and\ \citenamefont
  {H{\"{y}}tch}}]{Huee2005}%
  \BibitemOpen
  \bibfield  {author} {\bibinfo {author} {\bibfnamefont {F.}~\bibnamefont
  {H{\"{u}}e}}, \bibinfo {author} {\bibfnamefont {C.~L.}\ \bibnamefont
  {Johnson}}, \bibinfo {author} {\bibfnamefont {S.}~\bibnamefont
  {Lartigue-Korinek}}, \bibinfo {author} {\bibfnamefont {G.}~\bibnamefont
  {Wang}}, \bibinfo {author} {\bibfnamefont {P.~R.}\ \bibnamefont {Buseck}},\
  and\ \bibinfo {author} {\bibfnamefont {M.~J.}\ \bibnamefont {H{\"{y}}tch}},\
  }\bibfield  {title} {\bibinfo {title} {{Calibration of projector lens
  distortions}},\ }\href {https://doi.org/10.1093/jmicro/dfi042} {\bibfield
  {journal} {\bibinfo  {journal} {Microscopy}\ }\textbf {\bibinfo {volume}
  {54}},\ \bibinfo {pages} {181} (\bibinfo {year} {2005})}\BibitemShut
  {NoStop}%
\bibitem [{\citenamefont {H{\"{y}}tch}\ and\ \citenamefont
  {Plamann}(2001)}]{Hytch2001}%
  \BibitemOpen
  \bibfield  {author} {\bibinfo {author} {\bibfnamefont {M.}~\bibnamefont
  {H{\"{y}}tch}}\ and\ \bibinfo {author} {\bibfnamefont {T.}~\bibnamefont
  {Plamann}},\ }\bibfield  {title} {\bibinfo {title} {{Imaging conditions for
  reliable measurement of displacement and strain in high-resolution electron
  microscopy}},\ }\href {https://doi.org/10.1016/s0304-3991(00)00099-1}
  {\bibfield  {journal} {\bibinfo  {journal} {Ultramicroscopy}\ }\textbf
  {\bibinfo {volume} {87}},\ \bibinfo {pages} {199} (\bibinfo {year}
  {2001})}\BibitemShut {NoStop}%
\end{thebibliography}%


\begin{thebibliography}{13}%
\makeatletter
\providecommand \@ifxundefined [1]{%
 \@ifx{#1\undefined}
}%
\providecommand \@ifnum [1]{%
 \ifnum #1\expandafter \@firstoftwo
 \else \expandafter \@secondoftwo
 \fi
}%
\providecommand \@ifx [1]{%
 \ifx #1\expandafter \@firstoftwo
 \else \expandafter \@secondoftwo
 \fi
}%
\providecommand \natexlab [1]{#1}%
\providecommand \enquote  [1]{``#1''}%
\providecommand \bibnamefont  [1]{#1}%
\providecommand \bibfnamefont [1]{#1}%
\providecommand \citenamefont [1]{#1}%
\providecommand \href@noop [0]{\@secondoftwo}%
\providecommand \href [0]{\begingroup \@sanitize@url \@href}%
\providecommand \@href[1]{\@@startlink{#1}\@@href}%
\providecommand \@@href[1]{\endgroup#1\@@endlink}%
\providecommand \@sanitize@url [0]{\catcode `\\12\catcode `\$12\catcode
  `\&12\catcode `\#12\catcode `\^12\catcode `\_12\catcode `\%12\relax}%
\providecommand \@@startlink[1]{}%
\providecommand \@@endlink[0]{}%
\providecommand \url  [0]{\begingroup\@sanitize@url \@url }%
\providecommand \@url [1]{\endgroup\@href {#1}{\urlprefix }}%
\providecommand \urlprefix  [0]{URL }%
\providecommand \Eprint [0]{\href }%
\providecommand \doibase [0]{https://doi.org/}%
\providecommand \selectlanguage [0]{\@gobble}%
\providecommand \bibinfo  [0]{\@secondoftwo}%
\providecommand \bibfield  [0]{\@secondoftwo}%
\providecommand \translation [1]{[#1]}%
\providecommand \BibitemOpen [0]{}%
\providecommand \bibitemStop [0]{}%
\providecommand \bibitemNoStop [0]{.\EOS\space}%
\providecommand \EOS [0]{\spacefactor3000\relax}%
\providecommand \BibitemShut  [1]{\csname bibitem#1\endcsname}%
\let\auto@bib@innerbib\@empty
\bibitem [{\citenamefont {H{\"{y}}tch}\ \emph {et~al.}(1998)\citenamefont
  {H{\"{y}}tch}, \citenamefont {Snoeck},\ and\ \citenamefont
  {Kilaas}}]{Hytch1998}%
  \BibitemOpen
  \bibfield  {author} {\bibinfo {author} {\bibfnamefont {M.}~\bibnamefont
  {H{\"{y}}tch}}, \bibinfo {author} {\bibfnamefont {E.}~\bibnamefont
  {Snoeck}},\ and\ \bibinfo {author} {\bibfnamefont {R.}~\bibnamefont
  {Kilaas}},\ }\bibfield  {title} {\bibinfo {title} {{Quantitative measurement
  of displacement and strain fields from HREM micrographs}},\ }\href
  {https://doi.org/10.1016/s0304-3991(98)00035-7} {\bibfield  {journal}
  {\bibinfo  {journal} {Ultramicroscopy}\ }\textbf {\bibinfo {volume} {74}},\
  \bibinfo {pages} {131} (\bibinfo {year} {1998})}\BibitemShut {NoStop}%
\bibitem [{\citenamefont {H{\"{u}}e}\ \emph {et~al.}(2005)\citenamefont
  {H{\"{u}}e}, \citenamefont {Johnson}, \citenamefont {Lartigue-Korinek},
  \citenamefont {Wang}, \citenamefont {Buseck},\ and\ \citenamefont
  {H{\"{y}}tch}}]{Huee2005}%
  \BibitemOpen
  \bibfield  {author} {\bibinfo {author} {\bibfnamefont {F.}~\bibnamefont
  {H{\"{u}}e}}, \bibinfo {author} {\bibfnamefont {C.~L.}\ \bibnamefont
  {Johnson}}, \bibinfo {author} {\bibfnamefont {S.}~\bibnamefont
  {Lartigue-Korinek}}, \bibinfo {author} {\bibfnamefont {G.}~\bibnamefont
  {Wang}}, \bibinfo {author} {\bibfnamefont {P.~R.}\ \bibnamefont {Buseck}},\
  and\ \bibinfo {author} {\bibfnamefont {M.~J.}\ \bibnamefont {H{\"{y}}tch}},\
  }\bibfield  {title} {\bibinfo {title} {{Calibration of projector lens
  distortions}},\ }\href {https://doi.org/10.1093/jmicro/dfi042} {\bibfield
  {journal} {\bibinfo  {journal} {Microscopy}\ }\textbf {\bibinfo {volume}
  {54}},\ \bibinfo {pages} {181} (\bibinfo {year} {2005})}\BibitemShut
  {NoStop}%
\bibitem [{\citenamefont {H{\"{y}}tch}\ and\ \citenamefont
  {Plamann}(2001)}]{Hytch2001}%
  \BibitemOpen
  \bibfield  {author} {\bibinfo {author} {\bibfnamefont {M.}~\bibnamefont
  {H{\"{y}}tch}}\ and\ \bibinfo {author} {\bibfnamefont {T.}~\bibnamefont
  {Plamann}},\ }\bibfield  {title} {\bibinfo {title} {{Imaging conditions for
  reliable measurement of displacement and strain in high-resolution electron
  microscopy}},\ }\href {https://doi.org/10.1016/s0304-3991(00)00099-1}
  {\bibfield  {journal} {\bibinfo  {journal} {Ultramicroscopy}\ }\textbf
  {\bibinfo {volume} {87}},\ \bibinfo {pages} {199} (\bibinfo {year}
  {2001})}\BibitemShut {NoStop}%
\bibitem [{\citenamefont {Su}\ \emph {et~al.}(2021)\citenamefont {Su},
  \citenamefont {Ning}, \citenamefont {Cho},\ and\ \citenamefont
  {Ross}}]{Su2021}%
  \BibitemOpen
  \bibfield  {author} {\bibinfo {author} {\bibfnamefont {T.}~\bibnamefont
  {Su}}, \bibinfo {author} {\bibfnamefont {S.}~\bibnamefont {Ning}}, \bibinfo
  {author} {\bibfnamefont {E.}~\bibnamefont {Cho}},\ and\ \bibinfo {author}
  {\bibfnamefont {C.~A.}\ \bibnamefont {Ross}},\ }\bibfield  {title} {\bibinfo
  {title} {{Magnetism and site occupancy in epitaxial Y-rich yttrium iron
  garnet films}},\ }\href {https://doi.org/10.1103/physrevmaterials.5.094403}
  {\bibfield  {journal} {\bibinfo  {journal} {Phys. Rev. Mater.}\ }\textbf
  {\bibinfo {volume} {5}},\ \bibinfo {pages} {094403} (\bibinfo {year}
  {2021})}\BibitemShut {NoStop}%
\bibitem [{\citenamefont {Santiso}\ \emph {et~al.}(2023)\citenamefont
  {Santiso}, \citenamefont {Garc{\'{i}}a}, \citenamefont {Romanque},
  \citenamefont {Henry}, \citenamefont {Bernier}, \citenamefont {Bagu{\'{e}}s},
  \citenamefont {Caicedo}, \citenamefont {Valvidares},\ and\ \citenamefont
  {Sandiumenge}}]{Santiso2023}%
  \BibitemOpen
  \bibfield  {author} {\bibinfo {author} {\bibfnamefont {J.}~\bibnamefont
  {Santiso}}, \bibinfo {author} {\bibfnamefont {C.}~\bibnamefont
  {Garc{\'{i}}a}}, \bibinfo {author} {\bibfnamefont {C.}~\bibnamefont
  {Romanque}}, \bibinfo {author} {\bibfnamefont {L.}~\bibnamefont {Henry}},
  \bibinfo {author} {\bibfnamefont {N.}~\bibnamefont {Bernier}}, \bibinfo
  {author} {\bibfnamefont {N.}~\bibnamefont {Bagu{\'{e}}s}}, \bibinfo {author}
  {\bibfnamefont {J.~M.}\ \bibnamefont {Caicedo}}, \bibinfo {author}
  {\bibfnamefont {M.}~\bibnamefont {Valvidares}},\ and\ \bibinfo {author}
  {\bibfnamefont {F.}~\bibnamefont {Sandiumenge}},\ }\bibfield  {title}
  {\bibinfo {title} {{Antisite Defects and Chemical Expansion in Low‐damping,
  High‐magnetization Yttrium Iron Garnet Films}},\ }\href
  {https://doi.org/10.1002/cnma.202200495} {\bibfield  {journal} {\bibinfo
  {journal} {ChemNanoMat}\ }\textbf {\bibinfo {volume} {9}},\ \bibinfo {pages}
  {202200495} (\bibinfo {year} {2023})}\BibitemShut {NoStop}%
\bibitem [{\citenamefont {Judy}(1966)}]{Judy1966}%
  \BibitemOpen
  \bibfield  {author} {\bibinfo {author} {\bibfnamefont {J.~H.}\ \bibnamefont
  {Judy}},\ }\bibfield  {title} {\bibinfo {title} {{Anisotropic Magnetic
  Resonance and Relaxation in Silicon-Substituted YIG}},\ }\href
  {https://doi.org/10.1063/1.1708455} {\bibfield  {journal} {\bibinfo
  {journal} {J. Appl. Phys.}\ }\textbf {\bibinfo {volume} {37}},\ \bibinfo
  {pages} {1328} (\bibinfo {year} {1966})}\BibitemShut {NoStop}%
\bibitem [{\citenamefont {Epstein}\ and\ \citenamefont
  {Tocci}(1967)}]{Epstein1967}%
  \BibitemOpen
  \bibfield  {author} {\bibinfo {author} {\bibfnamefont {D.~J.}\ \bibnamefont
  {Epstein}}\ and\ \bibinfo {author} {\bibfnamefont {L.}~\bibnamefont
  {Tocci}},\ }\bibfield  {title} {\bibinfo {title} {{High Temperature Resonance
  Losses in Silicon-Doped Yttrium--Iron Garnet (YIG)}},\ }\href
  {https://doi.org/10.1063/1.1755026} {\bibfield  {journal} {\bibinfo
  {journal} {Appl. Phys. Lett.}\ }\textbf {\bibinfo {volume} {11}},\ \bibinfo
  {pages} {55} (\bibinfo {year} {1967})}\BibitemShut {NoStop}%
\bibitem [{\citenamefont {Rosenberg}\ \emph {et~al.}(2021)\citenamefont
  {Rosenberg}, \citenamefont {Litzius}, \citenamefont {Shaw}, \citenamefont
  {Riley}, \citenamefont {Beach}, \citenamefont {Nembach},\ and\ \citenamefont
  {Ross}}]{Rosenberg2021}%
  \BibitemOpen
  \bibfield  {author} {\bibinfo {author} {\bibfnamefont {E.~R.}\ \bibnamefont
  {Rosenberg}}, \bibinfo {author} {\bibfnamefont {K.}~\bibnamefont {Litzius}},
  \bibinfo {author} {\bibfnamefont {J.~M.}\ \bibnamefont {Shaw}}, \bibinfo
  {author} {\bibfnamefont {G.~A.}\ \bibnamefont {Riley}}, \bibinfo {author}
  {\bibfnamefont {G.~S.~D.}\ \bibnamefont {Beach}}, \bibinfo {author}
  {\bibfnamefont {H.~T.}\ \bibnamefont {Nembach}},\ and\ \bibinfo {author}
  {\bibfnamefont {C.~A.}\ \bibnamefont {Ross}},\ }\bibfield  {title} {\bibinfo
  {title} {{Magnetic Properties and Growth-Induced Anisotropy in Yttrium
  Thulium Iron Garnet Thin Films}},\ }\href
  {https://doi.org/10.1002/aelm.202100452} {\bibfield  {journal} {\bibinfo
  {journal} {Adv. Electron. Mater.}\ }\textbf {\bibinfo {volume} {7}},\
  \bibinfo {pages} {2100452} (\bibinfo {year} {2021})}\BibitemShut {NoStop}%
\bibitem [{\citenamefont {Manuilov}\ \emph {et~al.}(2009)\citenamefont
  {Manuilov}, \citenamefont {Khartsev},\ and\ \citenamefont
  {Grishin}}]{Manuilov2009}%
  \BibitemOpen
  \bibfield  {author} {\bibinfo {author} {\bibfnamefont {S.~A.}\ \bibnamefont
  {Manuilov}}, \bibinfo {author} {\bibfnamefont {S.~I.}\ \bibnamefont
  {Khartsev}},\ and\ \bibinfo {author} {\bibfnamefont {A.~M.}\ \bibnamefont
  {Grishin}},\ }\bibfield  {title} {\bibinfo {title} {{Pulsed laser deposited
  Y3Fe5O12 films: Nature of magnetic anisotropy I}},\ }\href
  {https://doi.org/10.1063/1.3272731} {\bibfield  {journal} {\bibinfo
  {journal} {J. Appl. Phys.}\ }\textbf {\bibinfo {volume} {106}},\ \bibinfo
  {pages} {123917} (\bibinfo {year} {2009})}\BibitemShut {NoStop}%
\bibitem [{\citenamefont {Manuilov}\ and\ \citenamefont
  {Grishin}(2010)}]{Manuilov2010}%
  \BibitemOpen
  \bibfield  {author} {\bibinfo {author} {\bibfnamefont {S.~A.}\ \bibnamefont
  {Manuilov}}\ and\ \bibinfo {author} {\bibfnamefont {A.~M.}\ \bibnamefont
  {Grishin}},\ }\bibfield  {title} {\bibinfo {title} {{Pulsed laser deposited
  Y3Fe5O12 films: Nature of magnetic anisotropy II}},\ }\href
  {https://doi.org/10.1063/1.3446840} {\bibfield  {journal} {\bibinfo
  {journal} {J. Appl. Phys.}\ }\textbf {\bibinfo {volume} {108}},\ \bibinfo
  {pages} {013902} (\bibinfo {year} {2010})}\BibitemShut {NoStop}%
\bibitem [{\citenamefont {Rosenberg}\ \emph {et~al.}(2023)\citenamefont
  {Rosenberg}, \citenamefont {Bauer}, \citenamefont {Cho}, \citenamefont
  {Kumar}, \citenamefont {Pelliciari}, \citenamefont {Occhialini},
  \citenamefont {Ning}, \citenamefont {Kaczmarek}, \citenamefont {Rosenberg},
  \citenamefont {Freeland}, \citenamefont {Chen}, \citenamefont {Wang},
  \citenamefont {LeBeau}, \citenamefont {Comin}, \citenamefont {de~Groot},\
  and\ \citenamefont {Ross}}]{Rosenberg2023}%
  \BibitemOpen
  \bibfield  {author} {\bibinfo {author} {\bibfnamefont {E.}~\bibnamefont
  {Rosenberg}}, \bibinfo {author} {\bibfnamefont {J.}~\bibnamefont {Bauer}},
  \bibinfo {author} {\bibfnamefont {E.}~\bibnamefont {Cho}}, \bibinfo {author}
  {\bibfnamefont {A.}~\bibnamefont {Kumar}}, \bibinfo {author} {\bibfnamefont
  {J.}~\bibnamefont {Pelliciari}}, \bibinfo {author} {\bibfnamefont {C.~A.}\
  \bibnamefont {Occhialini}}, \bibinfo {author} {\bibfnamefont
  {S.}~\bibnamefont {Ning}}, \bibinfo {author} {\bibfnamefont {A.}~\bibnamefont
  {Kaczmarek}}, \bibinfo {author} {\bibfnamefont {R.}~\bibnamefont
  {Rosenberg}}, \bibinfo {author} {\bibfnamefont {J.~W.}\ \bibnamefont
  {Freeland}}, \bibinfo {author} {\bibfnamefont {Y.}~\bibnamefont {Chen}},
  \bibinfo {author} {\bibfnamefont {J.}~\bibnamefont {Wang}}, \bibinfo {author}
  {\bibfnamefont {J.}~\bibnamefont {LeBeau}}, \bibinfo {author} {\bibfnamefont
  {R.}~\bibnamefont {Comin}}, \bibinfo {author} {\bibfnamefont {F.~M.~F.}\
  \bibnamefont {de~Groot}},\ and\ \bibinfo {author} {\bibfnamefont {C.~A.}\
  \bibnamefont {Ross}},\ }\bibfield  {title} {\bibinfo {title} {{Revealing Site
  Occupancy in a Complex Oxide: Terbium Iron Garnet}},\ }\href
  {https://doi.org/10.1002/smll.202300824} {\bibfield  {journal} {\bibinfo
  {journal} {Small}\ }\textbf {\bibinfo {volume} {19}},\ \bibinfo {pages}
  {2300824} (\bibinfo {year} {2023})}\BibitemShut {NoStop}%
\bibitem [{\citenamefont {Arias}\ and\ \citenamefont
  {Mills}(1999)}]{Arias1999a}%
  \BibitemOpen
  \bibfield  {author} {\bibinfo {author} {\bibfnamefont {R.}~\bibnamefont
  {Arias}}\ and\ \bibinfo {author} {\bibfnamefont {D.~L.}\ \bibnamefont
  {Mills}},\ }\bibfield  {title} {\bibinfo {title} {Extrinsic contributions to
  the ferromagnetic resonance response of ultrathin films},\ }\href
  {https://doi.org/10.1103/physrevb.60.7395} {\bibfield  {journal} {\bibinfo
  {journal} {Phys. Rev. B}\ }\textbf {\bibinfo {volume} {60}},\ \bibinfo
  {pages} {7395} (\bibinfo {year} {1999})}\BibitemShut {NoStop}%
\bibitem [{\citenamefont {Jermain}\ \emph {et~al.}(2016)\citenamefont
  {Jermain}, \citenamefont {Paik}, \citenamefont {Aradhya}, \citenamefont
  {Buhrman}, \citenamefont {Schlom},\ and\ \citenamefont
  {Ralph}}]{Jermain2016}%
  \BibitemOpen
  \bibfield  {author} {\bibinfo {author} {\bibfnamefont {C.~L.}\ \bibnamefont
  {Jermain}}, \bibinfo {author} {\bibfnamefont {H.}~\bibnamefont {Paik}},
  \bibinfo {author} {\bibfnamefont {S.~V.}\ \bibnamefont {Aradhya}}, \bibinfo
  {author} {\bibfnamefont {R.~A.}\ \bibnamefont {Buhrman}}, \bibinfo {author}
  {\bibfnamefont {D.~G.}\ \bibnamefont {Schlom}},\ and\ \bibinfo {author}
  {\bibfnamefont {D.~C.}\ \bibnamefont {Ralph}},\ }\bibfield  {title} {\bibinfo
  {title} {{Low-damping sub-10-nm thin films of lutetium iron garnet grown by
  molecular-beam epitaxy}},\ }\href {https://doi.org/10.1063/1.4967695}
  {\bibfield  {journal} {\bibinfo  {journal} {Appl. Phys. Lett.}\ }\textbf
  {\bibinfo {volume} {109}},\ \bibinfo {pages} {192408} (\bibinfo {year}
  {2016})}\BibitemShut {NoStop}%
\end{thebibliography}%

\end{document}


\title{Supplementary Materials for \\ -- \\ Lattice-tunable Substituted Iron Garnets for Low-temperature Magnonics}

\author{William Legrand}
\email{william.legrand@neel.cnrs.fr}
\author{Yana Kemna}
\author{Stefan Sch{\"{a}}ren}
\author{Hanchen Wang}
\author{Davit Petrosyan}
\affiliation{Department of Materials, ETH Zurich, H{\"{o}}nggerbergring 64, 8093 Zurich, Switzerland}
\author{Luise Holder}
\author{Richard Schlitz}
\affiliation{Department of Physics, University of Konstanz, 78457 Konstanz, Germany}
\author{Myriam H.~Aguirre}
\affiliation{Department of Condensed Matter Physics, University of Zaragoza, E-50009 Zaragoza, Spain}
\affiliation{Institute of Nanoscience and Materials of Arag{\'{o}}n, UNIZAR-CSIC, E-50018 Zaragoza, Spain}
\affiliation{Laboratory of Advanced Microscopy, University of Zaragoza, E-50018 Zaragoza, Spain}
\author{Michaela Lammel}
\affiliation{Department of Physics, University of Konstanz, 78457 Konstanz, Germany}
\author{Pietro Gambardella}
\email{pietro.gambardella@mat.ethz.ch}
\affiliation{Department of Materials, ETH Zurich, H{\"{o}}nggerbergring 64, 8093 Zurich, Switzerland}

\maketitle

\tableofcontents

\clearpage

\section{X-ray reflectivity data}\label{sec:XRR}

\begin{figure*}[h]
    \centering
    \includegraphics[clip,width=7.0in,trim=0 0.18in 0 0]{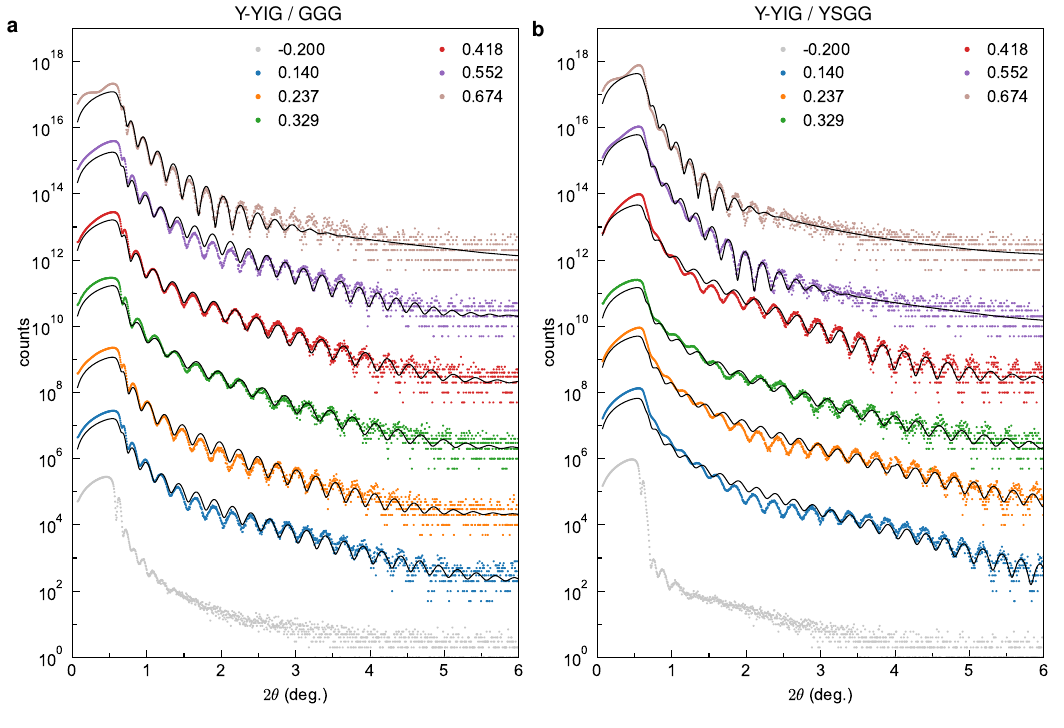}
    \caption{X-ray reflectivity curves for \SI{30}{\nano\meter}-thick \ce{Y_{3+x}Fe_{5-x}O_{12}} films, labeled Y-YIG-1--7 in main manuscript, on \pan{a}\ GGG and \pan{b}\ YSGG substrates. The legend displays the estimated substitution $x$ for each composition. Black lines are the reflectivity curves fit to the data. The curve for each sample is offset by a factor $10^2$.}
    \label{fig:XRR}
\end{figure*}

X-ray reflectivity curves for films Y-YIG-1--7 have been acquired on a Panalytical X'pert MRD thin film diffractometer, equipped with an Eulerian cradle, a Cu-tube, a parallel beam X-ray mirror, and a Ge(220) four-bounce Cu(K$\alpha$1) monochromator to suppress the Cu(K$\alpha$2,K$\beta$) radiation lines. The X-ray reflectivity curves are included in Fig.~\ref{fig:XRR}, and are fitted to extract film thickness values listed in the main manuscript. The roughness parameter probed by the X-ray in such reflectivity measurements in found systematically at 0.5$\pm$\SI{0.1}{\nano\meter}, except for samples with substitution $x=0.552$ on YSGG and with substitution $x=0.674$ on both substrates, for which this roughness reaches 1--\SI{2}{\nano\meter}. All the analysis of the present X-ray reflectivity curves has been performed in Python, relying on the package xrayutilities.

\clearpage

\section{Extended angular range X-ray diffraction data}\label{sec:XRDextended}

\begin{figure*}[h]
    \centering
    \includegraphics[clip,width=7.0in]{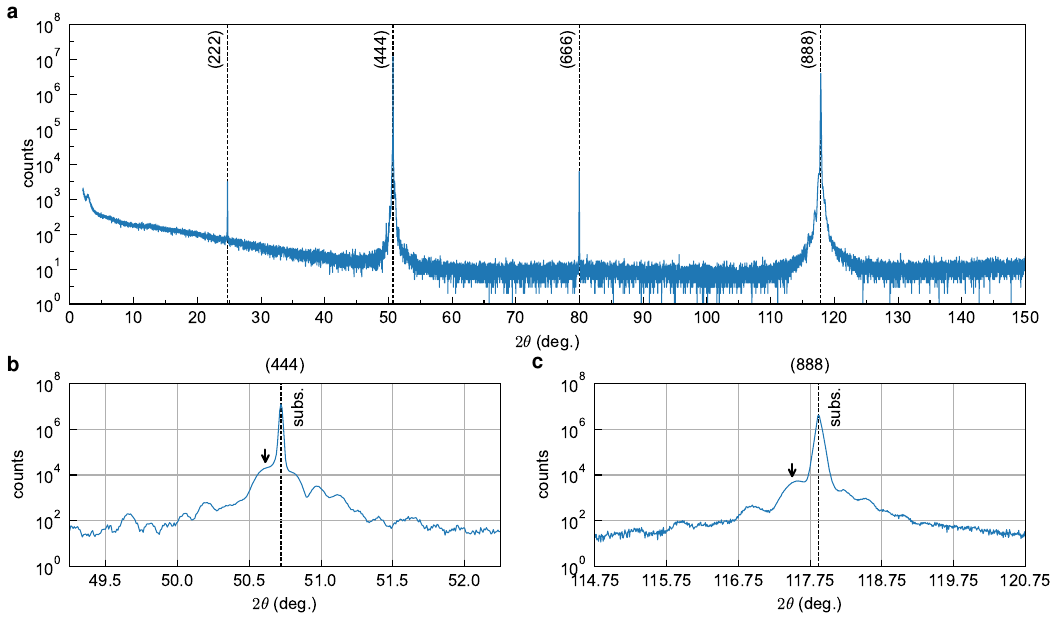}
    \caption{High-resolution XRD for a \SI{90}{\nano\meter}-thick \ce{Y_{3+x}Fe_{5-x}O_{12}} film, $x\approx0.418$, nominally identical to Y-YIG-5. \pan{a}\ Full angular range $2\theta-\omega$ diffractogram for $2\theta=$ 2--\SI{150}{deg}. No other peaks than garnet peaks are present. Magnified plots around \pan{b}\ the YSGG and Y-YIG (444) and \pan{c}\ the YSGG and Y-YIG (888) diffraction peaks. Vertical dashed lines locate substrate peaks. The arrows indicate the main film peak position.}
    \label{fig:extendedXRD}
\end{figure*}

High-resolution X-ray diffraction data for a film nominally identical in composition to Y-YIG-5 grown on YSGG, and a thickness of \SI{90}{\nano\meter} as well as a larger substrate area, has been acquired on a Panalytical X'pert MRD thin film diffractometer, equipped with an Eulerian cradle, a Cu-tube, a parallel beam X-ray mirror, and a Ge(220) four-bounce Cu(K$\alpha$1) monochromator to suppress the Cu(K$\alpha$2,K$\beta$) radiation lines. Scans are performed in symmetric $2\theta-\omega$ geometry and probe the out-of-plane lattice vectors. The X-ray diffractogram is presented in Fig.~\ref{fig:extendedXRD}. As appears in Fig.~\ref{fig:extendedXRD}a, the substrate (444) and (888) peaks are visible, as well as (222) and (666) peaks, which are forbidden in the bulk but appear because of multiple diffraction, possibly combined with surface imperfections in the substrate. The Y-YIG film (444) and (888) peaks and their Laue oscillations are present, with a shorter angular periodicity due to a larger thickness in comparison to films Y-YIG-1--7, and partially overlapping with the substrate peaks, see arrows in the zoomed-in plots of Figs.~\ref{fig:extendedXRD}b,c. Only garnet peaks from the substrate and film are seen in this full-range scan ($2\theta=$ 2--\SI{150}{deg}), notably without any peaks corresponding to other crystalline phases such as orthoferrite, yttria, etc. This shows that in the present growth conditions, an epitaxial Y-YIG film is formed with a coherent interface, ruling out secondary phases.

\clearpage

\section{X-ray reciprocal space maps}\label{sec:RSM}

\begin{figure*}[h]
    \centering
    \includegraphics[clip,width=7.0in]{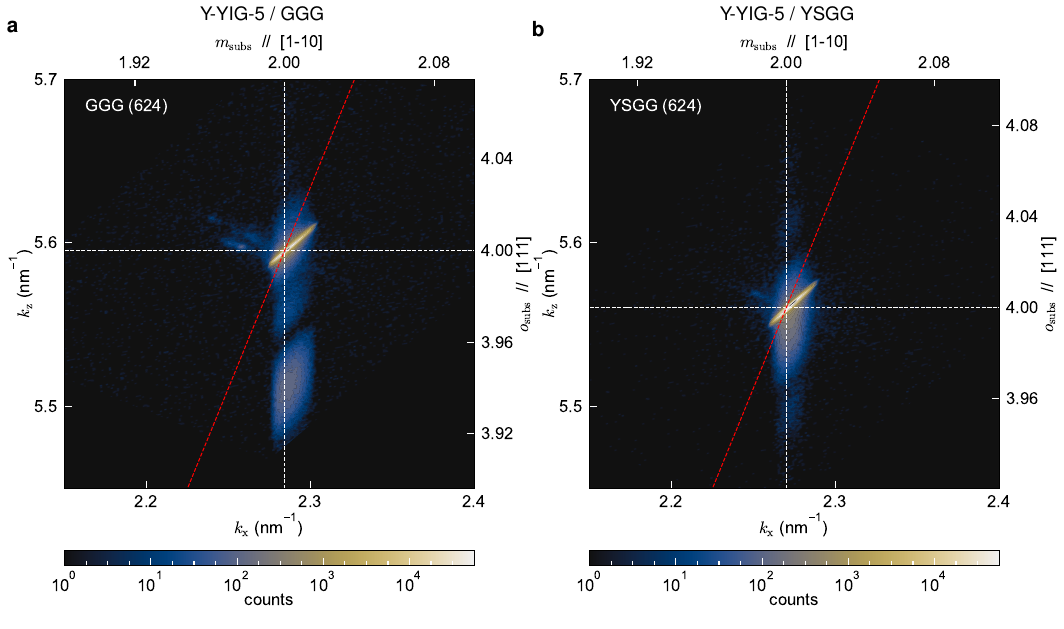}
    \caption{X-ray reciprocal space maps around the (624) substrate peaks for \SI{30}{\nano\meter}-thick \ce{Y_{3+x}Fe_{5-x}O_{12}} films, labeled Y-YIG-5 in main manuscript, on \pan{a}\ GGG and \pan{b}\ YSGG substrates. Vertical and horizontal white dashed lines indicate alignment with substrate (624) peak within the film plane and along the film normal. Red dashed lines indicate the peak position expected in case of relaxed growth. In each panel, the film peaks are aligned with the substrate peak at same $k_{\rm{x}}$, indicating a pseudomorphic growth.}
    \label{fig:RSM}
\end{figure*}

X-ray reciprocal space maps (RSMs) for films Y-YIG-5 have been acquired on a Panalytical X'pert MRD thin film diffractometer, equipped with an Eulerian cradle, a Cu-tube, a parallel beam X-ray mirror, and a Ge(220) four-bounce Cu(K$\alpha$1) monochromator to suppress the Cu(K$\alpha$2,K$\beta$) radiation lines. Maps around the (624) diffraction peak of the cubic structure of the garnet, shown in Fig.~\ref{fig:RSM}, are obtained by varying $\omega$ and $2\theta$ angles, keeping the incident and diffracted beams in the plane formed by the substrate dicing axis along [1-10] and substrate surface normal along [111]. The in-plane reciprocal vector $k_{\rm{x}}$ hence lies along the [1-10] axis and the out-of-plane reciprocal vector $k_{\rm{z}}$ along the [111] axis. The (hkl) reciprocal lattice coordinates can thus be converted into (mno) coordinates aligned with substrate edges, with notably $m_{\rm{subs}}=k_{\rm{x}}d_{1-10}$ and $o_{\rm{subs}}=k_{\rm{z}}d_{111}$, where $d_{1-10}$ and $d_{111}$ are corresponding plane spacings in the substrate, differing between Figs.~\ref{fig:RSM}a,b. In each case, the alignment of the substrate peak with the Laue oscillations from the film along a fixed in-plane $k_{\rm{x}}$ vector (vertical white dashed line) demonstrates the pseudomorphic growth of the Y-YIG films, free of strain relaxation. By contrast, a relaxed growth would feature film peaks located along the red dashed lines. In particular for Y-YIG-5 grown on YSGG, the film and substrate lattice parameter are nearly matched, as evidenced by the strong overlap of the main film peak with the substrate (624) peak located at $(m_{\rm{subs}},o_{\rm{subs}})=(2,4)$.

\clearpage

\section{Additional transmission electron microscopy images and analysis}\label{sec:tem}

Cross-sectional electron-transparent samples have been produced from a sample nominally identical to Y-YIG-3, with estimated $x=0.230$, priorly characterized by FMR and XRD to display very similar structural and magnetic properties, summarized in Table \ref{tab:TEMsamples}. In the following, we refer to these samples as Y-YIG-TEM. The lamellas for electron spectro-microscopy examination were prepared by Focused Ion Beam (FIB), on a Dual Beam Helios 650 equipment (FIB-SEM) from Thermofisher. The structural characterization is performed relying on High-Resolution Scanning Transmission Electron Microscopy with High Angular Annular dark field detector (HRSTEM-HAADF). The measurements are performed in a probe-corrected FEI Titan ranging 80--\SI{300}{\kilo\electronvolt}.

\begin{table*}[h]
    \caption{Summary of the properties of the additional Y-YIG epitaxial films investigated in TEM. Sputtering plasma rectification voltages $U_{\ce{Y2O3}}$ and $U_{\ce{Fe2O3}}$ for each target, relative proportions of deposited oxides $y$ and corresponding stoichiometry $x$ in \ce{Y_{3+x}Fe_{5-x}O_{12}}; film thickness $t$ obtained from X-ray reflectivity measurements; and thickness $t$, out-of-plane lattice parameter $a_{\perp}$, corresponding unstrained film lattice parameter $a$ and relative lattice parameter change $\Delta{}a/a_0$, extracted from fits to the XRD measurements.}
    \label{tab:TEMsamples}
    \begin{tabular*}{\textwidth}{@{\extracolsep{\fill}}lS[table-format=3.1]S[table-format=3.1]S[table-format=1.3]S[table-format=1.2]S[table-format=2.1]S[table-format=2.1]S[table-format=1.4]S[table-format=1.4]S[table-format=1.2]}
        \hline\hline
        & \multicolumn{4}{c}{{Sputtering}} & {XRR} & \multicolumn{4}{c}{{XRD}} \\
        \cline{2-5}\cline{6-6}\cline{7-10}
        & {$U_{\ce{Y2O3}}$} & {$U_{\ce{Fe2O3}}$} & {$y$} & {$x$} & {$t$} & {$t$} & {$a_{\perp}$} & {$a$} & {$\Delta{}a/a_0$} \\
        Samples & {(\si{\volt})} & {(\si{\volt})} & {($\pm$0.020)} & {($\pm$0.040)} & {(\si{\nano\meter})} & {(\si{\nano\meter})} & {(\si{\nano\meter})} & {(\si{\nano\meter})} & {(\si{\percent})} \\
        \hline
        Y-YIG-TEM/GGG   & 198.7 & 103.3 & 1.002 & 0.230 & {\multirow{2}{*}{30.5}}      & {\multirow{2}{*}{30.6}} & 1.2467 & 1.2427 & 0.41 \\
        Y-YIG-TEM/YSGG & 198.7 & 103.3 & 1.002 & 0.230 &                                               &                                         & 1.2396 & 1.2425 & 0.40 \\
        \hline\hline
    \end{tabular*}

\end{table*}

\begin{figure*}[h]
    \centering
    \includegraphics[clip,width=7.0in]{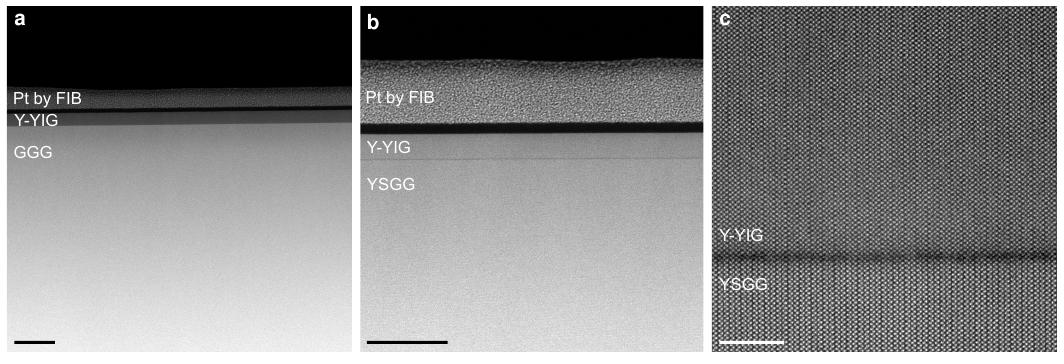}
    \caption{Extended STEM-HAADF data. \pan{a}\ Low-magnification view of Y-YIG-TEM/GGG. The intermediate gray contrast area above the layer is the Pt deposited for FIB preparation. Scale bar is \SI{100}{\nano\meter}. \pan{b}\ Low-magnification view of Y-YIG-TEM/YSGG.  The intermediate gray contrast area above the layer is the Pt deposited for FIB preparation. Scale bar is \SI{100}{\nano\meter}. \pan{c}\ Large-magnification view of the YSGG and Y-YIG interface in HRSTEM-HAADF, exhibiting the particularly small interdiffusion extent of $\pm$\SI{1}{\nano\meter}. Scale bar is \SI{5}{\nano\meter}.}
    \label{fig:TEMlowmag}
\end{figure*}

In Figs.~\ref{fig:TEMlowmag}a,b, low-magnification images obtained in High Angular Annular dark-field mode show the overall aspect of the epitaxial films. For both films grown on GGG and YSGG, observation at various locations only reveals flat and excellently defined interfaces, with an estimated roughness that of the polishing of the starting substrate. The Y-YIG layers exhibit perfectly coherent growth and the thickness estimated from the STEM-HAADF images is 31~$\pm$~\SI{0.5}{\nano\meter}, in perfect agreement with X-ray structural characterization. Note the difference of contrast between substrate and Y-YIG, varying in the case of either GGG or YSGG. Since the STEM-HAADF contrast of each atom increases with their atomic number approximately as $Z^{1.8}$, Gd atoms provide a brighter contrast. Due to the absence of a heavier rare-earth in YSGG, this substrate appears with a STEM-HAADF contrast very similar to that of Y-YIG. In Fig.~\ref{fig:TEMlowmag}c, the YSGG/Y-YIG interface is clearly visible in HRSTEM-HAADF at larger magnifications and reflects the surface roughness of the polished substrates.

\begin{figure*}[h]
    \centering
    \includegraphics[clip,width=7.0in]{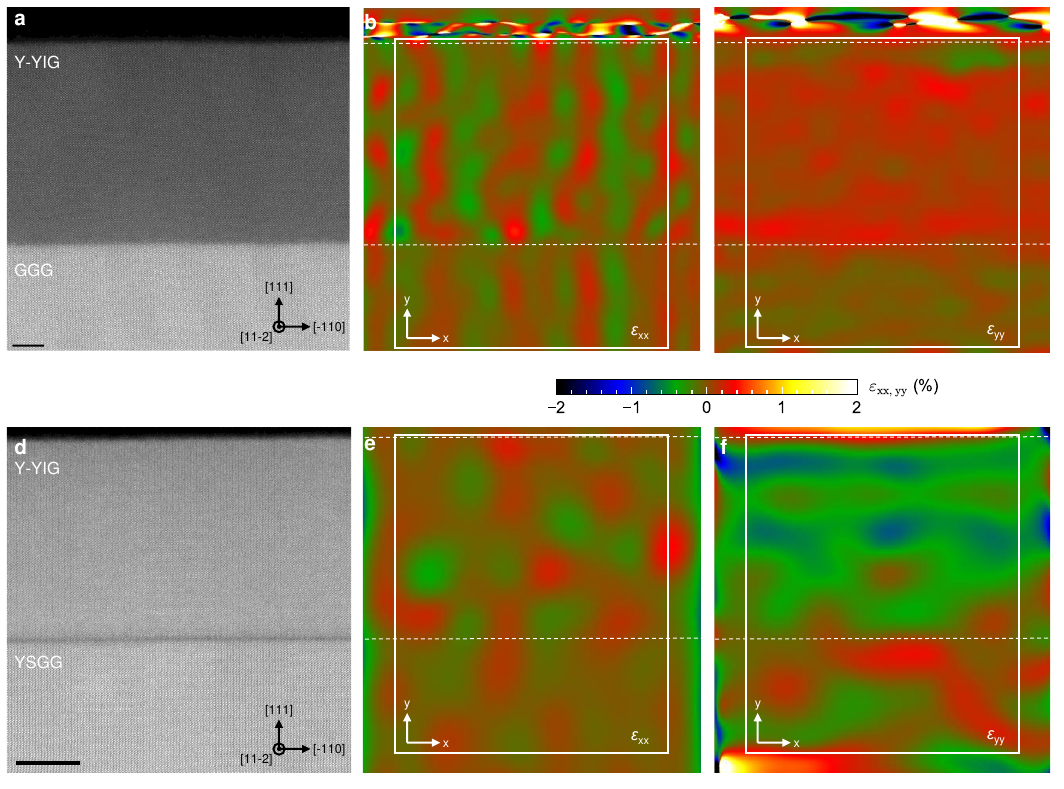}
    \caption{Geometric phase analysis data leading to the thickness profiles of Fig.~2 in main. \pan{a}\ STEM-HAADF image of Y-YIG-TEM on GGG. Scale bar is \SI{5}{\nano\meter}. Strain maps of Y-YIG-TEM on GGG along $x=[-110]$ and $y=[111]$ with \pan{b}\ $\varepsilon_{\rm{xx}}$ and \pan{c}\ $\varepsilon_{\rm{yy}}$ components. \pan{d}\ STEM-HAADF image of Y-YIG-TEM on YSGG. Scale bar is \SI{10}{\nano\meter}. Strain maps of Y-YIG-TEM on YSGG along $x=[-110]$ and $y=[111]$ with \pan{e}\ $\varepsilon_{\rm{xx}}$ and \pan{f}\ $\varepsilon_{\rm{yy}}$ components. White frames indicate the central part used for averaging in the thickness profiles along $y$. The Y-YIG interfaces are located at the dashed lines.}
    \label{fig:GPAimages}
\end{figure*}

We detail further here the geometric phase analysis (GPA) analysis of the strain in the samples, performed following the procedures reported in Refs.\ \cite{Hytch1998,Huee2005,Hytch2001}. The two-dimensional maps resulting from the transformation are provided in Fig.~\ref{fig:GPAimages}. Because the growth is coherent from the substrate, we expect $\varepsilon_{\rm{xx}}=0$ for both samples on GGG and YSGG. From the XRD fit parameters listed in Table \ref{tab:TEMsamples}, we expect $\varepsilon_{\rm{yy}}=(1.2467-1.2382)/1.2382=$~\SI{0.69}{\percent} on GGG, and $\varepsilon_{\rm{yy}}=(1.2396-1.2460)/1.2460=$\SI{-0.52}{\percent} on YSGG. The strain profiles displayed in main are obtained by averaging  the strain maps $\varepsilon_{\rm{xx}}$ and $\varepsilon_{\rm{yy}}$ in the film direction, on a width indicated by the white frames in Figs.~\ref{fig:GPAimages}b,c,e,f. The small amplitude fluctuations of the $\varepsilon_{\rm{xx,yy}}$ values are related to resolution limits and noise. The values of $\varepsilon_{\rm{xx,yy}}$ are quantitatively consistent with our interpretation of coherent growth, free of structural relaxation.

\clearpage

\section{Magnetometry and substrate susceptibility analysis}\label{sec:susceptibility}

\begin{figure*}[h]
    \centering
    \includegraphics[clip,width=7.0in]{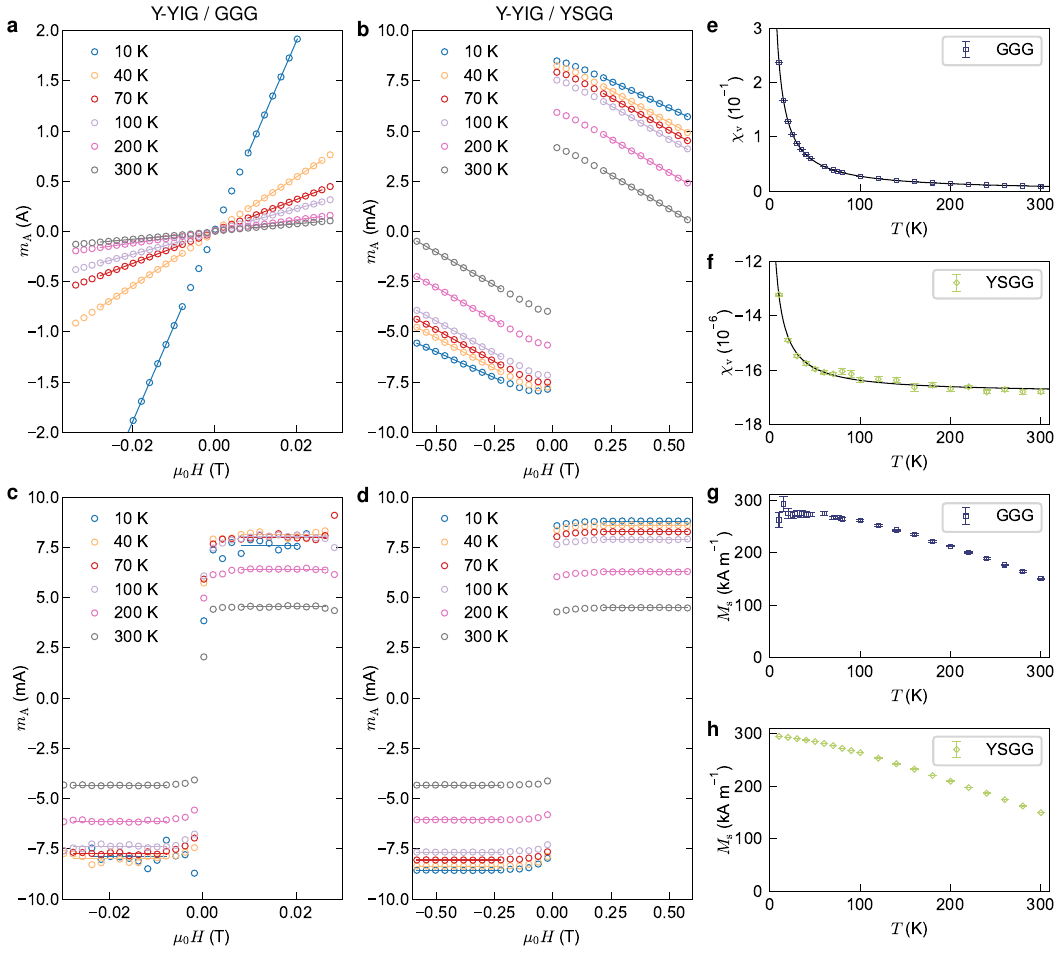}
    \caption{SQUID magnetometry for Y-YIG-5 on GGG and YSGG. Magnetic moment per unit area $m_{\rm{A}}$ as a function of applied field $\mu_0H$ for the sample with \pan{a}\ GGG and \pan{b}\ YSGG substrate. Magnetic moment per unit area $m_{\rm{A}}$ as a function of applied field $\mu_0H$ for film grown on \pan{c}\ GGG and \pan{d}\ YSGG, after subtraction of the substrate magnetic moment. Temperature dependence of the volumic susceptibility $\chi_{\rm{V}}$ of the \pan{e}\ GGG and \pan{f}\ YSGG substrate. Solid line is a fit to $\chi_{\rm{V}}=\chi_{\rm{P}}/T-\chi_{\rm{D}}$. Temperature dependence of the volumic saturation magnetization $M_{\rm{s}}$ of Y-YIG-5 on the \pan{g}\ GGG and \pan{h}\ YSGG substrate.}
    \label{fig:SQUID}
\end{figure*}

An example of magnetization data acquired by SQUID magnetometry for films Y-YIG-5 are reported in this supplementary section. The untreated magnetization data are shown in Figs.~\ref{fig:SQUID}a,b, while Figs.~\ref{fig:SQUID}c,d show the data from which the linear contribution due to substrate susceptibility has been subtracted, at selected temperatures 10, 40, 70, 100, 200 and \SI{300}{\kelvin}. Figs.~\ref{fig:SQUID}e--h compare the temperature dependences of the substrate susceptibility and of the saturation magnetization of the Y-YIG.

Paramagnetic impurities are essentially absent from the YSGG substrates in comparison to GGG substrates. In terms of a volumic DC susceptibility $\chi_{\rm{V}}$, evolving with temperature as $\chi_{\rm{V}}=\chi_{\rm{P}}/T-\chi_{\rm{D}}$, where $\chi_{\rm{P}}$ is the paramagnetic component with an inverse temperature dependence and $\chi_{\rm{D}}$ the diamagnetic component, we find that substrate paramagnetism reduces from $\chi_{\rm{P}}=$ \SI{2.5}{\kelvin} for GGG to $\chi_{\rm{P}}=$ \SI{3.6e-5}{\kelvin} for YSGG. Paramagnetic impurities in similar amounts are also found in the pristine YSGG substrates. Small residual fractions of paramagnetic impurities, in particular rare-earths, are indeed expected to be present in the diamagnetic substrate, but considering their low concentration levels, this is expected to have a negligible impact on magnetic couplings with the epitaxial iron garnet films. The amount of impurities present in the Y-YIG films, for instance, originating from the target material impurities, are not accessible to our measurements since it is dominated by the paramagnetic defects present in the substrate.

\clearpage

\section{Indeterminacy of the defect chemistry of YIG and Y-YIG thin films}\label{sec:indeterminacy}

In general, three kinds of substitutions and defects are expected in iron garnet thin films departing from a nominal YIG unit cell: (a) inclusion of excess non-magnetic atoms (here Y) replacing Fe in octahedral sites and in tetrahedral sites, which are off-stoichiometric antisite defects we aim for; (b) formation of antisite defects of disorder type, swapping Y and Fe between dodecahedral and octahedral sites; (c) formation of Fe and oxygen vacancies. If vacancies do not occur in 2:3 relative proportions, cationic charge defects are consequently formed, with presence of either \ce{Fe^{2+}} or \ce{Fe^{4+}} cations. This corresponds to a general formula \ce{\{Y_{3-c}Fe_{c}\}[Y_{a+c}Fe_{$2-a-c-\delta_{\rm{a}}$}](Y_{d}Fe_{$3-d-\delta_{\rm{d}}$})O_{$12-\delta_{\rm{o}}$}}, with $\{.\}$, $[.]$ and $(.)$ representing dodecahedral, octahedral and tetrahedral sites, respectively, and $c$ corresponding to disorder-type antisite defects, $a$ the excess antisite Y, $d$ the excess antisite Y, $\delta_{\rm{a}}$ octahedral vacancies, $\delta_{\rm{d}}$ tetrahedral vacancies and $\delta_{\rm{O}}$ oxygen vacancies.

Here, the fraction of octahedral substitution is $k_{\rm{a}}=(a+c+\delta_{\rm{a}})/2$ and of tetrahedral substitution is $k_{\rm{d}}=(d+\delta_{\rm{d}})/3$, and the iron garnet formula is also written \ce{\{Y_{3-c}Fe_{c}\}[Y_{a+c}Fe_{$2(1-k_{\rm{a}})$}](Y_{d}Fe_{$3(1-k_{\rm{d}})$})O_{$12-\delta_{\rm{o}}$}}. The best fits with molecular-field-coefficients model to the magnetometry data presented in main actually provide $k_{\rm{a}}-c/2$ and $k_{\rm{d}}$,  and suggest that off-stoichiometric Y does not occupy tetrahedral sites, with $d\approx0$. This is consistent with previous reports \cite{Su2021,Santiso2023}. In the absence of vacancies and disorder-type antisite defects, octahedral substitution would then follow $k_{\rm{a}}=a/2=x/2$, which is not verified in the fits to the magnetometry data. Instead, $2k_{\rm{a}}-c-x$ is found $\approx0.19$ for Y-YIG-3--6. We now discuss vacancies and disorder-type antisite defects.

Already $\approx$~\SI{1}{\percent} of non-trivalent Fe atoms ($x\approx0.05$) in \ce{Y_{3}Fe_{5-x}Si_{x}O_{12}} is sufficient to cause a considerable increase of 1 or \SI{2}{\milli\tesla} in room-temperature FMR linewidth at \SI{10}{\giga\hertz} \cite{Judy1966,Epstein1967}. Consequently, any imbalance in the proportion of iron and oxygen vacancies would lead to large magnetic dissipation. Because Y-YIG-2--6 achieve a room-temperature FMR linewidth better than 1 or \SI{2}{\milli\tesla}, we can bound vacancies imbalance to $|3\delta_{\rm{a}}+3\delta_{\rm{d}}-2\delta_{\rm{o}}|<0.05$.

Considering the antiferromagnetic coupling between magnetic atoms located in tetrahedral sites and both octahedral and dodecahedral sites, and the absence of significant amounts of \ce{Fe^{2+}} or \ce{Fe^{4+}}, the total zero-temperature magnetic moment per formula unit is approximately $5\mu_{\rm{B}}|c+(2-2k_{\rm{a}})-(3-3k_{\rm{d}})|=5\mu_{\rm{B}}|1+(a+\delta_{\rm{a}})-(d+\delta_{\rm{d}})|$. It is notably not sensitive to $c$, and only the temperature variations of the magnetization might provide a hint of antisite defects of disorder type, which seems challenging to obtain quantitatively.  It is proposed to assign the difference between $a$ and $x$ to the formation of iron and oxygen vacancies in 2:3 proportions, while the presence of antisite defects of disorder type is expected but not presently quantified.

To further verify the presence of vacancies and the evolution of the Y:Fe ratio in the present series of sputtered epitaxial films, X-ray photoelectron spectroscopy has been acquired on samples Y-YIG-1--7. After cleaning of the surfaces with a soft Ar plasma, the samples are immediately introduced in a ultrahigh vacuum chamber ($10^{-9}\,$\si{\milli\bar} range). The X-ray source is Al K$\alpha$ radiation (\SI{1486.6}{\electronvolt}) and a flood gun is employed to minimize the effects of surface charging, providing spectra without charging artifacts. The ejected electrons are collected by a hemispherical analyzer at \SI{100}{\electronvolt} constant pass energy for survey scans and \SI{30}{\electronvolt} for detailed peak scans aimed for quantification. Detailed scans are repeated 10 times with a step size of \SI{0.1}{\electronvolt}. The energy scale is calibrated with the C 1s line from the adventitious carbon contamination at \SI{285.0}{\electronvolt}. Measurements are done at a constant angle, with normal incidence \SI{90}{deg} between the sample surface and the analyzer. 

\begin{figure*}
    \centering
    \includegraphics[clip,width=7.0in]{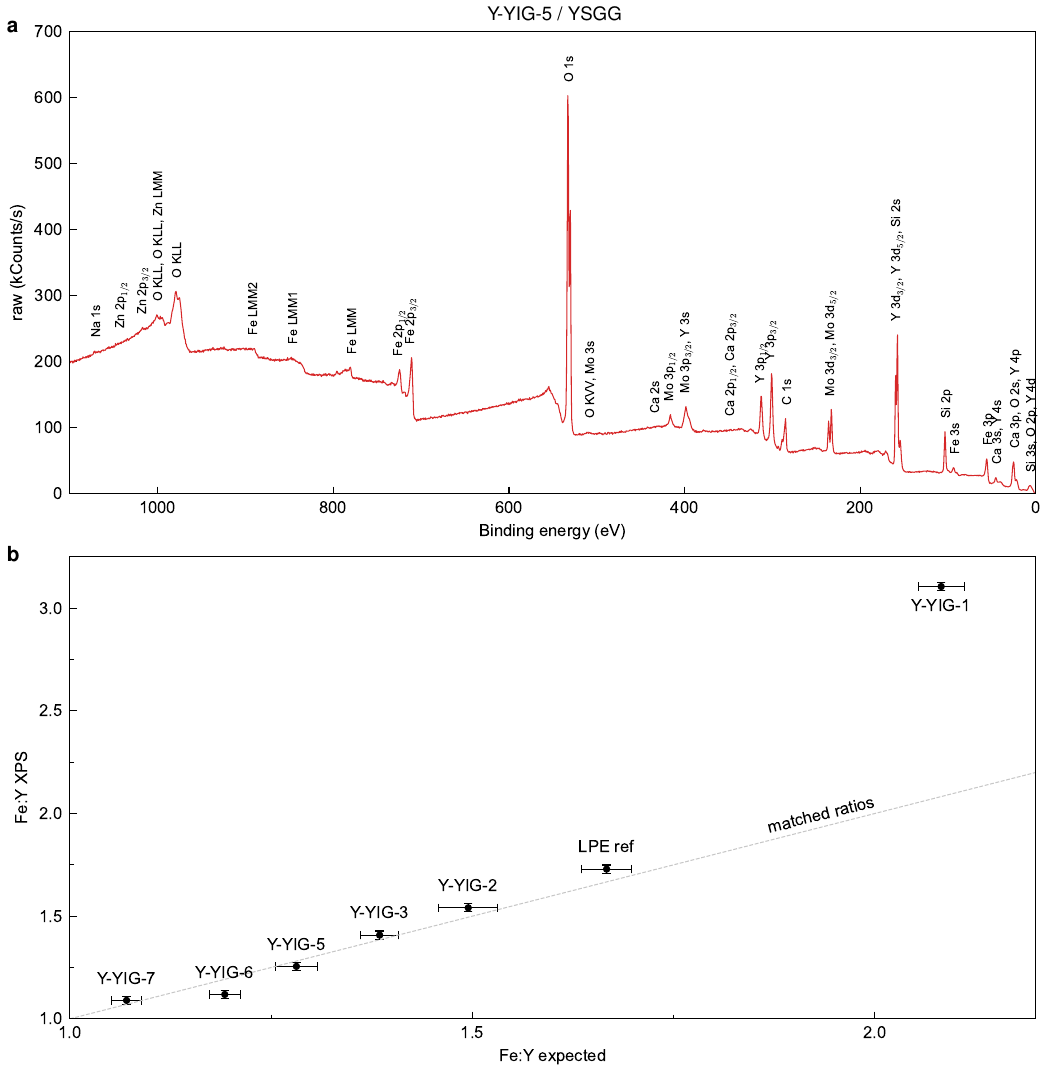}
    \caption{X-ray photoelectron spectroscopy for Y-YIG series. \pan{a}\ Broad survey scan for film Y-YIG-5 and identification of peaks by binding energies and splittings. \pan{b}\ Evolution of adjusted Fe:Y equivalent homogeneous atomic fraction for samples Y-YIG-1--7, excluding Y-YIG-4, and reference sample obtained by liquid phase epitaxy. The dashed line identifies an agreement between Fe:Y deduced from magnetometry ($k_{\rm{a}}$, $k_{\rm{d}}$) plus sputter rates or XRD lattice parameter ($x$), and Fe:Y deduced from relative peak areas in X-ray photoelectron spectroscopy analyzed as in Fig.~\ref{fig:XPSpeaks}.}
    \label{fig:XPS}
\end{figure*}

\begin{figure*}
    \centering
    \includegraphics[clip,width=7.0in]{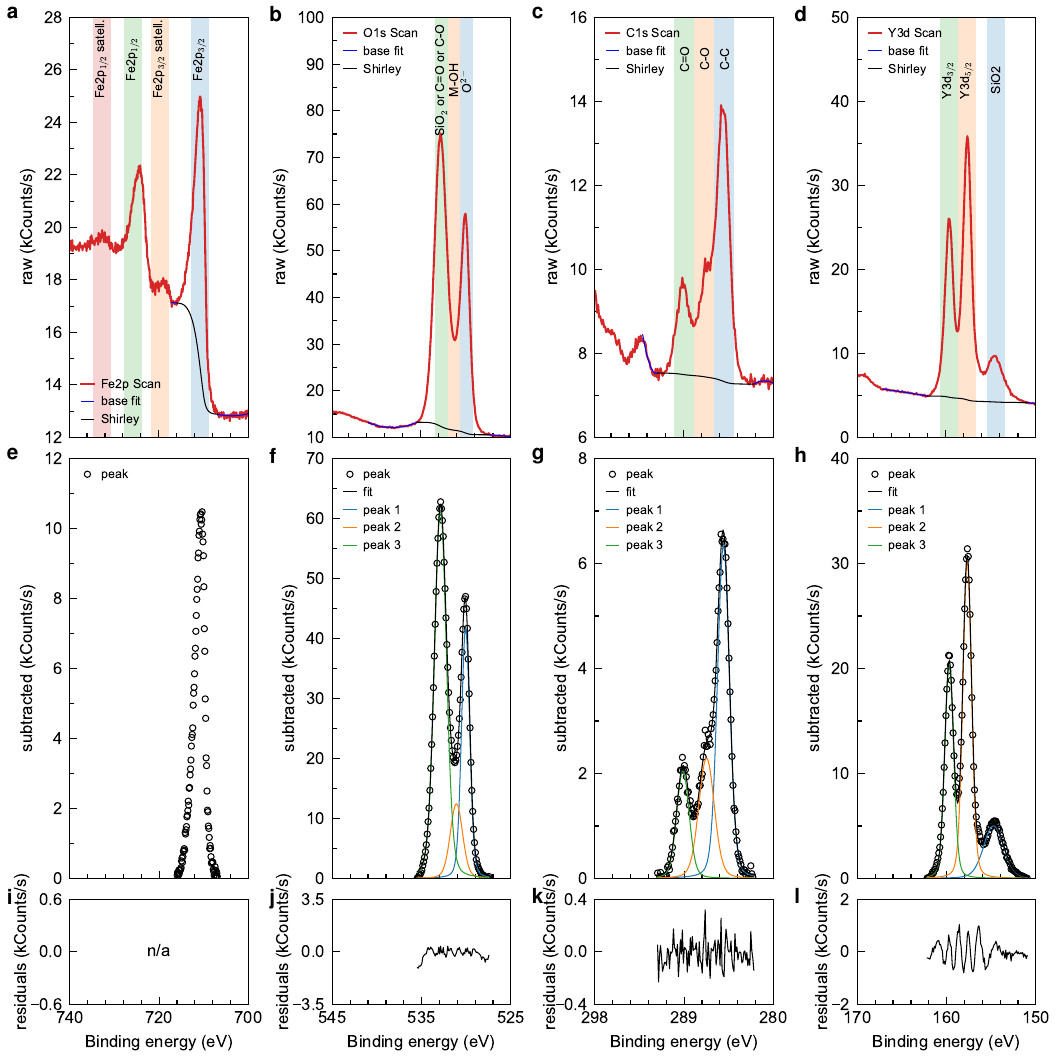}
    \caption{X-ray photoelectron peaks of interest for quantification. Detailed scans on regions of interest for \pan{a}\ Fe 2p, \pan{b}\ O 1s, \pan{c}\ C 1s and \pan{d}\ Y 3d. Likely chemical shifts and satellite peaks are identified by the background colors. The blue lines identify baseline estimations on each side of the peaks by polynomial fits of order 1--3. The black lines identify the background deduced from iterative Shirley method, proportional at each energy to the integrated photoelectron intensity from lower binding energies (to higher kinetic energies). Photoelectron signal with subtracted background for \pan{e}\ Fe 2p$_{3/2}$, \pan{f}\ O 1s, \pan{g}\ C 1s and \pan{h}\ Y 3d. The Fe 2p$_{3/2}$ peak is directly integrated, while O 1s, C 1s and Y 3d are fit with three peaks corresponding to the chemical shifts identified above in panels b--d. Black lines correspond to the sum fits, while colored lines correspond to each individual peak. Residuals of the multiple-peak fits for \pan{j}\ O 1s, \pan{k}\ C 1s and \pan{l}\ Y 3d. The numerous Fe 2p peaks are not fit individually.}
    \label{fig:XPSpeaks}
\end{figure*}

A broad survey scan in binding energies up to \SI{1100}{\electronvolt} reveals a small amount of contamination by C, Ca, Na, as well as Zn, Mo (sputter sample holder), and Si (silicon grease used for fixing samples in various prior measurements). A representative scan is shown for Y-YIG-5 in Fig.~\ref{fig:XPS}a, on which all peaks are identified. A quantitative analysis is conducted on Fe 2p$_{3/2}$, O 1s, C 1s and Y 3d$_{5/2}$ peaks, shown for Y-YIG-5 in Fig.~\ref{fig:XPSpeaks}. To this end, a Shirley background is subtracted to each peak, relying then on direct integration in the case of Fe 2p$_{3/2}$, or multiple-peak fitting for O 1s, C 1s and Y 3d$_{5/2}$ peaks, using Voigt profiles with \SI{80}{\percent} Gaussian and \SI{20}{\percent} Lorentzian. Specifically, the Fe 2p region presents a consistent doublet with satellite peaks, typical of \ce{Fe^{3+}}, whereas it shows no traces of metallic Fe. The O 1s region is fit with three partially overlapping peaks corresponding to \ce{SiO2} or \ce{C-O} or \ce{C=O} bonds, \ce{M-OH}, and \ce{O^{2-}}, respectively. The C 1s region is fit with three partially overlapping peaks corresponding to \ce{C=O} bonds, \ce{C-O} bonds and \ce{C-C} bonds (\SI{285}{\electronvolt} reference), respectively. The Y 3d region presents a consistent doublet and a Si 2s peak corresponding to the presence of \ce{SiO2}, leading to three partially overlapping peaks. The presence of many satellite peaks for Fe 2p and O 1s, and the partial contamination of the sample surfaces with some carbon and silicon compounds hinder a precise background deduction and the direct use of standard sensitivity factors for peak quantification. Standard tabulated sensitivity factors yield indeed a very unlikely Fe:Y ratio below 1 for all samples, consistent with previous investigations \cite{Rosenberg2021}. However, we may compare the peak areas to those obtained on a \ce{Y3Fe5O12} sample grown by liquid phase epitaxy, considered to feature a known reference composition. Adjusting the ratio of Fe:Y sensitivity factors to match with this reference and present scans requires the relative sensitivity factors of Fe 2p$_{3/2}$ to Y 3d$_{5/2}$ to be $\approx0.65$, when it is usually the opposite, as the relative sensitivity factor of Fe 2p is higher than Y 3d when all satellites are included and no overlapping peaks are present. Nevertheless, a general trend is obtained in terms of a relative composition between samples. Figure \ref{fig:XPS}b presents the adjusted Fe:Y equivalent homogeneous atomic fraction deduced from X-ray photoelectron spectroscopy, as a function of the expected Fe:Y stoichiometry deduced from sputter rates (or X-ray diffraction) and from magnetometry, using $x$, $k_{\rm{a}}$, $k_{\rm{d}}$ in Fe:Y~$=(5-2k_{\rm{a}}-3k_{\rm{d}}+c)/(3+x)$. Y-YIG-1 is confirmed to be Fe-rich compared to our reference sample, whereas Y-YIG-2--7 are found increasingly Y-rich compared to the reference (Y-YIG-4 is excluded from the analysis due to excessive surface pollution hindering a quantitative analysis of the Y 3d$_{5/2}$ peak). The overall agreement is therefore very good, and highlights the broad control and tunability of the actual Fe:Y stoichiometry in the co-sputtered epitaxial films of Y-YIG, while the amount of vacancies and antisites appears not to vary appreciably in the series, except for Y-YIG-1.

\begin{figure*}[h]
    \centering
    \includegraphics[clip,width=7.0in]{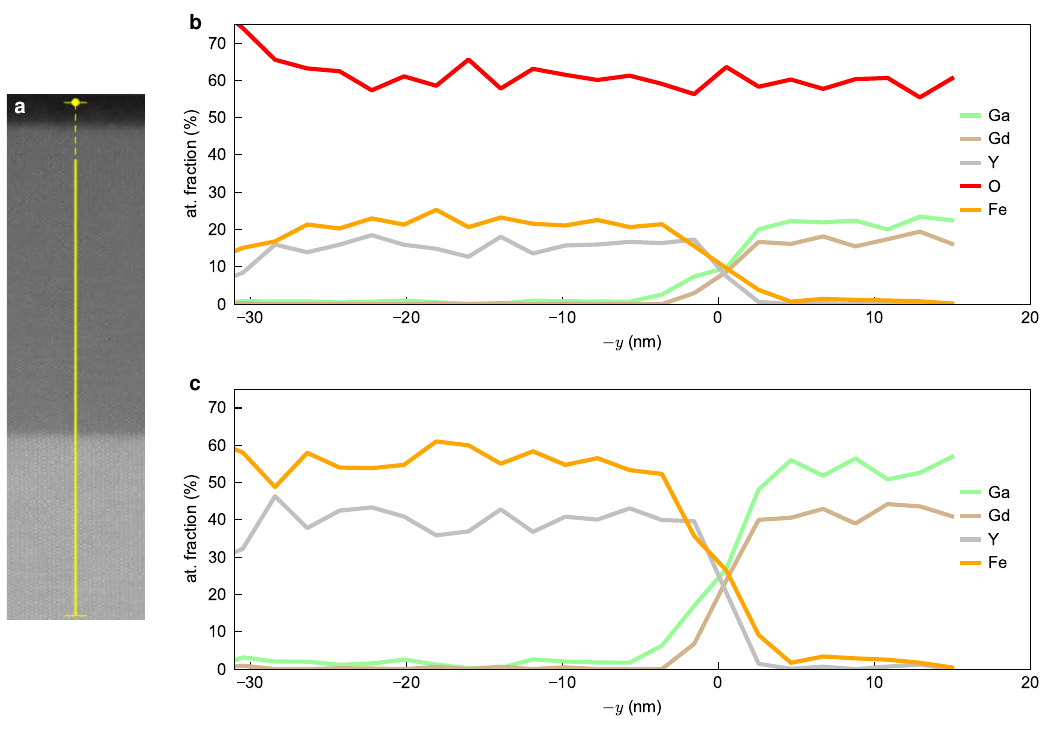}
    \caption{Energy-dispersive X-ray spectroscopy scans for Y-YIG-TEM on GGG. \pan{a}\ Detail of the STEM-HAADF contrast indicating where EDS is scanned (yellow line), \pan{b}\ atomic fractions by element, with oxygen quantification, and \pan{c}\ atomic fractions by element, without oxygen quantification.}
    \label{fig:TEM-EDS-GGG}
\end{figure*}

\begin{figure*}[h]
    \centering
    \includegraphics[clip,width=7.0in]{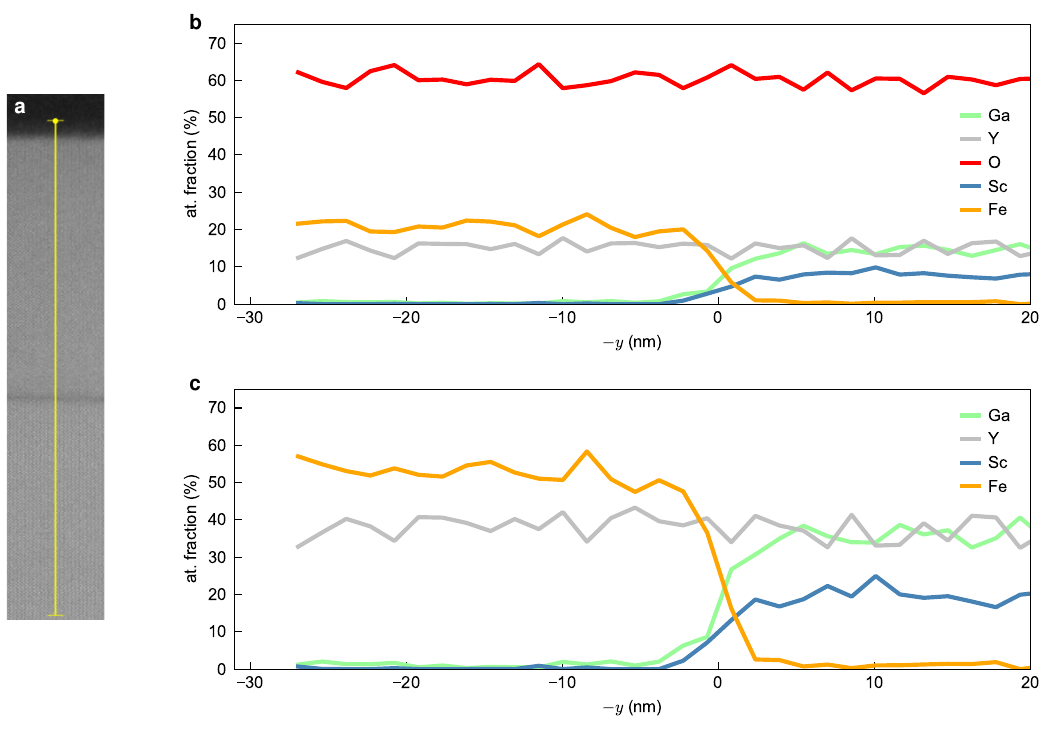}
    \caption{Energy-dispersive X-ray spectroscopy scans for Y-YIG-TEM on YSGG. \pan{a}\ Detail of the STEM-HAADF contrast indicating where EDS is scanned (yellow line), \pan{b}\ atomic fractions by element, with oxygen quantification, and \pan{c}\ atomic fractions by element, without oxygen quantification.}
    \label{fig:TEM-EDS-YSGG}
\end{figure*}

\begin{figure*}[h]
    \centering
    \includegraphics[clip,width=7.0in]{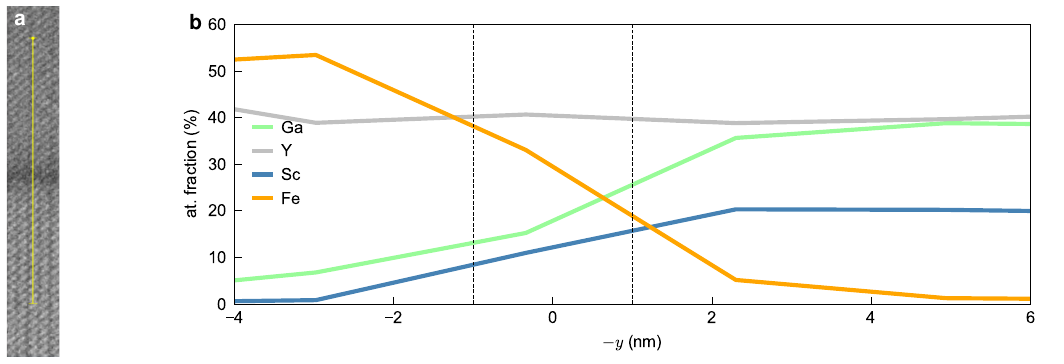}
    \caption{Energy-dispersive X-ray spectroscopy scans at the Y-YIG-TEM / YSGG interface. \pan{a}\ Detail of the HRSTEM-HAADF contrast indicating where EDS is scanned (yellow line), and \pan{b}\ atomic fractions by element, without oxygen quantification. A $\pm$~\SI{1}{\nano\meter} interdiffusion/roughness region across the interface is highlighted by the dashed lines.}
    \label{fig:TEM-EDS-interf}
\end{figure*}

To complement this investigation of stoichiometry on the whole series, we have also performed compositional analysis on the TEM cross-sectional samples Y-YIG-TEM, relying on an energy-dispersive spectrometer (EDS) Oxford Instruments Ultim Max TLE 100. The analysis of the EDS data is performed with the Oxford Aztec software. Different parts of the samples are analyzed: across the Y-YIG layer on GGG in Fig.~\ref{fig:TEM-EDS-GGG}, across the Y-YIG layer on YSGG in Fig.~\ref{fig:TEM-EDS-YSGG}, and in a narrow extent across the Y-YIG/YSGG interface in Fig.~\ref{fig:TEM-EDS-interf}. The interdiffusion region is limited to $\pm$~\SI{1}{\nano\meter} from the interface on YSGG (Figs.~\ref{fig:TEM-EDS-YSGG}b,c and \ref{fig:TEM-EDS-interf}b), mostly reflecting the substrate roughness, whereas it is slightly larger with $\pm$~\SI{2}{\nano\meter} for the interface on GGG (Figs.~\ref{fig:TEM-EDS-GGG}b,c).

Regarding the accuracy of the composition quantification, the K line emission of oxygen is generally dependent on the orientation of the sample with respect to the detector, which often leads to an overestimation or underestimation of its atomic concentration. Nevertheless, the atomic concentration of oxygen in the substrate is precisely known to be \SI{60}{\percent} and can be used as a reference. In the linescans including oxygen quantification, we find it overestimated, and then correct the sensitivities of all elements such that the oxygen concentration in the substrate is \SI{60}{\percent}. We compare in the following the compositional analysis results that either correct or simply exclude oxygen quantification. Due to the high brilliance and fluorescence of Gd, its atomic concentration also tends to be overestimated.  As expected, we find an overestimation of the concentration of Gd in the GGG substrate ($\approx$~\SI{18}{\percent} instead of \SI{15}{\percent}), and conversely, the concentration of Ga is underestimated. Comparison with other measurements on stoichiometric YIG samples confirmed that the Y:Fe ratio in the presence of Y, Fe and O is itself correctly estimated, within \SI{5}{\percent}. In summary, we expect EDS sensitivity factors to be overestimated for O, overestimated for Gd, correct for Fe and Y, likely correct for Ga, and thus for the present Y-YIG-TEM samples, the Y:Fe ratio is satisfactorily estimated.  Both methods of correcting oxygen sensitivity or excluding it from the analysis provide identical results, and lead to the estimate that Y:Fe = 1.35--1.40 in samples Y-YIG-TEM. This is in excellent agreement with Y:Fe = 1.38 expected from XPS for both Y-YIG-3 and Y-YIG-TEM, and cross-validates the stoichiometry estimations. 

The lack of precise knowledge about the amounts of Fe vacancies and Y-Fe disorder-type antisite vacancies is not specific to Y-YIG but is a general feature of all \si{\nano\meter}-thick iron garnet films grown by physical vapor deposition processes \cite{Manuilov2009,Manuilov2010,Santiso2023,Rosenberg2023}. Very little direct information is usually available regarding the trivalence of iron and the presence of vacancies in garnet films, essentially because the analysis of oxygen stoichiometry and cationic charge of iron in nanometer-thin films is challenging. Despite the indeterminacy of some point- and disorder-type defects, present in the large majority of works on iron garnet thin films, our present results demonstrate a precise and continuous tuning of the Y:Fe garnet composition, to be matched with arbitrary substrates.

\clearpage

\section{Optimization of sputter growth vs.\ oxygen flow}\label{sec:SputterOxygen}

\begin{figure*}[h]
    \centering
    \includegraphics[clip,width=7.0in]{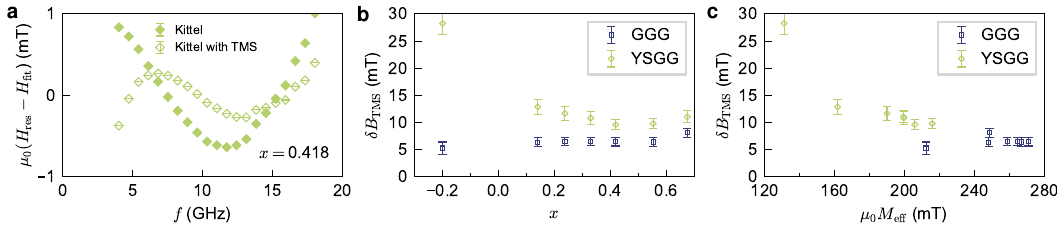}
    \caption{Additional data related to ferromagnetic resonance (FMR) at room temperature. \pan{a}\ Residuals for fits of the resonance field $\mu_0H_{\rm{res}}$ as a function of frequency. Filled symbols correspond to Kittel law, open symbols to Kittel law corrected for two-magnon scattering, for Y-YIG-5 on YSGG. \pan{b}\ Two-magnon scattering renormalization shift $\delta{}B_{\rm{TMS}}$ as a function of the substitution parameter $x$ and \pan{c}\ two-magnon scattering renormalization shift $\delta{}B_{\rm{TMS}}$ as a function of the effective anisotropy parameter $\mu_0M_{\rm{eff}}$, for films Y-YIG-1--7 films on GGG and YSGG substrates.}
    \label{fig:FMR_resids_Bk}
\end{figure*}

We begin this section by discussing the model used to fit the room-temperature FMR data. The usual form of the Kittel equation for in-plane magnetized isotropic thin film samples is given by $\omega=\gamma\mu_0\sqrt{H_{\rm{ext}}(H_{\rm{ext}}+M_{\rm{eff}})}$. In our experimental conditions, however, a fit to this equation yields significant residuals, see for instance Fig.~\ref{fig:FMR_resids_Bk}a for Y-YIG-5 on YSGG. Especially for films grown by physical vapor deposition techniques, the effects of intermixing at interfaces, sample roughness and surface contributions to the magnetic anisotropy are expected to combinely reinforce two-magnon scattering processes. These two-magnon interactions renormalize the magnon energies, causing the FMR frequency to shift downwards, and the FMR linewidth to broaden \cite{Arias1999a,Jermain2016}. The impact of two-magnon scattering on the in-plane FMR is reinforced in thin films with small exchange stiffness \cite{Arias1999a}.

In the presence of two-magnon scattering, we have \cite{Arias1999a,Jermain2016}
\begin{align}
\label{eq:invKittel_TMS}
\mu_0H_{\rm{res}}&=\sqrt{\left(\frac{\mu_0M_{\rm{eff}}}{2}\right)^2+\left(\frac{\omega}{\gamma}+\delta{}B_{\rm{TMS}}\right)^2}-\frac{\mu_0M_{\rm{eff}}}{2} \\
\label{eq:Kittel_TMS}
\omega&=\gamma\mu_0\sqrt{H_{\rm{ext}}(H_{\rm{ext}}+M_{\rm{eff}})}-\gamma\delta{}B_{\rm{TMS}}.
\end{align}
The different measurements performed at room temperature and low temperatures support the presence of a non-negligible two-magnon scattering in the samples of this Y-YIG series. Consistently, the residuals of the Kittel fit are significantly reduced after inclusion of a two-magnon scattering shift following Eqs.~\eqref{eq:invKittel_TMS} and \eqref{eq:Kittel_TMS}, see Fig.~\ref{fig:FMR_resids_Bk}a. The values of $\delta{}B_{\rm{TMS}}$ for Y-YIG-1--7 on GGG and YSGG are reported as a function of $x$ in Fig.~\ref{fig:FMR_resids_Bk}b and as a function of $\mu_0M_{\rm{eff}}$ in Fig.~\ref{fig:FMR_resids_Bk}c. The two-magnon scattering shift is expected to obey $\delta{}B_{\rm{TMS}}=\mu_0r(H_{\rm{S}})^2$ \cite{Jermain2016}, where $r$ is a substrate-dependent parameter characterizing the intensity of two-magnon scattering and notably related to roughness, and $H_{\rm{S}}$ is the interfacial- and surface-induced component of the magnetic anisotropy field $H_{\rm{ani}}$. Consistently, the shift is found different on GGG and YSGG, and is minimized for lattice-matched conditions, yielding the least strain-induced contributions to the magnetic anisotropy.

\begin{figure*}
    \centering
    \includegraphics[clip,width=7.0in]{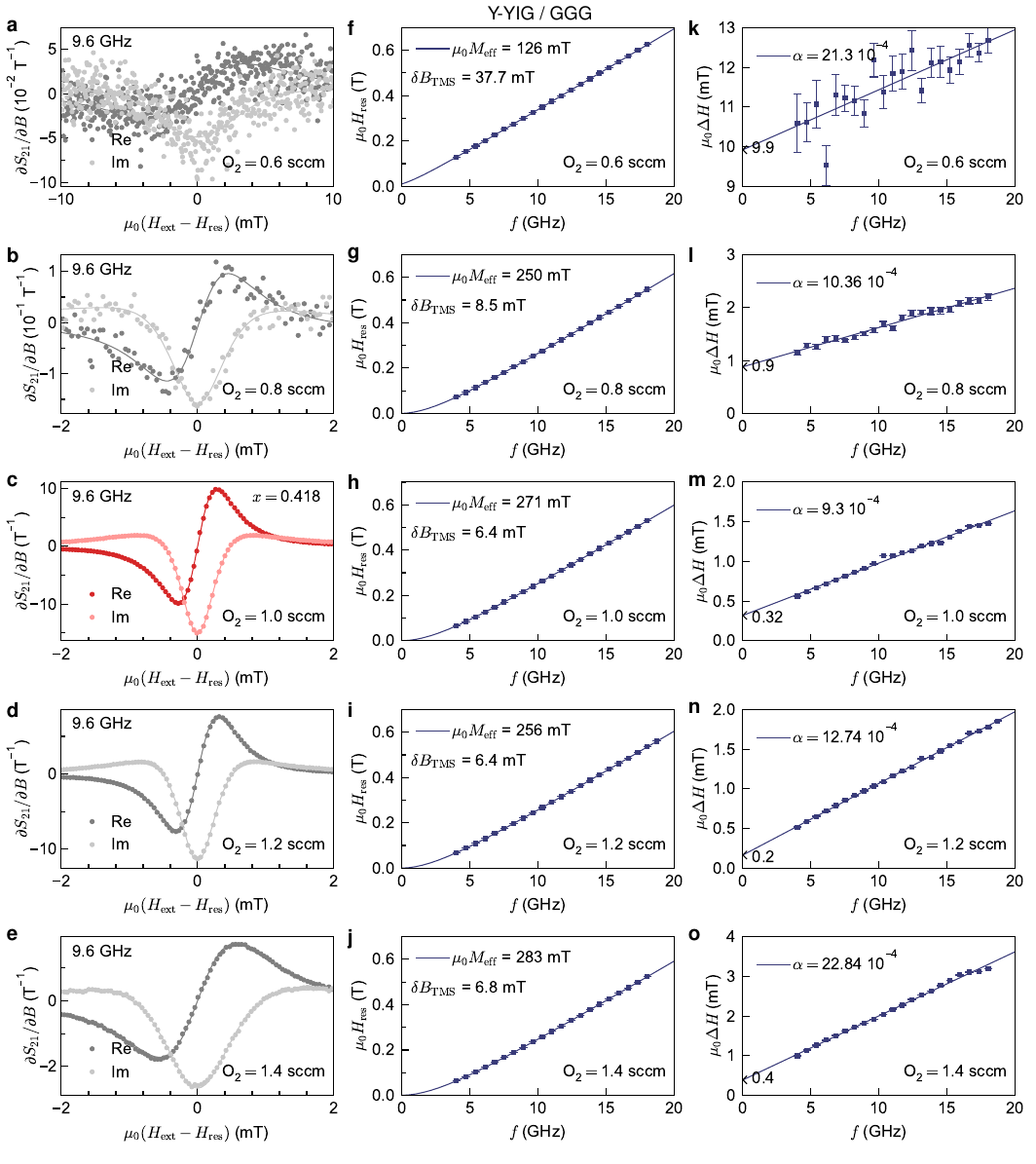}
    \caption{Ferromagnetic resonance at room temperature for films grown with 0.6--\SI{1.4}{sccm} \ce{O2} on GGG. \pan{a--e}\ Derivative of the complex microwave transmission $\partial{}S_{21}/\partial{}B$ at \SI{9.6}{\giga\hertz} as a function of external magnetic field around resonance. Solid lines are fits to in-plane magnetic susceptibility, with $\real{\partial{}S_{21}/\partial{}B}$ and $\imag{\partial{}S_{21}/\partial{}B}$ proportional to $\partial{}\chi''/\partial{}B$ and $\partial{}\chi'/\partial{}B$, respectively. \pan{f--j}\ Resonance field $\mu_0H_{\rm{res}}$ as a function of frequency, fitted to Kittel law with two-magnon scattering (solid lines) to extract $\mu_0M_{\rm{eff}}$ and $\delta{}B_{\rm{TMS}}$. \pan{k--o}\ Resonance linewidth $\mu_0\Delta{}H$ as a function of frequency and fit to $\mu_0\Delta{}H=\mu_0\Delta{}H_0+2\alpha{}\omega/\gamma$ (solid lines). \ce{O2} flow is \pan{a,f,k}\ \SI{0.6}{sccm}, \pan{b,g,l}\ \SI{0.8}{sccm}, \pan{c,h,m}\ \SI{1.0}{sccm}, \pan{d,i,n}\ \SI{1.2}{sccm}, \pan{e,j,o}\ \SI{1.4}{sccm}.}
    \label{fig:FMR_O2_GGG}
\end{figure*}

\begin{figure*}
    \centering
    \includegraphics[clip,width=7.0in]{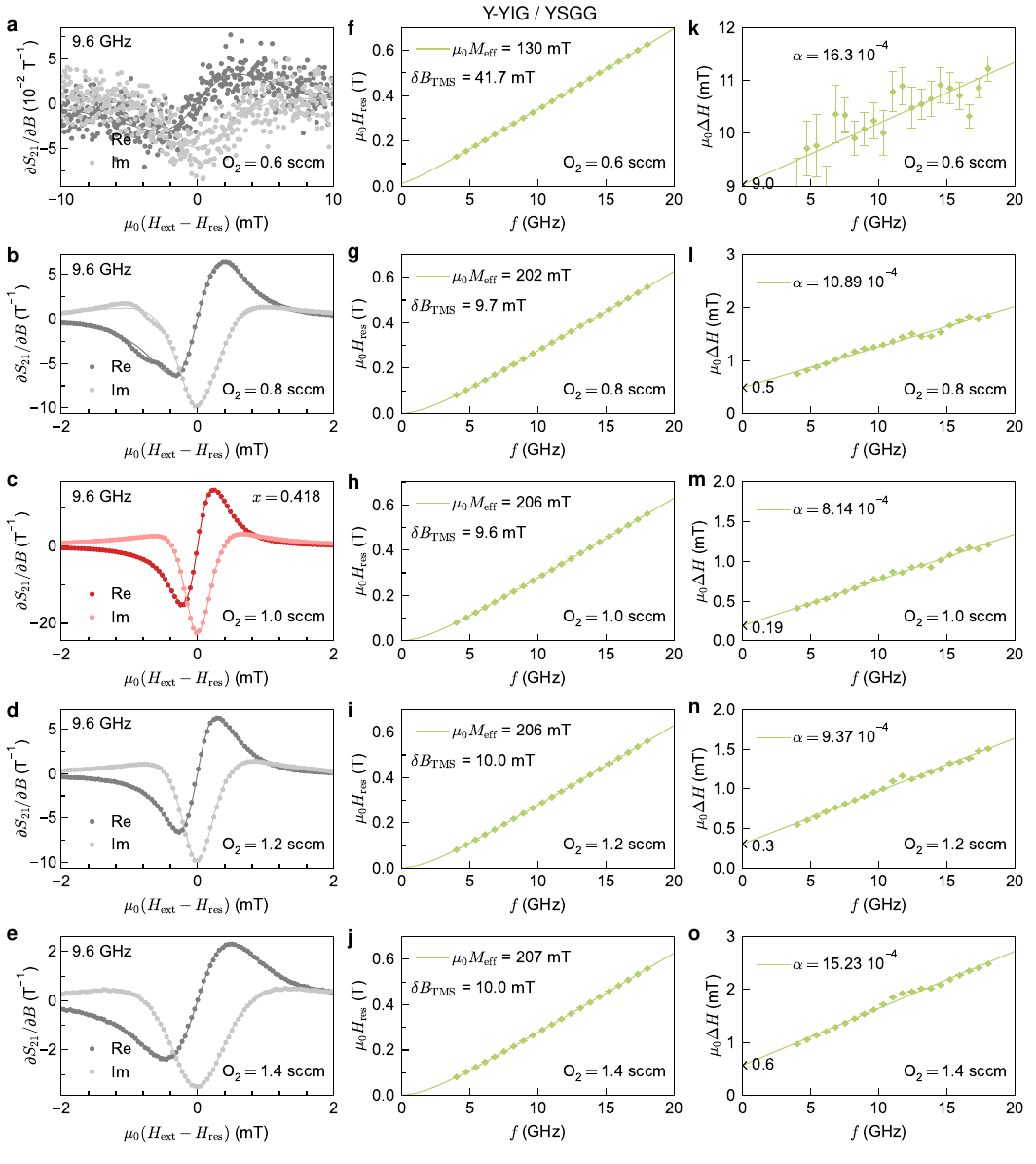}
    \caption{Ferromagnetic resonance at room temperature for films grown with 0.6--\SI{1.4}{sccm} \ce{O2} on YSGG. \pan{a--e}\ Derivative of the complex microwave transmission $\partial{}S_{21}/\partial{}B$ at \SI{9.6}{\giga\hertz} as a function of external magnetic field around resonance. Solid lines are fits to in-plane magnetic susceptibility, with $\real{\partial{}S_{21}/\partial{}B}$ and $\imag{\partial{}S_{21}/\partial{}B}$ proportional to $\partial{}\chi''/\partial{}B$ and $\partial{}\chi'/\partial{}B$, respectively. \pan{f--j}\ Resonance field $\mu_0H_{\rm{res}}$ as a function of frequency, fitted to Kittel law with two-magnon scattering (solid lines) to extract $\mu_0M_{\rm{eff}}$ and $\delta{}B_{\rm{TMS}}$. \pan{k--o}\ Resonance linewidth $\mu_0\Delta{}H$ as a function of frequency and fit to $\mu_0\Delta{}H=\mu_0\Delta{}H_0+2\alpha{}\omega/\gamma$ (solid lines). \ce{O2} flow is \pan{a,f,k}\ \SI{0.6}{sccm}, \pan{b,g,l}\ \SI{0.8}{sccm}, \pan{c,h,m}\ \SI{1.0}{sccm}, \pan{d,i,n}\ \SI{1.2}{sccm}, \pan{e,j,o}\ \SI{1.4}{sccm}.}
    \label{fig:FMR_O2_YSGG}
\end{figure*}

\begin{figure*}
    \centering
    \includegraphics[clip,width=7.0in]{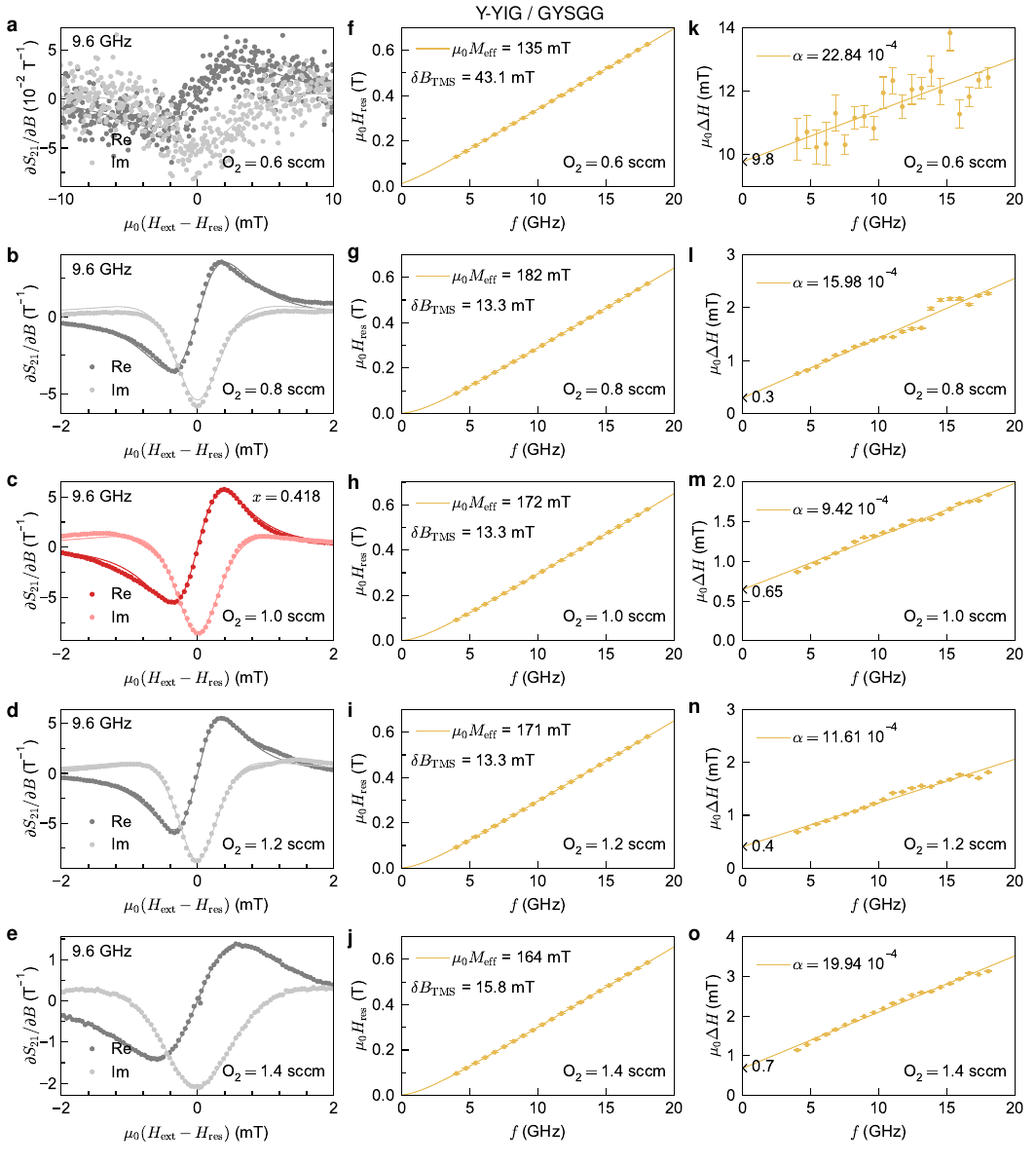}
    \caption{Ferromagnetic resonance at room temperature for films grown with 0.6--\SI{1.4}{sccm} \ce{O2} on GYSGG. \pan{a--e}\ Derivative of the complex microwave transmission $\partial{}S_{21}/\partial{}B$ at \SI{9.6}{\giga\hertz} as a function of external magnetic field around resonance. Solid lines are fits to in-plane magnetic susceptibility, with $\real{\partial{}S_{21}/\partial{}B}$ and $\imag{\partial{}S_{21}/\partial{}B}$ proportional to $\partial{}\chi''/\partial{}B$ and $\partial{}\chi'/\partial{}B$, respectively. \pan{f--j}\ Resonance field $\mu_0H_{\rm{res}}$ as a function of frequency, fitted to Kittel law with two-magnon scattering (solid lines) to extract $\mu_0M_{\rm{eff}}$ and $\delta{}B_{\rm{TMS}}$. \pan{k--o}\ Resonance linewidth $\mu_0\Delta{}H$ as a function of frequency and fit to $\mu_0\Delta{}H=\mu_0\Delta{}H_0+2\alpha{}\omega/\gamma$ (solid lines). \ce{O2} flow is \pan{a,f,k}\ \SI{0.6}{sccm}, \pan{b,g,l}\ \SI{0.8}{sccm}, \pan{c,h,m}\ \SI{1.0}{sccm}, \pan{d,i,n}\ \SI{1.2}{sccm}, \pan{e,j,o}\ \SI{1.4}{sccm}.}
    \label{fig:FMR_O2_GYSGG}
\end{figure*}

To then provide an example of the procedure to optimize the FMR linewidth resulting from the growth of Y-YIG by rf co-sputtering, we report below our study based on room-temperature FMR of films grown with varied oxygen flows in the chamber, in the range 0.6--\SI{1.4}{sccm}, while \ce{Ar} flow is maintained at \SI{100}{sccm}. The rf powers for \ce{Y2O3} and \ce{Fe2O3} are fixed at the ratio providing $x\approx0.42$, in the aim of obtaining films that are lattice-matched with YSGG. We proceed with depositions on GGG, YSGG and GYSGG in order to establish a more complete picture and to observe potential differences in the growth on different substrates. This study is summarized by Figs.~\ref{fig:FMR_O2_GGG}--\ref{fig:FMR_O2_GYSGG}, which present FMR peaks measured at \SI{9.6}{\giga\hertz}, resonance field $\mu_0H_{\rm{res}}$ as a function of frequency, and resonance linewidth $\mu_0\Delta{}H$ as a function of frequency, for growth on GGG, YSGG and GYSGG substrates, respectively. The graphs also display best fit results. Beyond the optimal linewidth at \SI{1.0}{sccm} \ce{O2} for \SI{100}{sccm} \ce{Ar} gas flows, there are variations of the two-magnon scattering frequency shift extracted from Kittel fits, which again is minimized in the best growth conditions that reduce sample roughness and inhomogeneities.

\clearpage

\section{Low-temperature FMR data for all recorded temperatures}\label{sec:lowTFMR}

\begin{figure*}[h]
    \centering
    \includegraphics[clip,width=7.0in]{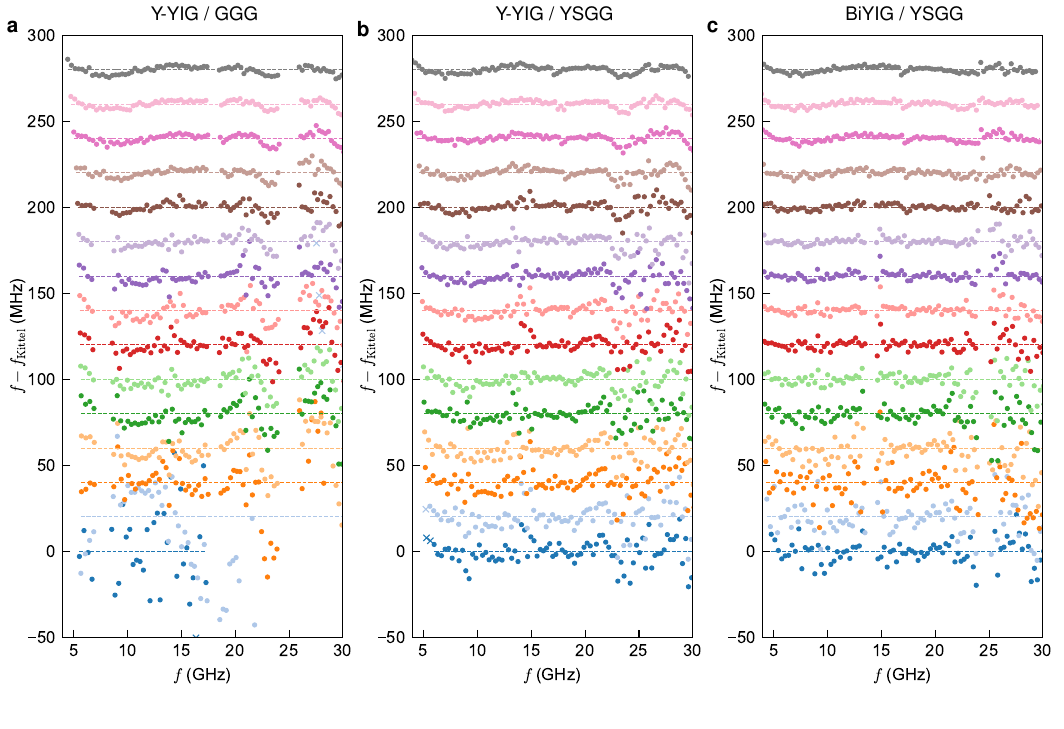}
    \caption{Analysis of residuals for ferromagnetic resonance (FMR) at all recorded temperatures within 10--\SI{300}{\kelvin} for films Y-YIG-5 on GGG and YSGG and BiYIG on YSGG. Resonance frequency offset from Kittel fit with $H_{\rm{K}}$ term as a function of resonance frequency for \pan{a}\ Y-YIG-5 on GGG, \pan{b}\ Y-YIG-5 on YSGG and \pan{c}\ BiYIG on YSGG. Each temperature is offset by \SI{20}{\mega\hertz} (dashed lines). Color scheme is identical to Fig.~\ref{fig:lowTFMRextended}.}
    \label{fig:lowTFMRresiduals}
\end{figure*}

\begin{figure*}
    \centering
    \includegraphics[clip,width=7.0in]{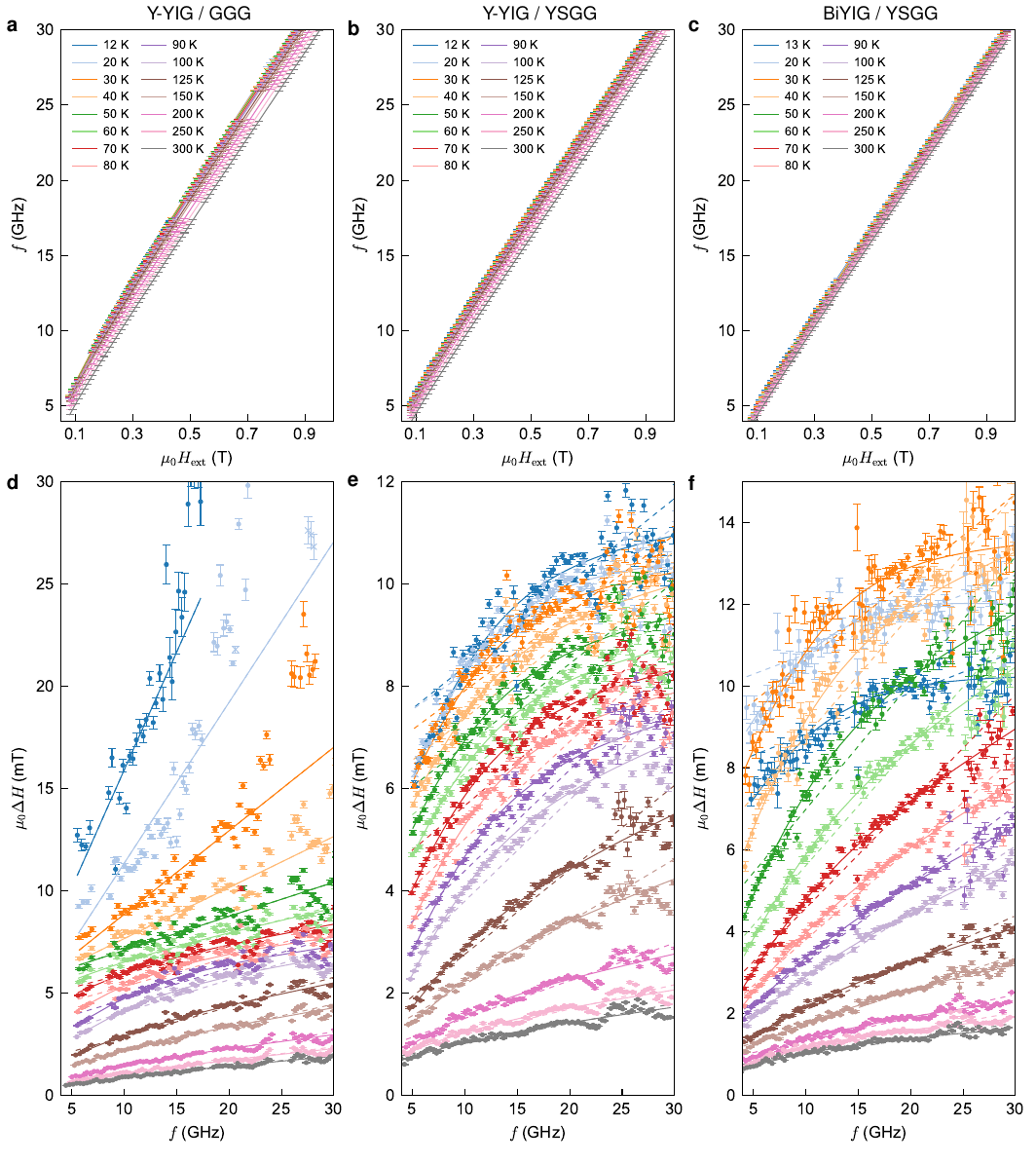}
    \caption{Ferromagnetic resonance (FMR) at all recorded temperatures within 10--\SI{300}{\kelvin} for films Y-YIG-5 on GGG and YSGG and BiYIG on YSGG. Resonance frequency as a function of applied magnetic field for \pan{a}\ Y-YIG-5 on GGG, \pan{b}\ Y-YIG-5 on YSGG and \pan{c}\ BiYIG on YSGG. Solid lines are fits to Kittel law with in-plane anisotropy term. FMR linewidth as a function of frequency for \pan{d}\ Y-YIG-5 on GGG, \pan{e}\ Y-YIG-5 on YSGG and \pan{f}\ BiYIG on YSGG. Solid lines are fits to a slow-relaxer impurities model, dashed lines are fits to $\mu_0\Delta{}H=\mu_0\Delta{}H_0+2\alpha{}\omega/\gamma$.}
    \label{fig:lowTFMRextended}
\end{figure*}

This supplementary section contains the extended versions of the graphs displayed in Figs.~5 and 6 of the main. As in the previous section, we first motivate the model used for the FMR frequency analysis. Equations \eqref{eq:invKittel_TMS} and \eqref{eq:Kittel_TMS}, including two-magnon scattering renormalization shift, fit well the data at room-temperature but cannot be used for the whole temperature range. The renormalization shift extracted from the fits changes sign from positive to negative between room temperature and low temperatures, for both for Y-YIG on YSGG and BiYIG on YSGG. However, a negative shift is not accounted for by the two-magnon scattering model \cite{Arias1999a}, and at least one other mechanism contributes to this frequency shift, for instance, a magneto-crystalline anisotropy component within the plane, along the measurement direction. We model this phenomenologically by including an additional in-plane anisotropy term along the measurement direction, in the following modified Kittel law $\omega=\gamma\mu_0\sqrt{(H_{\rm{ext}}+H_{\rm{K}})(H_{\rm{ext}}+H_{\rm{K}}+M_{\rm{eff}})}$. We highlight that this form is very close to Eqs.\ \eqref{eq:invKittel_TMS} and \eqref{eq:Kittel_TMS}, and provides nearly identical fit residuals, where a positive $\delta{}B_{\rm{TMS}}$ corresponds to a negative value of $\mu_0H_{\rm{K}}$. Because the modified Kittel laws with two-magnon scattering and with anisotropy are mathematically very similar, it is not possible to separate both contributions in our experimental conditions. As such, the $\mu_0H_{\rm{K}}$ term combines two-magnon scattering with other possible magnetic anisotropy contributions. Figure~\ref{fig:lowTFMRresiduals} presents the residuals of the fit at all temperatures for each system, which display no significant deviations from this phenomenological Kittel law with added in-plane anisotropy term. Figure~\ref{fig:lowTFMRextended} combines the dependence of the resonance frequency on the applied field (with fits to Kittel law) and the dependence of the FMR linewidth on frequency (with fits to Gilbert and slow-relaxer impurities models), for all temperatures investigated, with 12, 20, 30, 40, 50, 60, 70, 80, 90, 100, 125, 150, 200, \SI{300}{\kelvin}. 

In the case of Y-YIG-5 on GGG, the very large increase of magnetic linewidth visible at lowest temperatures $T\leq$~\SI{40}{\kelvin} cannot be explained by either model of two-magnon scattering or slow-relaxer impurities, see the zero values for $A(T)$ and $\Gamma(T)$ in Fig.~\ref{fig:correlation}e, because the slope becomes steeper above a finite frequency, which is not compatible with either model. Here, the linewidth does not follow a simplified $\omega\tau/[1+(\omega\tau)^2]$ dependence on frequency like what is expected for slow-relaxer impurities, suggesting an interfacial coupling effect instead. Nevertheless, the peaked behavior of the linewidth, related to the paramagnetic moment of the GGG, appears as a shift of the linewidth maximum with temperature, which is found at $\approx$~15--\SI{20}{\giga\hertz} at \SI{12}{\kelvin}, $\approx$~20--\SI{25}{\giga\hertz} at \SI{20}{\kelvin} and $>$~\SI{30}{\giga\hertz} at \SI{30}{\kelvin}. In the case of BiYIG on YSGG, the cancellation of the slope in the high-frequency part of the fit shows that two-magnon scattering alone is not sufficient to explain the data.

\clearpage

\section{Apparent correlation between FMR parameters}\label{sec:SOC}

\begin{figure*}[h]
    \centering
    \includegraphics[clip,width=7.0in]{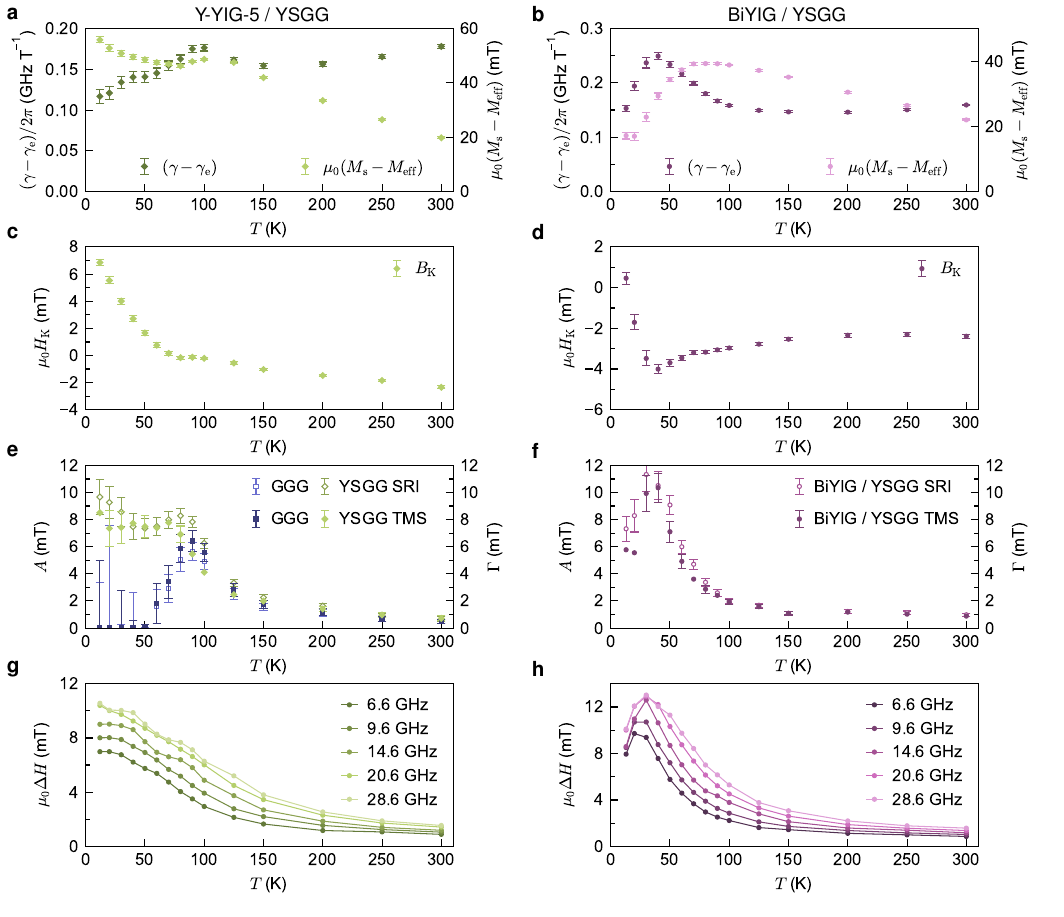}
    \caption{For \pan{a}\ Y-YIG-5 and \pan{b}\ BiYIG on YSGG, difference to electron gyromagnetic ratio $(\gamma-\gamma_{\rm{e}})/(2\pi)$ and effective out-of-plane uniaxial magnetic anisotropy term $\mu_0\left(M_{\rm{s}}-M_{\rm{eff}}\right)$, which excludes both anisotropy contribution due to the shape anisotropy of the film (or equivalently, dipolar interactions) and in-plane component of (magnetocrystalline) anisotropy terms. For \pan{c}\ Y-YIG-5 and \pan{d}\ BiYIG on YSGG, phenomenological in-plane anisotropy term $\mu_0H_{\rm{K}}$ obtained from Kittel fit to the field dependence of the resonance frequency. For \pan{e}\ Y-YIG-5 and \pan{f}\ BiYIG on YSGG, slow-relaxer impurities linewidth pre-factor $A(T)$ and two-magnon scattering linewidth pre-factor $\Gamma(T)$, obtained from the fits to each of these two models. Panel e also displays data for Y-YIG-5 on GGG for comparison. For \pan{g}\ Y-YIG-5 and \pan{h}\ BiYIG on YSGG, FMR linewidth $\mu_0\Delta{}H$ at several frequencies, as a function of temperature (reported from Figs.~5 and 6 of the main).}
    \label{fig:correlation}
\end{figure*}

This supplementary section provides additional graphs to discuss the apparent correlation between gyromagnetic ratio, uniaxial and magnetocrystalline anisotropy, and magnetic resonance linewidth in the temperature dependence of the films of both \ce{Y_{3}(Y_{x}Fe_{5-x})O_{12}} and \ce{Bi_{0.8}Y_{2.2}Fe_{5}O_{12}} under investigation. The gyromagnetic ratio of the ferromagnetic resonance is defined as $\gamma=g\mu_{\rm{B}}/\hbar$, where $g$ is the spectroscopic g-factor. The difference with $\gamma_{\rm{e}}=g_{\rm{e}}\mu_{\rm{B}}/\hbar$ is assigned to an orbital component of the magnetization coupling to the electron spin magnetization. Besides, we obtain the out-of-plane anisotropy field $\mu_0H_{\rm{ani}}=\mu_0\left(M_{\rm{s}}-M_{\rm{eff}}\right)$ by interpolation from the saturation magnetization $M_{\rm{s}}$ and effective magnetization $M_{\rm{eff}}$ shown in main. These are compared to each other in Figs.~\ref{fig:correlation}a,b for \ce{Y_{3}(Y_{x}Fe_{5-x})O_{12}} and \ce{Bi_{0.8}Y_{2.2}Fe_{5}O_{12}}, respectively. The phenomenological in-plane anisotropy field $\mu_0H_{\rm{K}}$ introduced in the Kittel relation is shown in Figs.~\ref{fig:correlation}c,d. 

Figs.~\ref{fig:correlation}e,f summarize the temperature dependence of the prefactors $A(T)$ for the slow-relaxer impurities and $\Gamma(T)$ for the two-magnon scattering contribution, obtained from fits to models on both systems. Analyses with both models feature very similar prefactors and cannot directly discriminate between them. We note however that an almost zero damping at low temperatures is deduced from fits to the two-magnon scattering model, which shows it cannot be the only contribution in this temperature range. In the case of BiYIG, $A(T)$ or $\Gamma(T)$ appear correlated with $\gamma-\gamma_{\rm{e}}$. We note that a peaked contribution of $\mu_0\Delta{}H\approx$ 8--\SI{12}{\milli\tesla} to the linewidth in BiYIG is compatible with the \SI{99.9}{\percent} quoted purity of the targets used for this work, of which not all \SI{0.1}{\percent} impurities would be rare-earths, referring to the linewidth contribution of \SI{80}{\milli\tesla} for \SI{0.1}{\percent} Tb, found by Dillon and Nielsen.

For comparison with Figs.~\ref{fig:correlation}a--f, similar graphs as in main for the corresponding FMR linewidth $\mu_0\Delta{}H$ are reported in Figs.~\ref{fig:correlation}g,h. The linewidth noticeably increases where $(\gamma-\gamma_{\rm{e}})$ and $\mu_0H_{\rm{ani}}$ show the largest deviations, and the respectively almost monotonic and peaked shapes of the curves in Figs.~\ref{fig:correlation}a,b are consistently found in Figs.~\ref{fig:correlation}e,f and \ref{fig:correlation}g,h. In addition to possible effects of strain and variations in the intersite spin-spin couplings, it seems that a spin-orbit related parameter may explain at least a common part in the dependences with temperature. Further studies distinguishing quantitatively the different contributions to the FMR parameters and relating them to impurity composition and precise oxygen stoichiometry in these systems shall provide more understanding of these observations and a means to mitigate the increase in the FMR linewidth.

\clearpage

\bibliography{supplement}